\documentclass[fleqn,usenatbib, onecolumn]{mnras}

\usepackage{newtxtext,newtxmath}

\usepackage[T1]{fontenc}

\DeclareRobustCommand{\VAN}[3]{#2}
\let\VANthebibliography\thebibliography
\def\thebibliography{\DeclareRobustCommand{\VAN}[3]{##3}\VANthebibliography}

\usepackage{graphicx}	
\usepackage{amsmath}	
\usepackage{orcidlink}

\usepackage{epstopdf}
\usepackage{ulem}
\usepackage{times}
\usepackage{graphicx}
\usepackage[usenames,dvipsnames]{pstricks}
\usepackage{psfrag}
\usepackage{lipsum}
\usepackage{natbib}
\usepackage{textcase}

\usepackage[lofdepth,lotdepth]{subfig} 

\def\aj{AJ}
\def\apj{ApJ}
\def\apjl{ApJ}
\def\apjs{ApJS}
\def\ao{Appl.~Opt.}
\def\aap{A\&A}
\def\mnras{MNRAS}
\def\nat{Nature}
\def\physrep{Phys.~Rep.}

\newcommand{\be}{\begin{equation}}
\newcommand{\ee}{\end{equation}}
\newcommand{\bary}{\begin{eqnarray}}
\newcommand{\eary}{\end{eqnarray}}

\title[Fermi-LAT Flares in Stratified Afterglows]{An Explanation of GRB Fermi-LAT Flares and High-Energy Photons in Stratified Afterglows}

\author[N. Fraija et al.]{
Nissim Fraija\,\orcidlink{0000-0002-0173-6453},$^{1}$\thanks{E-mail: nifraija@astro.unam.mx}
Boris Betancourt Kamenetskaia\,\orcidlink{0000-0002-2516-5739},$^{2,3}$,
Antonio Galv\'an-G\'amez\,\orcidlink{0000-0001-5193-3693},$^{1}$,
Peter Veres\,\orcidlink{0000-0002-2149-9846$},$^{4,5}$,
\newauthor Rosa L. Becerra\,\orcidlink{0000-0002-0216-3415},$^{6}$,
Simone Dichiara\,\orcidlink{0000-0001-6849-1270}$^{7}$,
Maria G. Dainotti\,\orcidlink{0000-0003-4442-8546},$^{8,9,10}$,
Francisco Lizcano$^{1}$,
Edilberto Aguilar-Ruiz$^{1}$
\\
$^{1}$Instituto de Astronom\' ia, Universidad Nacional Aut\'onoma de M\'exico,\\ Circuito Exterior, C.U., A. Postal 70-264, 04510 M\'exico City, M\'exico\\
$^{2}$Technical University of Munich, TUM School of Natural Sciences, Physics Department, James-Franck-Str 1, 85748 Garching, Germany\\
$^{3}$Max-Planck-Institut f\"ur Physik (Werner-Heisenberg-Institut), F\"ohringer Ring 6, 80805 Munich, Germany\\
$^{4}$Department of Space Science, University of Alabama in Huntsville, Huntsville, AL 35899, USA\\
$^{5}$Center for Space Plasma and Aeronomic Research, University of Alabama in Huntsville, Huntsville, AL 35899, USA\\
$^{6}$Instituto de Ciencias Nucleares, Universidad Nacional Aut\'onoma de M\'exico, Apartado Postal 70-264, 04510 M\'exico, CDMX, M\'exico\\
$^{7}$Department of Astronomy and Astrophysics, The Pennsylvania State University, 525 Davey Lab, University Park, PA 16802, USA\\
$^{8}$National Astronomical Observatory of Japan, 2-21-1 Osawa, Mitaka, Tokyo 181-8588, Japan\\
$^{9}$Space Science Institute, Boulder, CO, USA\\
$^{10}$The Graduate University for Advanced Studies, SOKENDAI, Shonankokusaimura, Hayama, Miura District, Kanagawa 240-0193, Japan\\
}

\date{Accepted XXX. Received YYY; in original form ZZZ}

\pubyear{2023}

\begin{document}
\label{firstpage}
\pagerange{\pageref{firstpage}--\pageref{lastpage}}
\maketitle

\begin{abstract}
The second {\itshape Fermi}/LAT gamma-ray burst (GRB) catalog (2FLGC) spanning the first decade of operations by the LAT collaboration was recently released. The closure relations of the synchrotron forward shock (FS) model are not able to reproduce a sizeable portion of the afterglow-phase light curves in this collection, indicating that there may be a large contribution from some other mechanism. Recently,  synchrotron self-Compton (SSC) light curves from the reverse shock (RS) regions were derived in the thick- and thin-shell regime for a constant-density medium, and it was demonstrated that analytical light curves could explain the~GeV  flare observed in several bursts from 2FLGC, including GRB 160509A. Here, we generalise the SSC RS scenario from the constant density to a stratified medium, and show that this contribution helps to describe the early light curves exhibited in some {\itshape Fermi}/LAT-detected bursts. As a particular case, we model a sample of eight bursts that exhibited a short-lasting emission with the synchrotron and SSC model from FS and RS regions, evolving in a stellar-wind environment, constraining the microphysical parameters, the circumburst density, the bulk Lorentz factor, and the fraction of shock-accelerated electrons. We demonstrate that the highest-energy photons can only be described by the SSC from the forward-shock region.
\end{abstract}




\section{Introduction}

Gamma-ray bursts (GRBs) are among the most powerful short-duration explosions in the universe. These sources are often discovered as transient gamma-ray flashes that last from milliseconds to hours and emit $\sim 10^{51} - 10^{54}$ erg \citep[for a review, see][]{2015PhR...561....1K}. Most GRBs are usually detected in the keV-MeV energy range, and their classification is determined by the duration of the first event called "prompt emission" ($T_{90}$).\footnote{$T_{90}$ represents the time  during which a GRB emits between $5\%$ and $95\%$ of the total observed counts from its prompt emission.}  Typically, GRBs are modelled by the empirical Band function \citep{1993ApJ...413..281B}. After the first event, a second event, known as afterglow, is observed from radio wavelengths up to TeV gamma-rays and is explained using the fireball scenario \citep[e.g., see][]{1978MNRAS.183..359C}. When a relativistic jet comes into contact with the circumburst medium, the basic model predicts the formation of two shock waves: a forward shock \citep[FS;][]{1999ApJ...513..669K, 2002ApJ...568..820G, 1998ApJ...497L..17S,Becerra2019c} and a reverse shock \citep[RS;][]{2000ApJ...545..807K, 2003ApJ...597..455K, 2016ApJ...818..190F,Becerra2019b}. Both shocks are predicted to produce synchrotron and synchrotron-self Compton (SSC) photons due to the acceleration and cool down of the non-thermal electron distribution. Synchrotron photons are scattered to energies larger than tens of~GeV  by the SSC process \citep{2001ApJ...559..110Z, 2019ApJ...883..162F}, even if the highest photon energy produced by the synchrotron process is $\sim 5 - 10 ~{\rm~GeV}$ \citep{2010ApJ...718L..63P, 2009ApJ...706L.138A, 2011MNRAS.412..522B}.

According to multi-wavelength data modelling during the afterglow phase, multiple GRBs are compatible with a circumburst medium with a density profile $\propto r^{-k}$ with a PL index in the range $0\leq k<3$. For example, \cite{2013ApJ...776..120Y} studied the evolution of the FS and RS emission in a circumstellar medium represented by $\propto r^{-k}$. They tested their afterglow model on 19 bursts and found that the density profile index ranged from $ 0.4 \leq {\rm k}\leq  1.4$. \cite{2013ApJ...774...13L}, who examined a larger sample of 146 bursts, reported a range consistent with Yi et al.'s result. Since black hole accretion of a stellar envelope produces the plateau phase in X-ray light curves, \cite{2008Sci...321..376K} constructed a density profile index with $k>2$.   Using a synchrotron afterglow model in the adiabatic domain without energy injection, \cite{2013ApJ...779L...1K} evaluated the multiwavelength data of GRB 130427A, one of the most intense bursts with redshift $z<1$. The authors concluded that these observations were consistent with a density profile index that ranged between $\rm k=0$ (constant-density medium) and $\rm k=2$ (wind-like medium). \cite{2020A&A...639L..11I} modelled the X-ray observations of the nearby SN 2020bvc. According to the authors, these afterglow observations were consistent with synchrotron emission evolving in a circumburst medium with a density profile index of ${\rm k}=1.5$. A complete analysis of {\itshape Fermi} and Swift observations by \cite{2018ApJ...863..138A, 2011ApJ...732...29C, 2013ApJ...763...71A} found that there is a preference for LAT-detected bursts to evolve in a wind-like circumstellar medium. This inference was due to the observed behaviour of the cooling frequency in the synchrotron spectrum in several GRBs. It was noted that it evolved in time to larger energies, a behaviour that contrasted with the predicted evolution to lower energies, as expected from a constant-density medium. As such, the afterglow was best fitted with a low-density wind-like model \citep{2018ApJ...863..138A}. The bursts in question were GRB 090323, GRB 090328, GRB 090902B, GRB 090926A \citep{2011ApJ...732...29C}, and GRB 110731A \citep{2013ApJ...763...71A}.

\cite{Ajello_2019} presented the Second \textit{Fermi}/LAT GRB Catalogue (2FLGC), covering the first decade of operations (from 2008 to 2018, August 4). The data set contains 169 bursts with high-energy emissions greater than $100$~MeV, including 29 bursts higher than $10$~GeV. Additionally, this data set includes a sample of 86 bursts with temporarily extended emission described with a power law (PL) function and a broken power law (BPL) function. A subset of these events (21 GRBs) showed a break in the LAT light curve between 63 and 1250~s. Although long-lasting emission is commonly interpreted as synchrotron radiation from external FSs \citep{2009MNRAS.400L..75K,2010MNRAS.409..226K}, not all LAT light curves fulfil the relation between the temporal and spectral indices' evolutions or the closure relations predicted if the FS dominates the emission \citep{1998ApJ...497L..17S}. Furthermore, data suggests that the entire~GeV  light curve for multiple bursts cannot be satisfactorily described by FS synchrotron radiation alone \citep{2011MNRAS.415...77M}.

\cite{2020ApJ...905..112F} derived the SSC light curves from the RS region for a constant-density medium. They demonstrated that analytical SSC light curves from the RS region could explain the~GeV  flares observed in several bursts from 2FLGC. The late-time steepening of LAT afterglow observations was caused by the synchrotron cooling break passing across the LAT energy band when the FS evolves in a uniform environment. Here, we generalise the SSC light curves from the RS for a uniform-density medium to a stratified environment with a profile density $\propto r^{\rm -k}$ with ${\rm k}$ lying in the range of $0\leq {\rm k} < 3$. The paper is structured as follows: in Section \S\ref{sec2}, we derive the SSC light curves evolving in a stratified medium. Section \S\ref{sec3} presents the analysis of {\itshape Fermi}/LAT data,  the derivation of the {\itshape Fermi}/LAT light curves and the modelling of the multiwavelength observations of a sample of eight GRBs with our current model for ${\rm k=2}$. In addition, we present the results and a discussion. Section \S\ref{sec4} provides a summary of our work and some concluding remarks. We use unprimed (primed) quantities in the observer (comoving) frame. The subindexes ``f" and ``r" refer to the quantities derived in the FS and RS regions, respectively.
\section{SSC light curves from the reverse shock evolving in a stratified environment}\label{sec2}
When the relativistic jet encounters the stratified medium, a RS is formed, and it propagates back into the outflow \citep{1997ApJ...476..232M, 1999ApJ...517L.109S, 2000ApJ...542..819K}. We consider a density profile of the stratified medium $n(r) =A_{\rm k_s} r^{-k}$ with $0\leq k < 3$, and use the value of $A_{\rm k_s}=\frac{\dot{M}_{\rm W}}{4\pi m_p\,v_{\rm W}}A_{\rm k}=3.0\times 10^{35}\,{\rm cm^{-1}}A_{\rm k}$ for ${\rm k=2}$ to report the quantities of the theoretical model, unless otherwise indicated. The terms $v_{\rm W}$ and $\dot{M}_{\rm W}$ correspond to the wind velocity and the mass-loss rate, respectively,  $m_p$ is the proton mass, and $A_{\rm k}$ is the density parameter.  The evolution of the bulk Lorentz factor is estimated as $\Gamma=\left(\frac{\ell}{\Delta} \right)^\frac{3-k}{2(4-k)}$, where $\ell=\left(\frac{(3-k)E}{4\pi  m_p c^2\,A_{\rm k_s}}\right)^{\frac{1}{3-k}}$ and $\Delta=2c(1+z)^{-1}\,t_{\rm x}$ are the Sedov length and the observed width of the shell, respectively.\footnote{The term $t_{\rm x}$ is the shock crossing time, $z$ is the redshift, and $c$ is the speed of light. The isotropic gamma-ray energy, $E_{\rm \gamma, iso}$, determines the isotropic equivalent kinetic energy, $E$, through the kinetic efficiency $\eta=E_{\rm \gamma, iso}/(E+E_{\rm \gamma, iso})$.} The dynamics of the RS is categorized into two distinct scenarios: the thick-shell regime for $\Gamma_c \lesssim \Gamma$ or $t_{\rm x} \lesssim T_{90}$, and in the thin-shell regime for $\Gamma < \Gamma_c$ or $T_{90} < t_{\rm x}$, where  $\Gamma_c$ is the critical Lorentz factor, which is calculated considering the evolution of the bulk Lorentz factor evaluated at the duration of the burst $\left(\Gamma_c\equiv\Gamma(t_{\rm x}=T_{90})\right)$.   In this case, the critical Lorentz factor becomes

{\small
\begin{eqnarray}
\Gamma_c\equiv  \left(\frac{3-k}{(4c)^{5-k} \pi m_p}\right)^{\frac{1}{2(4-k)}}  (1+z)^{\frac{3-k}{2(4-k)}} E^\frac{1}{2(4-k)}A_{\rm k_s}^{-\frac{1}{2(4-k)}} T_{90}^{-\frac{3-k}{2(4-k)}}\,.
\end{eqnarray}
}

\subsection{Dynamics of the thick-shell scenario}

The RS in the thick-shell scenario is ultra-relativistic and can decelerate the shell dramatically. Before the RS crosses the shell, the scale of the hydrodynamic quantities as a function of the observed time is $\gamma_3\propto t^{-\frac{2-k}{2(4-k)}}$, $n_{3}\propto t^{-\frac{k+6}{2(4-k)}}$, $p_{3}\propto t^{-\frac{k+2}{4-k}}$ and $N_{\rm e}\propto t$. Electrons are accelerated and described by a PL distribution ($\propto \gamma_e^{-p}\,d\gamma_e$ with $p>2$), and the magnetic field is magnified in the RS zone. The terms $\gamma_e$ and $p$ are the Lorentz factor of the electrons and the spectral index, respectively. We consider that a fraction $\varepsilon_{\rm e_r}$ and $\varepsilon_{\rm B_r}$ of the total energy goes to the electrons and the magnetic field, respectively. Therefore, the magnetic field, the minimum and cooling electron Lorentz factors evolve as $B'_{\rm r}\propto t^{-\frac{k+2}{2(4-k)}}$, $\gamma_{\rm m,r}\propto t^{\frac{2-k}{2(4-k)}}$ and $\gamma_{\rm c,r}\propto t^{\frac{3k-2}{2(4-k)}}$, respectively. Moreover, the synchrotron spectral breaks and the maximum synchrotron flux in terms of observed time are $\nu^{\rm syn}_{\rm m, r}\propto t^{-\frac{k}{4-k}}$ and $\nu^{\rm syn}_{\rm c, r}\propto t^{\frac{3k-4}{4-k}}$ and $F^{\rm syn}_{\rm max,r }\propto t^{\frac{2-k}{4-k}}$, respectively.

Once the RS crosses the shell, the scale of the hydrodynamic variables becomes $\gamma_3\propto t^{\frac{2k-7}{4(4-k)}}$, $n_{3}\propto t^{\frac{2k-13}{4(4-k)}}$, $p_{3}\propto t^{\frac{2k-13}{3(4-k)}}$, and $N_{\rm e}\propto t^0$. In this case, the magnetic field, the minimum and cooling electron Lorentz factors, the synchrotron spectral breaks and the maximum synchrotron flux in terms of the observed time are $B'\propto t^{\frac{2k-13}{6(4-k)}}$, $\gamma_{\rm m,r}\propto t^{\frac{2k-13}{12(4-k)}}$ and $\gamma_{\rm c,r}\propto t^{\frac{25-2k}{12(4-k)}}$, $\nu^{\rm syn}_{\rm m, r}\propto t^{-\frac{73-14k}{12(4-k)}}$ and $\nu^{\rm syn}_{\rm c, r}\propto t^{\frac{1+2k}{4(4-k)}}$ and $F^{\rm syn}_{\rm max,r}\propto t^{-\frac{47-10k}{12(4-k)}}$, respectively.
We derive the expected light curves before and after the shock crossing time. 
\subsubsection{SSC light curves for $t < t_{\rm x}$}
At the deceleration radius $r=4c(1+z)^{-1}\gamma_3^2\,t$, the same electron population can up-scatter synchrotron photons up to higher energies as $h\nu^{\rm ssc}_{\rm i, r}\sim\gamma^2_{\rm i, r} h\nu^{\rm syn}_{\rm i, r}$ with ${\rm i=m}$ and ${\rm c}$, reaching a maximum flux of $F^{\rm ssc}_{\rm max,r}\sim\, \frac{4\sigma_Tn r}{3g(p)}\,F^{\rm syn}_{\rm max,r}$ with $g(p)=\frac{p-1}{p-2}$.\footnote{The term $\sigma_T$ is the Thompson cross-section and $h$ is the Planck constant.} Therefore, the SSC spectral breaks and the maximum SSC flux are

{\small
\bary\label{ssc_before1}
h \nu^{\rm ssc}_{\rm m, r}&\simeq& 0.6\,{\rm~GeV}\,\left(\frac{1+z}{2}\right)^{\frac{3(k-2)}{4-k}} g^4(2.2) \zeta_{\rm e,-1}^{-4} \varepsilon^4_{\rm e_r,-1} \varepsilon^{\frac12}_{\rm B_r,-2}\,\Gamma^4_{2.5} A^{\frac{3}{4-k}}_{\rm k,-1}\Delta^{\frac{k+2}{2(4-k)}}_{11.8}\,E^{-\frac{k+2}{2(4-k)}}_{53} t^{\frac{2(1-k)}{4-k}}_1\,,\cr
h\nu^{\rm ssc}_{\rm c, r}&\simeq& 0.7\,{\rm eV}\,\left(\frac{1+z}{2}\right)^{\frac{2-5k}{4-k}}  \left(\frac{1+Y_{\rm r}}{2} \right)^{-4}\varepsilon^{-\frac72}_{\rm B_r,-2} A^{-\frac{9}{4-k}}_{\rm k,-1}\Delta^{\frac{10-7k}{2(4-k)}}_{11.8}\,E^{\frac{7k-10}{2(4-k)}}_{53}\,t^{\frac{6(k-1)}{4-k}}_1,\cr
F^{\rm ssc}_{\rm max,r} &\simeq& 6.8\,{\rm mJy}\, \left(\frac{1+z}{2} \right)^{\frac{2k}{4-k}}\,g^{-1}(2.2)\,\zeta_{\rm e,-1} \varepsilon^{\frac12}_{\rm B_r,-2}\,A_{\rm k,-1}^{\frac{4}{4-k}} \,d^{-2}_{\rm z,28.3}\,\Gamma^{-1}_{2.5}\,\Delta^{\frac{12-5k}{2(k-4)}}_{11.8}\,E^{\frac{12-5k}{2(4-k)}}_{53} t^{\frac{4-3k}{4-k}}_1\,,
\eary
}
where $\zeta_e$ is the fraction of electrons accelerated during the shock \citep{2006MNRAS.369..197F}, $Y_{\rm r}$ is the Compton parameter and {\small $d_{\rm z}=(1+z)\frac{c}{H_0}\int^z_0\,\frac{d\tilde{z}}{\sqrt{\Omega_{\rm M}(1+\tilde{z})^3+\Omega_\Lambda}}$}  \citep{1972gcpa.book.....W}  corresponds to the luminosity distance, with the constant values of $\Omega_m=1-\Omega_\Lambda=0.315$ and $H_0=67.4 ~{\rm km}\, {\rm s}^{-1} ~{\rm Mpc}^{-1}$ \citep{2020A&A...641A...6P}.
 We report the proportionality constants of the spectral breaks and maximum fluxes considering the values of $p=2.2$ and $t_{\rm x}=15\,{\rm s}$, unless otherwise stated.    At a particular observed energy ($h\nu$), the SSC light curves for the fast- and the slow-cooling regime evolve as
{\small
\begin{eqnarray}
\label{fast}
F^{\rm ssc}_{\rm \nu, r}(t < t_{\rm x}) \propto \begin{cases} 
 t^{-\frac{5k-6}{4-k}}\, \nu_{\rm }^{\frac13},\hspace{1.5cm} \nu<\nu^{\rm ssc}_{\rm c, r}, \cr
 t^{\frac{1}{4-k}}\, \nu_{\rm }^{-\frac12},\hspace{1.6cm} \nu^{\rm ssc}_{\rm c, r}<\nu<\nu^{\rm ssc}_{\rm m, r}, \cr
 t^{\frac{(1-k)p+k}{4-k}}\,\nu_{\rm }^{-\frac{p}{2}},\,\,\,\,\, \hspace{0.7cm} \nu^{\rm ssc}_{\rm m, r}<\nu < \nu^{\rm ssc}_{\rm KN, m, r}\,, \cr
\end{cases}
\end{eqnarray}
}
and
{\small
\begin{eqnarray}
\label{slow_before}
F^{\rm ssc}_{\rm \nu, r} (t < t_{\rm x})\propto \begin{cases} 
t^{-\frac{7k-10}{3(4-k)}}\, \nu_{\rm }^{\frac13},\hspace{1.5cm} \nu<\nu^{\rm ssc}_{\rm m, r}, \cr
t^{\frac{p(1-k)+3-2k}{4-k}}\, \nu_{\rm }^{-\frac{p-1}{2}},\hspace{0.5cm} \nu^{\rm ssc}_{\rm m, r}<\nu<\nu^{\rm ssc}_{\rm c, r}, \cr
t^{\frac{(1-k)p+k}{4-k}}\,\nu_{\rm }^{-\frac{p}{2}},\,\,\,\,\,\hspace{0.9cm}\nu^{\rm ssc}_{\rm c, r}<\nu < \nu^{\rm ssc}_{\rm KN, c, r}\,, \cr
\end{cases}
\end{eqnarray}
}
%
respectively.   Furthermore, the Klein-Nishina (KN) break must be considered since the SSC flux is strongly attenuated if the photon energy lies above this break.   For $\nu^{\rm ssc}_{\rm c, r}<\nu^{\rm ssc}_{\rm m, r}$ and $\nu^{\rm ssc}_{\rm m, r}<\nu^{\rm ssc}_{\rm c, r}$, the KN breaks ($h\nu^{\rm ssc}_{\rm KN, m, r} \simeq\frac{\gamma_3\gamma_{m,r}}{(1+z)}\,m_e c^2$ and $h\nu^{\rm ssc}_{\rm KN, c, r} \simeq\frac{\gamma_3\gamma_{c,r}}{(1+z)}\,m_e c^2$) become
{\small
\bary
h \nu^{\rm ssc}_{\rm KN,m,r}&\simeq& 3.7\times 10\,{\rm~GeV}\,\left(\frac{1+z}{2}\right)^{-1}g(2.2)\zeta_{e,-1}^{-1} \,\varepsilon_{\rm e_r,-1}\,\Gamma_{2.5}\,,\cr
h \nu^{\rm ssc}_{\rm KN,c,r}&\simeq& 0.2\,{\rm~GeV}\left(\frac{1+z}{2}\right)^{-\frac{k+2}{4-k}} \left(\frac{1+Y_{\rm r}}{2} \right)^{-1}\varepsilon^{-1}_{\rm B_r,-2}A^{-\frac{3}{4-k}}_{\rm k,-1}\,\Delta^{\frac{1-k}{4-k}}_{11.8}\,E^{\frac{k-1}{4-k}}_{53}\,t^{\frac{2(k-1)}{4-k}}_1\,,
\eary
}
with $m_e$ the electron mass. Following \cite{2009ApJ...703..675N} with the definition of a new spectral break ($h\nu^{\rm ssc}_{\rm KN, 0, r} \simeq\frac{\gamma_3\gamma_{0,r}}{(1+z)}\,m_e c^2$)\footnote{See eqs. 20 and 49 of \cite{2009ApJ...703..675N} for the definition of $\gamma_{\rm 0,r}$ in the fast ($\gamma_{\rm m, r} > \gamma_{\rm c, r}$) and slow ($\gamma_{\rm c, r} > \gamma_{\rm m, r}$) cooling regimes, respectively.} the additional PLs in the SSC light curves due to KN effects are
{\small
\begin{eqnarray}
\label{fast_KN}
F^{\rm ssc}_{\rm \nu, r}(t < t_{\rm x}) \propto \begin{cases} 
 t^{\frac{p-k(p-1)}{4-k}}\, \nu_{\rm }^{-(p-1)},\hspace{1.76cm} \nu^{\rm ssc}_{\rm KN, m, r}<\nu<\nu^{\rm ssc}_{\rm 0, r}, \cr
 t^{\frac{p-k(p-1)}{4-k}}\, \nu_{\rm }^{-(p-\frac12)},\hspace{1.69cm} \nu^{\rm ssc}_{\rm 0, r}<\nu<\nu^{\rm ssc}_{\rm KN, c, r}, \cr
 t^{-\frac{5-3p-k(8-3p)}{3(4-k)}}\,\nu_{\rm }^{-(p+\frac13)},\,\,\,\,\, \hspace{0.9cm} \nu^{\rm ssc}_{\rm KN, c, r}<\nu\,, \cr
\end{cases}
\end{eqnarray}
}

and

{\small
\begin{eqnarray}
\label{slow_before_KN}
F^{\rm ssc}_{\rm \nu, r} (t < t_{\rm x})\propto \begin{cases} 
t^{\frac{2-k}{4-k}}\, \nu_{\rm }^{-(p-1)},\hspace{1.5cm} \nu^{\rm ssc}_{\rm KN, c, r}<\nu<\nu^{\rm ssc}_{\rm 0, r}, \cr
t^{\frac{1}{4-k}}\, \nu_{\rm }^{-\frac{p+1}{2}},\hspace{1.69cm} \nu^{\rm ssc}_{\rm 0, r}<\nu<\nu^{\rm ssc}_{\rm m, r}, \cr
t^{\frac{1}{4-k}}\,\nu_{\rm }^{-(p+\frac{1}{3})},\,\,\,\,\,\hspace{1.2cm}\nu^{\rm ssc}_{\rm m, r}<\nu\,, \cr
\end{cases}
\end{eqnarray}
}
for the fast (weak) - and slow-cooling regimes, respectively.
\subsubsection{SSC light curves for $t > t_{\rm x}$}
The SSC spectral breaks and the maximum SSC flux are given by
{\small
\bary\label{ssc_after1}
h \nu^{\rm ssc}_{\rm m, r}&\simeq&  2.7\times 10^{-3}\,{\rm eV}  \, \left(\frac{1+z}{2} \right)^{\frac{17-2k}{4(4-k)}}\,g^4(2.2)\,\zeta_{\rm e,-1}^{-4}\, \,\varepsilon^4_{\rm e_r,-1}\varepsilon^{\frac12}_{\rm B_r,-2} A^{-\frac{3}{k-4}}_{\rm k,-1}\,\Gamma^4_{2.5}\Delta_{11.8}^{\frac{3(4k-15)}{4(k-4)}}\,E^{\frac{k+2}{2(k-4)}}_{53}t^{\frac{3(11-2k)}{4(k-4)}}_{2},\cr
h \nu^{\rm ssc}_{\rm cut, r}&\simeq& 1.3\,{\rm eV}\, \left(\frac{1+z}{2} \right)^{\frac{17-2k}{4(4-k)}}\,\varepsilon^{-\frac72}_{\rm B_r,-2} A^{\frac{9}{k-4}}_{\rm k,-1}\, \left(\frac{1+Y_{\rm r}}{2}\right)^{-4}\,\Delta^{\frac{29+4k}{4(4-k)}}_{11.8}\,E^{\frac{10-7k}{2(k-4)}}_{53}\,t^{\frac{3(11-2k)}{4(k-4)}}_{2},\cr
F^{\rm ssc}_{\rm max, r}&\simeq& 0.3\,{\rm mJy}\,\left(\frac{1+z}{2} \right)^{\frac{89-16k}{12(4-k)}}\,g^{-1}(2.2)\,\zeta_e\, \varepsilon^{\frac12}_{\rm B_r,-2}\, \Gamma^{-1}_{2.5}\,A^{\frac{4}{4-k}}_{\rm k,-1}\, \Delta_{11.8}^{\frac{17-10k}{12(4-k)}} \,d^{-2}_{\rm z,28.3} \,E^{\frac{12-5k}{2(4-k)}}_{53}\,t^{\frac{41-4k}{12(k-4)}}_{2}, \,\,\,\,\,\,\,\,
\eary
}
where the cutoff  frequency $\nu^{\rm ssc}_{\rm cut, r}=\nu^{\rm ssc}_{\rm c, r}(t_{\rm x})\,\left(\frac{t}{t_{\rm x}} \right)^{\frac{33-6k}{4(k-4)}}$ is estimated  by the requirement that no electrons are shocked anymore and the fluid expands adiabatically. For $\nu^{\rm ssc}_{\rm c, r}<\nu^{\rm ssc}_{\rm m, r}$ and $\nu^{\rm ssc}_{\rm m, r}<\nu^{\rm ssc}_{\rm c, r}$, the spectral breaks in the KN regime are
{\small
\bary
h \nu^{\rm ssc}_{\rm KN, m, r}&\simeq& 8.9\,{\rm~GeV}\,\left(\frac{1+z}{2}\right)^{-\frac{7-2k}{6(4-k)}} g(2.2)\,\zeta_{e,-1}^{-1}\,\varepsilon_{\rm e_r,-1}\,\Gamma_{2.5}\,\Delta^{\frac{17-4k}{6(4-k)}}_{11.8}\,t^{-\frac{17-4k}{6(4-k)}}_{2},\cr
h \nu^{\rm ssc}_{\rm KN, c, r}&\simeq& 0.8\,{\rm~GeV}\left(\frac{1+z}{2}\right)^{\frac{13-2k}{3(k-4)}} \left(\frac{1+Y_{\rm r}}{2} \right)^{-1}\varepsilon^{-1}_{\rm B_r,-2}A^{\frac{3}{k-4}}_{\rm k,-1}\,E^{\frac{k-1}{4-k}}_{53}\,\Delta^{\frac{2(k-2)}{3(4-k)}}_{11.8}\,t^{\frac{k+1}{3(4-k)}}_{2}.\,\,
\eary
}
The SSC light curves for the slow-cooling regime evolve as  
%
%
%
{\small
\begin{eqnarray}\nonumber
\label{slow_after}
F^{\rm ssc}_{\rm \nu, r} (t > t_{\rm x}) \propto \begin{cases}
t^{\frac{k+4}{6(k-4)}}_1\, \nu_{\rm}^{\frac13},\hspace{2.23cm} \nu<\nu^{\rm ssc}_{\rm m, r}, \cr
t^{\frac{17-10k-99p+18kp}{24(4-k)}}\, \nu_{\rm }^{-\frac{p-1}{2}},\hspace{0.4cm} \nu^{\rm ssc}_{\rm m, r}<\nu<\nu^{\rm ssc}_{\rm cut, r}, \cr
0,\,\,\,\,\hspace{3.3cm}   \nu^{\rm ssc}_{\rm cut, r}<\nu\, . \cr
\end{cases}
\end{eqnarray}
}
The SSC emission generated in the RS region could decay faster because of the angular time delay effect \citep{2000ApJ...543...66P, 2003ApJ...597..455K}. During this phase, the evolution of the SSC flux due to high-latitude afterglow emission is described by $F^{\rm ssc}_{\rm \nu, r}(t > t_{\rm x})\propto t^{-(\beta+2)}$, where $\beta$ is the spectral index with $\beta=1/2,\,(p-1)/2$ or $p/2$ for the fast and slow cooling regimes, respectively.
%

\subsection{Dynamics of the thin-shell scenario}
In the thin-shell regime, the RS becomes mildly relativistic and hence cannot decelerate the shell. In this case, the shock crossing time is longer than the duration of the burst ($T_{90}< t_{\rm x}$) with {\small $t_{\rm x} =\left(\frac{(3-k)E\,(1+z)^{3-k}}{2^{4-k} c^{5-k}\pi\,m_p\,A_k}\right)^\frac{1}{3-k}\,\Gamma^{-\frac{8-2k}{3-k}}$}. Before the RS crosses the shell, the scaling of the hydrodynamic quantities as a function of the observed time is $\gamma_3\propto t^0$, $n_{3}\propto t^{-3}$, $p_{3}\propto t^{-k}$, and $N_{\rm e}\propto t^{\frac{3-k}{2}}$. In this regime, the magnetic field, the minimum, and the cooling electron Lorentz factors evolve as $B'_{\rm r}\propto t^{-\frac{k}{2}}$, $\gamma_{\rm m,r}\propto t^{-(k-3)}$ and $\gamma_{\rm c,r}\propto t^{k-1}$, respectively. Moreover, the synchrotron spectral breaks and the maximum synchrotron flux in terms of the observed time are $\nu^{\rm syn}_{\rm m, r}\propto t^{\frac{12-5k}{2}}$ and $\nu^{\rm syn}_{\rm c, r}\propto t^{\frac{3k-4}{2}}$ and $F^{\rm syn}_{\rm max,r }\propto t^{\frac{3-2k}{2}}$, respectively.

Once the RS crosses the shell, for ${\rm k=2}$, the scaling of the hydrodynamic variables becomes $\gamma_3\propto t^{-\frac13}$, $n_{3}\propto t^{-\frac87}$, $p_{3}\propto t^{-\frac{32}{21}}$ and $N_{\rm e}\propto t^0$. In this case, the magnetic field, the minimum and cooling electron Lorentz factors, the synchrotron spectral breaks and the maximum synchrotron flux as function of the observed time are $B'\propto t^{-\frac{16}{21}}$, $\gamma_{\rm m,r}\propto t^{-\frac{8}{21}}$ and $\gamma_{\rm c,r}\propto t^\frac67$, $\nu^{\rm syn}_{\rm m, r}\propto t^{-\frac{13}{7}}$ and $\nu^{\rm syn}_{\rm c, r}\propto t^{\frac{13}{21}}$ and $F^{\rm syn}_{\rm max,r}\propto t^{-\frac{23}{21}}$, respectively.
We derive the expected light curves before and after the shock crossing time. 

\subsubsection{SSC light curves for $t < t_{\rm x}$}
For this time interval, the SSC spectral breaks and the maximum SSC flux are
{\small
\bary\label{ssc_after2}
h \nu^{\rm ssc}_{\rm m, r}&\simeq&  0.2\,{\rm eV}  \, \left(\frac{1+z}{2}\right)^{\frac{9k-26}{2}} g^4(2.2) \zeta_{\rm e,-1}^{-4} \varepsilon^4_{\rm e_r,-1} \varepsilon^{\frac12}_{\rm B_r,-2} \,A^\frac92_{\rm k,-1}\,\Gamma^{34-9k}_{1.5}\,E^{-4}_{52}\,t^{\frac{24-9k}{2}}_{2.5},\cr
h \nu^{\rm ssc}_{\rm c, r}&\simeq& 2.7\times 10^{-6}\,{\rm eV}\, \left(\frac{1+z}{2} \right)^{\frac{6-7k}{2}}\,\varepsilon^{-\frac72}_{\rm B_r,-2} A^{-\frac72}_{\rm k,-1} \left(\frac{1+Y_{\rm r}}{2}\right)^{-4}\,\Gamma^{7k-10}_{1.5} \,t^{\frac{7k-8}{2}}_{2.5},\cr
F^{\rm ssc}_{\rm max, r}&\simeq& 0.2\,{\rm mJy}\,\left(\frac{1+z}{2} \right)^{\frac{4k-3}{2}}\,g^{-1}(2.2)\,\zeta_{\rm e,-1}\, \varepsilon^{\frac12}_{\rm B_r,-2}\,\Gamma^{7-4k}_{1.5}\,A^2_{\rm k,-1}\,d^{-2}_{\rm z,28.3}\,E^{\frac12}_{52}\,t^{\frac {5-4k}{2}}_{2.5}.
\eary
}
%
%
In analogy to the description of the SSC light curves for the thick-shell regime, the SSC light curves before the shock-crossing time for the fast- and the slow-cooling regimes at a particular observed energy are
{\small
\begin{eqnarray}
\label{fast_before_thin}
F^{\rm ssc}_{\rm \nu, r} (t < t_{\rm x}) \propto \begin{cases} 
 t^{\frac{23-19k}{6}}\, \nu_{\rm }^{\frac13},\hspace{2.4cm} \nu<\nu^{\rm ssc}_{\rm c, r}, \cr
 t^{\frac{2-k}{4}}\, \nu_{\rm }^{-\frac12},\hspace{2.4cm} \nu^{\rm ssc}_{\rm c, r}<\nu<\nu^{\rm ssc}_{\rm m, r}, \cr
 t^{\frac{3p(8-3k)+2(4k-11)}{4}}\,\nu_{\rm }^{-\frac{p}{2}},\,\,\,\,\, \hspace{0.65cm} \nu^{\rm ssc}_{\rm m, r}<\nu < \nu^{\rm ssc}_{\rm KN, m, r}\,, \cr
\end{cases}
\end{eqnarray}
}
and
{\small
\begin{eqnarray}
\label{slow_before_thin1}
F^{\rm ssc}_{\rm \nu, r} (t < t_{\rm x}) \propto \begin{cases}
t^{-\frac{k+3}{2}}\, \nu_{\rm }^{\frac13},\hspace{2.4cm} \nu<\nu^{\rm ssc}_{\rm c, r}, \cr
t^{\frac{3p(8-3k)+k-14}{4}}\, \nu_{\rm }^{-\frac{p-1}{2}},\hspace{0.9cm} \nu^{\rm ssc}_{\rm m, r}<\nu<\nu^{\rm ssc}_{\rm c, r}, \cr
 t^{\frac{3p(8-3k) + 2(4k-11)}{4}}\,\nu_{\rm }^{-\frac{p}{2}},\,\,\,\,\,\hspace{0.6cm}\nu^{\rm ssc}_{\rm c, r}<\nu < \nu^{\rm ssc}_{\rm KN, c, r}\,, \cr
\end{cases}
\end{eqnarray}
}
respectively.  For $\nu^{\rm ssc}_{\rm c, r}<\nu^{\rm ssc}_{\rm m, r}$ and $\nu^{\rm ssc}_{\rm m, r}<\nu^{\rm ssc}_{\rm c, r}$, the spectral breaks in the KN regime are
{\small
\bary
h \nu^{\rm ssc}_{\rm KN, m, r}&\simeq& 6.1\times 10^{-2}\,{\rm~GeV}\,\left(\frac{1+z}{2}\right)^{k-4}g(2.2)\zeta_{e,-1}^{-1}\,\varepsilon_{\rm e_r,-1}\,A_{\rm k,-1}\,\Gamma^{9-2k}_{1.5}\,E_{52}^{-1}\,t^{3-k}_{2.5}\,,\cr
h \nu^{\rm ssc}_{\rm KN,c, r}&\simeq& 3.9\times 10^{-3}\,{\rm~GeV}\left(\frac{1+z}{2}\right)^{-k} \left(\frac{1+Y_r}{2}\right)^{-1}\varepsilon^{-1}_{\rm B_r,-2}\,A^{-1}_{\rm k,-1}\,\Gamma^{2(k-1)}_{1.5}\,t^{k-1}_{2.5}\,.
\eary
}

In analogy to the description of the SSC light curves for the thick-shell regime,  the additional PLs in the SSC light curves due to KN effects are \citep{2009ApJ...703..675N}
{\small
\begin{eqnarray}
\label{fast_before_thin_KN}
F^{\rm ssc}_{\rm \nu, r}(t < t_{\rm x}) \propto \begin{cases} 
 t^{-\frac{2(17-15p)-k(12-11p)}{4}}\, \nu_{\rm }^{-(p-1)},\hspace{1.5cm} \nu^{\rm ssc}_{\rm KN, m, r}<\nu<\nu^{\rm ssc}_{\rm 0, r}, \cr
 t^{-\frac{2(14-15p)-k(10-11p)}{4}}\, \nu_{\rm }^{-(p-\frac12)},\hspace{1.4cm} \nu^{\rm ssc}_{\rm 0, r}<\nu<\nu^{\rm ssc}_{\rm KN, c, r}, \cr
 t^{-\frac{2(47-45p)-k(40-33p)}{12}}\,\nu_{\rm }^{-(p+\frac13)},\,\,\,\,\, \hspace{1.22cm} \nu^{\rm ssc}_{\rm KN, c, r}<\nu\,, \cr
\end{cases}
\end{eqnarray}
}
and
{\small
\begin{eqnarray}
\label{slow_before_thin1_KN}
F^{\rm ssc}_{\rm \nu, r} (t < t_{\rm x})\propto \begin{cases} 
t^{-\frac{2(9-11p)-k(4-7p)}{4}}\, \nu_{\rm }^{-(p-1)},\hspace{1.5cm} \nu^{\rm ssc}_{\rm KN, c, r}<\nu<\nu^{\rm ssc}_{\rm 0, r}, \cr
t^{-\frac{8(1-2p)-k(2-5p)}{4}}\, \nu_{\rm }^{-\frac{p+1}{2}},\hspace{1.8cm} \nu^{\rm ssc}_{\rm 0, r}<\nu<\nu^{\rm ssc}_{\rm m, r}, \cr
t^{-\frac{6(5-11p)-k(8-21p)}{12}}\,\nu_{\rm }^{-(p+\frac{1}{3})},\,\,\,\,\,\hspace{1.15cm}\nu^{\rm ssc}_{\rm m, r}<\nu\,, \cr
\end{cases}
\end{eqnarray}
}
for the fast (weak) - and slow-cooling regimes, respectively.
\subsubsection{SSC light curves for $t > t_{\rm x}$ with $k=2$}
Following the previous process, the spectral breaks and the maximum flux of SSC emission are given by
{\small
\bary\label{ssc_before2}
h \nu^{\rm ssc}_{\rm m, r}&\simeq& 0.7\,{\rm eV} \left(\frac{1+z}{2}\right)^{\frac{34}{21}}\,g^4(2.2)\,\zeta_{\rm e,-1}^{-4}\,\varepsilon^4_{\rm e_r,-1} \varepsilon^{\frac12}_{\rm B_r,-2}\,A^{-\frac{47}{42}}_{\rm k,-1}\, \Gamma^{-\frac{136}{21}}_{1.5}\, E^{\frac{34}{21}}_{52}\,t^{-\frac{55}{21}}_{3.5}\, ,\cr
h \nu^{\rm ssc}_{\rm cut, r}&\simeq& 1.0\times 10^{-4}\,{\rm eV} \left(\frac{1+z}{2}\right)^{\frac{34}{21}}\,\left(\frac{1+Y_{\rm r}}{2}\right)^{-4}  \varepsilon^{-\frac72}_{\rm B_r,-2} \,A^{-\frac{383}{42}}_{\rm k,-1} \Gamma^{-\frac{388}{21}}_{1.5}\, E^{\frac{118}{21}}_{52}\, t^{-\frac{55}{21}}_{3.5}\, ,\cr
F^{\rm ssc}_{\rm max,r} &\simeq&  4.9\times 10^{-3}\,{\rm mJy}\, \left(\frac{1+z}{2}\right)^{\frac{17}{7}} \,g^{-1}(2.2)\,\zeta_{\rm e,-1}\,\varepsilon^{\frac12}_{\rm B_r,-2} A^{\frac{29}{14}}_{\rm k,-1}\, \Gamma^{-\frac{5}{7}}_{1.5}\,d^{-2}_{\rm z,28.3}E^{\frac{3}{7}}_{52}\, t^{-\frac{10}{7}}_{3.5}\,,\cr
&&\hspace{5.5cm}\eary
}
where the cutoff frequency $\nu^{\rm ssc}_{\rm cut, r}=\nu^{\rm ssc}_{\rm c, r}(t_{\rm x})\,\left(\frac{t}{t_{\rm x}} \right)^{-\frac{55}{21}}$ is estimated  by the requirement that no electrons are shocked anymore and the fluid expands adiabatically. For $\nu^{\rm ssc}_{\rm c, r}<\nu^{\rm ssc}_{\rm m, r}$ and $\nu^{\rm ssc}_{\rm m, r}<\nu^{\rm ssc}_{\rm c, r}$, the spectral breaks in the KN regime are given by
{\small
\bary
h\nu^{\rm ssc}_{\rm KN, m, r}&\simeq& 0.1\,{\rm~GeV}  \left(\frac{1+z}{2}\right)^{-\frac{2}{7}}g(2.2)\zeta_{e,-1}^{-1}\varepsilon_{\rm e_r,-1}A^{-\frac{5}{7}}_{\rm k,-1}\Gamma^{-\frac{13}{7}}_{1.5}\,E^{\frac{5}{7}}_{52}\,t^{-\frac{5}{7}}_{3.5},\cr
h\nu^{\rm ssc}_{\rm KN, c, r}&\simeq& 2.4\times 10^{-2}\,{\rm~GeV}  \left(\frac{1+z}{2}\right)^{-\frac{32}{21}} \left(\frac{1+Y_{\rm r}}{2} \right)^{-1}\varepsilon^{-1}_{\rm B_r,-2}\,A^{-\frac{31}{21}}_{\rm k,-1}\,\Gamma^{\frac{2}{21}}_{1.5}\,E^{\frac{10}{21}}_{52}\,t^{\frac{11}{21}}_{3.5}.\,\,\,
\eary
}
The SSC light curves for the slow-cooling regime are
%
%
%
{\small
\begin{eqnarray}\nonumber
\label{slow_before_thin2}
F^{\rm ssc}_{\rm \nu, r} (t > t_{\rm x}) \propto \begin{cases}
t^{-\frac{5}{9}}\, \nu_{\rm }^{\frac13},\hspace{1.7cm} \nu<\nu^{\rm ssc}_{\rm c, r}, \cr
t^{-\frac{5(11p+1)}{42}}\, \nu_{\rm }^{-\frac{p-1}{2}},\hspace{0.3cm} \nu^{\rm ssc}_{\rm m, r}<\nu<\nu^{\rm ssc}_{\rm cut, r}, \cr
 0,\,\,\,\,\,\hspace{2.25cm}\nu^{\rm ssc}_{\rm cut, r}<\nu\,. \cr
\end{cases}
\end{eqnarray}
}
The SSC emission from the RS region could decay faster due to the angular time delay effect. During the emission at high latitudes, the SSC flux evolves in the same manner as the flux mentioned above in the case of the thick-shell scenario.

\subsection{Synchrotron light curves from forward shocks}
The dynamics of the forward shocks for a relativistic outflow expanding into a constant-density (${\rm k=0}$) and stellar-wind environments (${\rm k=2}$) are treated in \cite{1998ApJ...497L..17S, 2000ApJ...536..195C, 2000ApJ...543...66P}. For a density profile $\propto r^{\rm -k}$, the evolution of the bulk Lorentz factor is $\Gamma\propto t^{-\frac{3-k}{2(4-k)}}$, and the electron Lorentz factors for the minimum and the cooling energy are $\gamma_{\rm m, f}\propto t^{-\frac{3-k}{2(4-k)}}$ and $\gamma_{\rm c,f}\propto t^{\frac{k+1}{2(4-k)}}$, respectively. Using the evolution of synchrotron energy breaks ($\nu^{\rm syn}_{\rm m, f}\propto t^{-\frac{3}{2}}$ and $\nu^{\rm syn}_{\rm c, f}\propto t^{\frac{4-3k}{2(k-4)}}$) and the maximum flux ($F^{\rm syn}_{\rm max,f} \propto t^{\frac{k}{2(k-4)}}$), the observed flux in the fast cooling regime is proportional to {\small $F^{\rm syn}_{\rm \nu, f}\propto t^{-\frac{1}{4}} \,\nu^{-\frac{1}{2}}$} for {\small $\nu<\nu^{\rm syn}_{\rm m,f}$} and {\small $\propto  t^{-\frac{3p-2}{4}}\,\nu^{-\frac{p}{2}}$} for {\small $\nu^{\rm syn}_{\rm m,f}<\nu$}. In the slow cooling regime, the observed flux is proportional to {\small $F^{\rm syn}_{\rm \nu, f}\propto t^{\frac{12-5k-3p(4-k)}{4(4-k)}}\,\nu^{-\frac{p-1}{2}}$ for $\nu <\nu^{\rm syn}_{\rm c,f}$} and {\small $\propto t^{-\frac{3p-2}{4}}\,\nu^{-\frac{p}{2}}$} for {\small $\nu^{\rm syn}_{\rm c,f}<\nu$}. 

\subsection{The Compton Y-parameter}
When KN effects are strong, the Compton parameter changes. In this case, this parameter in the slow-cooling regime can be estimated with 

{\small
\bary
\label{Yth_kn}
Y_j(Y_j+1) = Y_{0,j}\, \begin{cases}
1\, \hspace{3cm} {\rm for} \hspace{0.2cm} {\nu^{\rm syn}_{\rm c,j} < \nu^{\rm syn}_{\rm KN,c,j}}\cr
\left(\frac{\nu^{\rm syn}_{\rm KN,c,j}}{\nu^{\rm syn}_{c,j}}\right)^{-\frac{p-3}{2}} \, \hspace{0.3cm} {\rm for} \hspace{0.2cm} { \nu^{\rm syn}_{\rm m,j} < \nu^{\rm syn}_{\rm KN,c,j} < \nu^{\rm syn}_{\rm c,j}}\,\,\,\,\,\,\cr
\left( \frac{\nu^{\rm syn}_{\rm m,j}}{\nu^{\rm syn}_{\rm c,j}} \right)^{-\frac{p-3}{2}}\left( \frac{\nu^{\rm syn}_{\rm KN,c,j}}{\nu^{\rm syn}_{\rm m,j}}\right)^\frac43 \, \hspace{0.cm} {\rm for} \,\hspace{0 cm} {\nu^{\rm syn}_{\rm KN,c,j} <\nu^{\rm syn}_{\rm m,j} }\,,
\end{cases}
\eary
}
where $Y_{0,j}=\frac{\varepsilon_{e,j}}{\varepsilon_{B,j}}  \left(\frac{\gamma_{\rm m,j}}{\gamma_{\rm c,j}}\right)^{p-2}$ with ${\rm j=r}$ or ${\rm f}$. For the RS region, $h\nu^{\rm syn}_{\rm KN, c, r} \simeq\frac{\gamma_3}{(1+z)}\,\frac{m_e c^2}{\gamma_{c,r}}$ for $\nu^{\rm syn}_{\rm c,r} > \nu^{\rm syn}_{\rm m,r}$ \citep[see][]{2009ApJ...703..675N, 2010ApJ...712.1232W}. It is worth mentioning that in the fast cooling regime the synchrotron break in the KN regime becomes $h\nu^{\rm syn}_{\rm KN, m, r} \simeq\frac{\gamma_3}{(1+z)}\,\frac{m_e c^2}{\gamma_{m,r}}$ for $\nu^{\rm syn}_{\rm c,r} < \nu^{\rm syn}_{\rm m,r}$. For the FS region, $h\nu^{\rm syn}_{\rm KN, c, f} \simeq\frac{\Gamma}{(1+z)}\,\frac{m_e c^2}{\gamma_{c, f}}$ for $\nu^{\rm syn}_{\rm c,f} > \nu^{\rm syn}_{\rm m,f}$ and $h\nu^{\rm syn}_{\rm KN, m, f} \simeq\frac{\gamma_3}{(1+z)}\,\frac{m_e c^2}{\gamma_{m,f}}$ for $\nu^{\rm syn}_{\rm c,f} < \nu^{\rm syn}_{\rm m,f}$. To describe the LAT data above 100~MeV, we must determine the Lorentz factor of electrons that may produce high-energy photons via the synchrotron process, include a new spectrum break, and recalculate the Compton value. The new spectral break becomes $h\nu^{\rm syn}_{\rm KN,c,j}(\gamma_*)$, and the recalculated Compton parameter, $Y(\gamma_{*})$, becomes {\small $Y(\gamma_*)=Y(\gamma_{c,j}) \left(\frac{\nu_{*}}{\nu_{c,j}}\right)^{\frac{p-3}{4}}   \left(\frac{\nu^{\rm syn}_{\rm KN,c,j}(\gamma_{\rm c,j})}{\nu^{\rm syn}_{c,j}}\right)^{-\frac{p-3}{2}}$} for $ \nu^{\rm syn}_{\rm m,j} <  h\nu^{\rm syn}_{\rm KN, c, j}(\gamma_*)=100\,{\rm~MeV} < \nu^{\rm syn}_{\rm c,j} < \nu^{\rm syn}_{\rm KN, c, j}(\gamma_{\rm c,j})$ \citep[for details, see][]{2010ApJ...712.1232W}.

\subsection{The maximum photon energy}
\subsubsection{Reverse-shock region}
The maximum energy radiated by the SSC process can be estimated by equalising the acceleration ($\propto \gamma_e B'^{-1}_{\rm r}$) and the synchrotron ($\propto\gamma_e^{-1} B'^{-2}_{\rm r}$) timescales. The maximum Lorentz factor of the electron distribution is $\gamma_{\rm max, r}=\left(3q_e/\xi\sigma_T B'_{\rm r}\right)^{\frac12}$ with $\xi$ the Bohm parameter\footnote{In the Bohm limit, this parameter becomes $\xi\sim 1$.} and $q_e$ the elementary charge. Therefore, the maximum energy radiated by the synchrotron model is $h\nu^{\rm syn}_{\rm max,r}= 3q^2_e\gamma_3/2\pi \sigma_T m_e c(1+z)$. In this case, the maximum energy generated by the SSC mechanism for $t< t_x$ in the thick and thin regime is
{\small
\bary\label{ssc_before3}
h\nu^{\rm ssc}_{\rm max, r}&\simeq& 1.3\,{\rm~GeV}\,\left(\frac{1+z}{2}\right)^{-\frac{4}{4-k}} \varepsilon^{-\frac12}_{\rm B_r,-2} A^{-\frac{2}{4-k}}_{\rm k,-1}\, \Delta^{-\frac{k}{2(4-k)}}_{11.8}\,E^{\frac{k}{2(4-k)}}_{53}\,t^{\frac{k}{4-k}}_1\,,
\eary
}
and
{\small
\bary\label{ssc_before4}
h\nu^{\rm ssc}_{\rm max, r}\simeq 0.4\,{\rm~GeV}\,\left(\frac{1+z}{2}\right)^{-\frac{k+2}{2}} \varepsilon^{-\frac12}_{\rm B_r,-2} A^{-\frac{1}{2}}_{\rm k,-1}\,\Gamma^{k}_{1.5}\,t^{\frac{k}{2}}_{2.5},
\eary
}
respectively. For $t>t_{\rm x}$, the maximum energies are determined by $\nu^{\rm ssc}_{\rm cut, r}$.

\subsubsection{Forward-shock region}  

The maximum Lorentz factor of the electron distribution is $\gamma_{\rm max, f}=\left(3q_e/\xi\sigma_T B'_{\rm f}\right)^{\frac12}$ and the maximum energy radiated by the synchrotron yields 

{\small
\bary
\label{ene_max}
h\nu^{\rm syn}_{\rm max,f} &\approx&
0.2\, {\rm~GeV}\left(\frac{1+z}{2}\right)^{\frac{k-5}{2(4-k)}}\,  A_{\rm k,-1}^{-\frac{1}{2(4-k)}} E_{53}^{\frac{1}{2(4-k)}}t_{2}^{-\frac{3-k}{2(4-k)}}\,.
\eary
}



%

\subsection{The short-lasting bright peak and the long-lasting emission in the {\itshape Fermi}/LAT band}
\begin{figure}
{ \centering
\resizebox*{0.95\textwidth}{0.8\textheight}
{\includegraphics{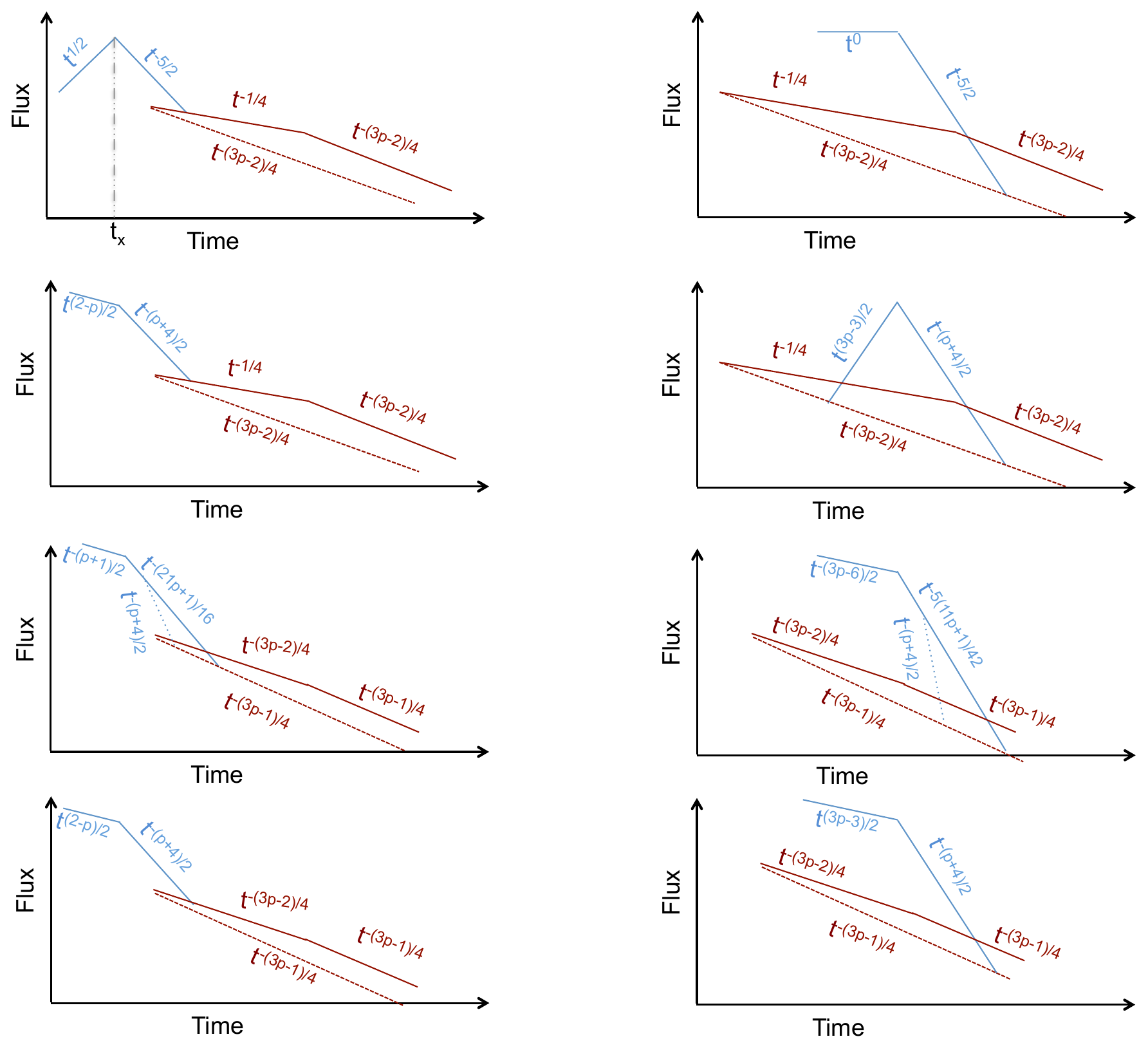}}
}
\caption{The  expected SSC flux from the RS region (blue lines) and the synchrotron flux from the FS region (red lines) when the outflow decelerates in the stellar wind medium. From top to bottom, the panels show the SSC RS light curves  for $\nu^{\rm ssc}_{\rm c,r}< \nu<\nu^{\rm ssc}_{\rm m,r}$, $\nu^{\rm ssc}_{\rm m,r}< \nu$, $\nu^{\rm ssc}_{\rm m,r}<\nu<\nu^{\rm ssc}_{\rm c,r}$ and $\nu^{\rm ssc}_{\rm c,r}< \nu$. The  expected SSC fluxes are presented in the thick- (left column) and thin- (right column) shell case. The double-dotted dashed black line corresponds to the shock crossing time. The breaks displayed in the solid red lines are the transitions between $\nu^{\rm syn}_{\rm m,f}<\nu<\nu^{\rm syn}_{\rm c,f}$ and $\nu^{\rm syn}_{\rm c,f}< \nu$.}
 \label{fig1:LC_LAT}
\end{figure}

We display in Figure~\ref{fig1:LC_LAT} the expected SSC and synchrotron light curves from the RS and FS model when the outflow decelerates in a stellar-wind density (${\rm k=2}$). They are presented in the fast and slow cooling regime, and when the RS lies in the thick (left column) and thin (right column) shell case. Possible transitions between the fast- and slow-cooling regimes and from constant-density to stellar-wind medium are not considered. The effects in the self-absorption regime are not considered because they are unimportant at the {\itshape Fermi}/LAT energy range \citep[e.g., see][]{2014ApJ...788...70P}. The SSC RS flux in blue lines illustrates that a peak is expected at $t=t_{\rm x}$ when it evolves in $ \nu^{\rm ssc}_{\rm c, r} <  \nu_{\rm LAT} <\nu^{\rm ssc}_{\rm m, r}$ for the thick-shell case and $ \nu^{\rm ssc}_{\rm m, r} <  \nu_{\rm LAT} <\nu^{\rm ssc}_{\rm c, r}$ for the thin-shell case. Otherwise, the flux decreases monotonically, exhibiting a break at $t=t_{\rm x}$. The temporal break is associated with the transition time (black line). Under the cooling conditions $ \nu^{\rm ssc}_{\rm m, r} <  \nu_{\rm LAT} <\nu^{\rm ssc}_{\rm cut, r}$ and $ t_{\rm x} < t$, the light curves in the thick and thin regime could exhibit a temporal break due to the passage of the cutoff break through the {\itshape Fermi}/LAT band ($\nu^{\rm ssc}_{\rm cut, r} < \nu_{\rm LAT} $).

The expected synchrotron FS fluxes in red lines are shown for a stellar-wind medium. Since the spectral breaks evolve as $\nu^{\rm syn}_{\rm m, f}\propto t^{-\frac32}$ and $\nu^{\rm syn}_{\rm c, f}\propto t^{\frac12}$, the temporal breaks displayed in the synchrotron FS fluxes correspond to the transitions from $\nu_{\rm LAT}<\nu^{\rm syn}_{\rm m, f}$ to $\nu^{\rm syn}_{\rm m, f}<\nu_{\rm LAT}$ for the fast cooling regime, and from $\nu^{\rm syn}_{\rm c, f} < \nu_{\rm LAT}$ to $\nu_{\rm LAT} < \nu^{\rm syn}_{\rm c, f}$ for the slow cooling regime. 

We argue that a LAT light curve with similar features to that shown in Figure~\ref{fig1:LC_LAT} can be described as a superposition of SSC RS and synchrotron FS emissions evolving in a stellar wind environment. It is worth noting that, depending on the value of the shock crossing time, the SSC emission would appear during the prompt emission (thick shell) or later (thin shell).

\section{Application: Second {\itshape Fermi}/LAT Catalog}\label{sec3}
\subsection{Our sample of GRBs}
\subsubsection{GRB 080916C}

GRB 080916C triggered the Gamma Burst Monitor (GBM) instrument on board the {\itshape Fermi} satellite at 00:12:45.613542 UT on September 16, 2008 \citep{2008GCN..8245....1G}. The duration of prompt emission measured by the GBM instrument was $T_{90}=62.98\,{\rm s}$ correspondent to a fluence of $(4.0\pm0.6)\times 10^{-5}\,{\rm erg\,cm^{-2}}$ and an isotropic energy of $E_{\rm \gamma, iso}=(1.7\pm 0.1)\times 10^{54}\, {\rm erg}$ \citep{Ajello_2019}. This burst was detected at a redshift $z= 4.35\pm 0.15$ \citep{2008GCN..8272....1C} at R.A.=$08^{\rm h}07^{\rm m}12^{\rm s}$ and Dec=-$61^\circ 18'00"$ with a $2.8^\circ$ uncertainty at 68 percent. In the energy range of 10 keV to 10~GeV, the apparent isotropic energy release was measured as $8.8 \times  10^{54}\, {\rm erg}$ \citep{2009Sci...323.1688A}. The {\itshape Swift}/XRT instrument began observing GRB 080916C at 17:08 UT, 17 hours after trigger time \citep{2008GCN..8261....1P}. This instrument monitored GRB 080916C in the  Photon Counting (PC) mode from $6.1\times 10^4$ to $\sim10^6\,{\rm s}$. The best-fit intrinsic absorption column density is $1.49\times 10^{21}\,{\rm cm^{-2}}$. GRB 080916C was detected by the Gamma-Ray Burst Optical/Near-Infrared Detector in optical bands \citep[GROND;][]{2008GCN..8272....1C}.

\subsubsection{GRB 090323}
At 00:02:42.63 UT on March 23, 2009, GRB 090323 triggered the {\itshape Fermi}/GBM instrument \citep{2009GCN..9021....1O}. The duration of the prompt emission  measured by the GBM instrument was $T_{90}=133.89\,{\rm s}$, corresponding to a fluence of $(0.26\pm0.07)\times 10^{-5}\,{\rm erg\,cm^{-2}}$ and an isotropic energy of $E_{\rm \gamma, iso}=(3.0\pm 2.0)\times 10^{53}\, {\rm erg}$ \citep{Ajello_2019}. The {\itshape Fermi}/LAT instrument also detected this burst \citep{2009GCN..9024....1K, 2009GCN..9031....1P}. The  isotropic energy in the LAT energy range was $\simeq 4.1 \times 10^{54}$ erg \citep{2013ApJS..209...11A}. The {\itshape Swift}/XRT instrument began observing GRB 090323 at 19:27 UT, 19.4 hours after the trigger. This instrument monitored GRB 090323 in the PC mode with a spectrum exposure of 6 ks. The best-fit intrinsic absorption column density is $2.20\times 10^{22}\,{\rm cm^{-2}}$ with a redshift $z=3.57$, which was measured by  \cite{2009GCN..9028....1C}. \cite{2010A&A...516A..71M} described the burst's optical and infrared observations in great detail.

\subsubsection{GRB 090902B}
GRB 090902B was detected by the {\itshape Fermi}/GBM instrument on 2009 September 2 at 11:05:08.31 UT with coordinates R.A.= 17$^{\rm h}$38$^{\rm m}$26$^{\rm s}$ and Dec: 26$^\circ$30'. The duration of the prompt emission in the energy band 50 - 300 keV was $T_{90}=19.33\,{\rm s}$, corresponding to a fluence $(7.0\pm1.0)\times 10^{-5}\,{\rm erg\,cm^{-2}}$, and an isotropic energy of $E_{\rm \gamma, iso}=(3.7\pm 0.3) \times 10^{53}\, {\rm erg}$ \citep{Ajello_2019}. The burst was within the {\itshape Fermi}/LAT field of view initially at an angle of $51^\circ$ from the sight of vision. The X-ray afterglow was detected within the LAT error circle by the {\itshape Swift}/XRT \citep{2009GCN..9868....1K}, and Ultraviolet/Optical Telescope (UVOT)   \citep{2009GCN..9869....1S}, and later by several other ground-based telescopes. The XRT instrument  monitored GRB 090902B in the PC mode with a spectrum exposure of 13.5 ks. The best-fit intrinsic absorption column density is $2.3^{+0.8}_{-0.6}\times 10^{22}\,{\rm cm^{-2}}$ with a redshift $z=1.822$, which was determined by the Gemini-North telescope \citep{2009GCN..9873....1C}.

\subsubsection{GRB 090926A}
GRB 090926A was discovered by the {\itshape Fermi}/GBM at 04:20:26.99 UT on 26 September 2009, with coordinates R.A.=354.5$^\circ$ and Dec=-64.2$^\circ$ \citep{2011ApJ...729..114A}. The duration of the prompt emission, fluence and isotropic energy measured by GBM instrument were $T_{90}=13.76\,{\rm s}$, $(2.9\pm 0.3)\times 10^{-5}\,{\rm erg\,cm^{-2}}$, and 
$E_{\rm \gamma, iso}=(5.3\pm 0.4)\times 10^{53}\, {\rm erg}$, respectively \citep{Ajello_2019}. GRB 090926A presented a characteristic high-energy PL component that is different from the known Band function, according to a joint LAT and GBM data analysis \citep{2011ApJ...729..114A}. {\itshape Swift}/XRT, INTEGRAL SPI-ACS \citep{2009GCN..9933....1B}, Suzaku/WAM \citep{2009GCN..9951....1N}, CORONAS-Photon \citep{2009GCN.10009....1C}, the Konus-wind experiment \citep{2009GCN..9959....1G} and the Ultra-violet Optical Telescope (UVOT) instrument on board {\it Neil Gehrels Swift Observatory}  \citep{2009GCN..9942....1M}, all identified this burst separately. The {\itshape Swift}/XRT instrument began observing GRB 090926A, 13.9 hours after the trigger. This instrument monitored GRB 090926A in the PC mode since $4.7\times 10^4$ to $1.8\times10^6\,{\rm s}$. The best-fit intrinsic absorption column density is $3.0^{+4.0}_{-3.0}\times 10^{21}\,{\rm cm^{-2}}$ with a redshift $z=2.1062$, which was estimated using X-shooter installed on the Very Large Telescope \citep{2009GCN..9942....1M}.

\subsubsection{GRB 110731A}
GRB 110731A triggered the {\itshape Fermi}/GBM instrument at 11:09:29.94 UT on 2011 July 31 \citep{2013ApJ...763...71A}. Approximately 30~s after the GBM trigger, the Burst Alert Telescope (BAT) on board {\it Neil Gehrels Swift Observatory} was also set off, and it located GRB 110731A at R.A.=$18^{\rm h}42^{\rm m}05^{\rm s}$ and Dec=-$28^\circ 32'44"$ with an uncertainty of 3 arcmin. The duration of the prompt emission  measured by GBM instrument was $T_{90}=7.49\,{\rm s}$, corresponding to a fluence and an isotropic energy of $(0.930\pm 0.07)\times 10^{-5}\,{\rm erg\,cm^{-2}}$ and $E_{\rm \gamma, iso}=(1.2\pm 0.3)\times 10^{53}\, {\rm erg}$, respectively \citep{Ajello_2019}. The isotropic energy measured using a  PL and a Band function was $(7.6\pm 0.2)\times 10^{53}\,{\rm erg}$ \citep{2013ApJ...763...71A}. The {\itshape Swift}/XRT instrument started detecting GRB 110731A at 11:10:36.9 UT, 66.4~s after the BAT trigger \citep{2011GCN.12215....1O}. This instrument monitored GRB 110731A in the windowed-timing (WT) mode with a spectrum exposure of 573~s and the PC mode with a spectrum exposure of 7.5 ks. The best-fit intrinsic absorption column density is $4.4^{+3.1}_{-3.0}\times 10^{21}\,{\rm cm^{-2}}$. The Swift UVOT instrument began observing the burst's position 75~s after the BAT trigger \citep{2011GCN.12215....1O}. Following the {\itshape Swift}/BAT trigger, several efforts were undertaken to observe the burst's position, such as those by the Faulkes Telescopes North and South \citep{2011GCN.12216....1B}, the Nordic Optical Telescope equipped with ALFOSC \citep{2011GCN.12220....1M}, Konus-Wind \citep{2011GCN.12223....1G}, the EVLA \citep{2011GCN.12227....1Z}, the Suzaku Wide-band All-sky Monitor (WAM) \citep{2011GCN.12244....1H} and the SAO RAS and Terskol observatories \citep{2011GCN.12333....1M}. \cite{2011GCN.12225....1T} obtained spectroscopic observations with the GMOS-N instrument on Gemini-North, which led to an estimation of a redshift of $z=2.83$.

\subsubsection{GRB 130427A}
GRB 130427A triggered the {\itshape Fermi}/GBM instrument at 07:47:06.42 UTC on April 27, 2013 \citep{2013GCN.14473....1V}. Approximately 50~s after the initial trigger, it also triggered the {\itshape Swift}/BAT instrument. Following the initial triggers, there was a substantial follow-up campaign by extraterrestrial observatories and ground-based experiments (from the ultra-violet, optical and X-ray telescopes \citep{2014Sci...343...48M};  SPI-ACS/\textit{INTEGRAL} \citep{2013GCN.14484....1P}; \textit{AGILE} \citep{2013GCN.14515....1V}; \textit{Konus-Wind} \citep{2013GCN.14487....1G}; \textit{NuSTAR} \citep{2013ApJ...779L...1K}; \textit{RHESSI} \citep{2013GCN.14590....1S}; MAXI/GSC \citep{2013GCN.14462....1K}; VLT/X-shooter \citep{2013GCN.14491....1F}). The duration of the prompt emission measured by GBM instrument was $T_{90}=138.24\,{\rm s}$ \citep{Ajello_2019} with an isotropic energy of $E_{\rm \gamma, iso}=(1.7\pm 0.2)\times 10^{52}\, {\rm erg}$ \citep{Ajello_2019}. The total fluence and the isotropic energy measured in the 10 keV - 100~GeV  range was $4.9\times 10^{-3}\,{\rm erg\, cm^{-2}}$ and $1.4\times 10^{54}\,{\rm erg}$, respectively \citep{2014Sci...343...42A}. The {\itshape Swift}/XRT instrument started observing GRB 130427A at 07:50:17.7 UT, 140.2~s after the BAT trigger. This instrument monitored GRB 130427A in the WT mode with a spectrum exposure of 1.9~ks and the PC mode with a spectrum exposure of 4.1 ks. The best-fitting absorption column (intrinsic) is $2.7^{+0.8}_{-0.8}\times 10^{20}\,{\rm cm^{-2}}$ and $1.1^{+0.4}_{-0.4}\times 10^{21}\,{\rm cm^{-2}}$ for WT and PC modes, respectively. The Combined Array for Research in Millimeter-wave Astronomy (CARMA) localized this burst to R.A. = $173.1367^{\circ}$, Dec. = $27.6989^{\circ}$ (J2000) with an uncertainty of 0.4 arc sec \citep{2013GCN.14494}. Optical spectroscopy from Gemini-North conducted by \cite{2013GCN.14455....1L} found the redshift of the GRB to be $z=0.34$.

\subsubsection{GRB 160625B}
{\itshape Fermi}/GBM triggered and located GRB 160625B at 22:40:16.28 UT, 2016 June 25 \citep{2016GCN..19581...1B}. Immediately, {\itshape Fermi}/LAT triggered this burst at 22:43:24.82 UT \citep{2016GCN..19586...1D}. The duration of the prompt emission measured by the GBM instrument was $T_{90}=453.38\,{\rm s}$, corresponding to a fluence and an isotropic energy of $(2.5\pm 0.3)\times 10^{-5}\,{\rm erg\,cm^{-2}}$ and $E_{\rm \gamma, iso}=(1.5\pm 0.1)\times 10^{53}\, {\rm erg}$, respectively \citep{Ajello_2019}. The {\itshape Swift}/XRT instrument monitored GRB 160625B in the PC mode from $9.8\times10^4$ to $4.1\times 10^6\,{\rm s}$. The best-fit absorption column (intrinsic) is $1.6\pm 0.6\times 10^{21}\,{\rm cm^{-2}}$, with a redshift $z=1.406$  \citep{2016GCN..19600...1X}. Several optical observations were performed with background telescopes \citep{2016arXiv161203089Z, 2017Natur.547..425T}.

\subsubsection{GRB 180720B}
GRB 180720B was identified and immediately followed by GBM and LAT instruments, the two instruments onboard the {\itshape Fermi} satellite \citep{2018GCN.22981....1R, 2018GCN.22980....1B}, and BAT, XRT, and UVOT instruments onboard the Swift satellite \citep{2018GCN.22973....1P, 2018GCN.22998....1B}. The BAT instrument triggered this burst on July 20, 2018, at 14:21:44 UT and located it with coordinates: ${\rm R.A.=00^{\rm h} 02^{\rm m} 07^{\rm s}}\,$ and ${\rm Dec}= -02^{\rm d} 56' 00''\, (J2000)$ with an uncertainty of 3 arcmin. The duration of the prompt emission measured by the GBM instrument was $T_{90}=48.90\,{\rm s}$ correspondent to a fluence $(0.19\pm 0.05)\times 10^{-5}\,{\rm erg\,cm^{-2}}$, and an isotropic energy of $E_{\rm \gamma, iso}=(0.39\pm 0.09)\times 10^{52}\, {\rm erg}$ \citep{Ajello_2019}. The {\itshape Swift}/XRT instrument started observing GRB 180720B at 14:23:11.0 UT, 86.5~s after the BAT trigger. This instrument monitored GRB 180720B in the WT mode with a spectrum exposure of 2.2~ks and the PC mode with a spectrum exposure of 3.9 ks. The best-fitting absorption column (intrinsic) is $3.72^{+0.11}_{-0.11}\times 10^{21}\,{\rm cm^{-2}}$ and $3.4^{+0.6}_{-0.5}\times 10^{21}\,{\rm cm^{-2}}$ for WT and PC modes, respectively. GRB 180720B started to be monitored in the optical and near-infrared (NIR) bands on July 20, 2018, at 14:22:57 UT, 73~s after the trigger time \citep{2018GCN.22977....1S}. \cite{2018GCN.22996....1V} detected the optical emission of GRB 180720B using the VLT/X-shooter spectrograph, associating a redshift of $z=0.654$.

\subsection{Data analysis}

The data files used for the {\itshape Fermi}/LAT analysis were obtained from the online data website.\footnote{https://fermi.gsfc.nasa.gov/cgi-bin/ssc/LAT/LATDataQuery.cgi} {\itshape Fermi}/LAT data was analyzed in the 0.1-100~GeV  energy range and within the time-resolved likelihood analysis for each burst ($\rm{t_{LAT,0}}$, $\rm{t_{LAT,1}}$, see Table~1 from \cite{Ajello_2019} for the definition) with the {\itshape Fermi} Science tools\footnote{https://fermi.gsfc.nasa.gov/ssc/data/analysis/software/} in their conda-based installation.\texttt{ScienceTools 2.2.0}\footnote{https://github.com/{\itshape Fermi}/LAT/Fermitools-conda/wiki} For this analysis we adopt the responses reported for each burst by \cite{Ajello_2019}, following the unbinned likelihood analysis presented by the {\itshape Fermi}/LAT team.\footnote{https://fermi.gsfc.nasa.gov/ssc/data/analysis/scitools/likelihood\_tutorial.html}  Using the \texttt{gtselect} tool, we select, with an event class 32 nor 8 depending on the response chosen\footnote{https://fermi.gsfc.nasa.gov/ssc/data/analysis/documentation/Cicerone/\\Cicerone\_Data/LAT\_DP.html}, a region of interest (ROI) around the position of the burst within a radius of 15$^{\circ}$. We apply a cut to the zenith angle above 100$^{\circ}$. Then, we select the appropriate time intervals (GTIs) using the \texttt{gtmktime} tool on the selected data before considering the ROI cut. To define the model needed to describe the source, the diffuse components and point sources of 4FGL-DR3 reported by \cite{2022ApJS..260...53A} embedded in ROI \texttt{ make4FGLxml}\footnote{https://fermi.gsfc.nasa.gov/ssc/data/analysis/user/} were used. We define a point source at the position of this burst, assuming a PL spectrum, and we define a diffuse galactic component using GALPROP \texttt{gll\_iem\_v07} as well as the extragalactic background \texttt{iso\_P8R3\_SOURCE\_V3\_v1}\footnote{https://fermi.gsfc.nasa.gov/ssc/data/access/lat/BackgroundModels.html}. For each burst, the spectral index is fixed with the value reported by \cite{Ajello_2019} in their Table~4, letting free the normalisation and the normalisation of the diffuse component. We use \texttt{gtdiffrsp} to take into account all of these components. Following the likelihood procedure, we produce a lifetime cube with the tool \texttt{gtltcube}, using a step $\delta \theta=0.025$, a bin size of 0.5 and a maximum zenith angle of 100$^{\circ}$. The exposure map was created using \texttt{gtexpmap}, considering a region of 30$^{\circ}$ around the GRB position and defining 100 spatial bins in longitude/latitude and 50 energy bins, and we perform the likelihood analysis with \texttt{pyLikelihood}\footnote{https://fermi.gsfc.nasa.gov/ssc/data/analysis/scitools/python\_tutorial.html}. Finally, we obtain the photons with a probability greater than 90$\%$ to be associated to each burst with the \texttt{gtsrcprob} tool.

\begin{figure}

{\centering

\subfloat[Sub a) GRB 080916C][\centering{{\itshape Fermi}/LAT light curve for GRB 080916C.}]{
\includegraphics[scale=0.39]{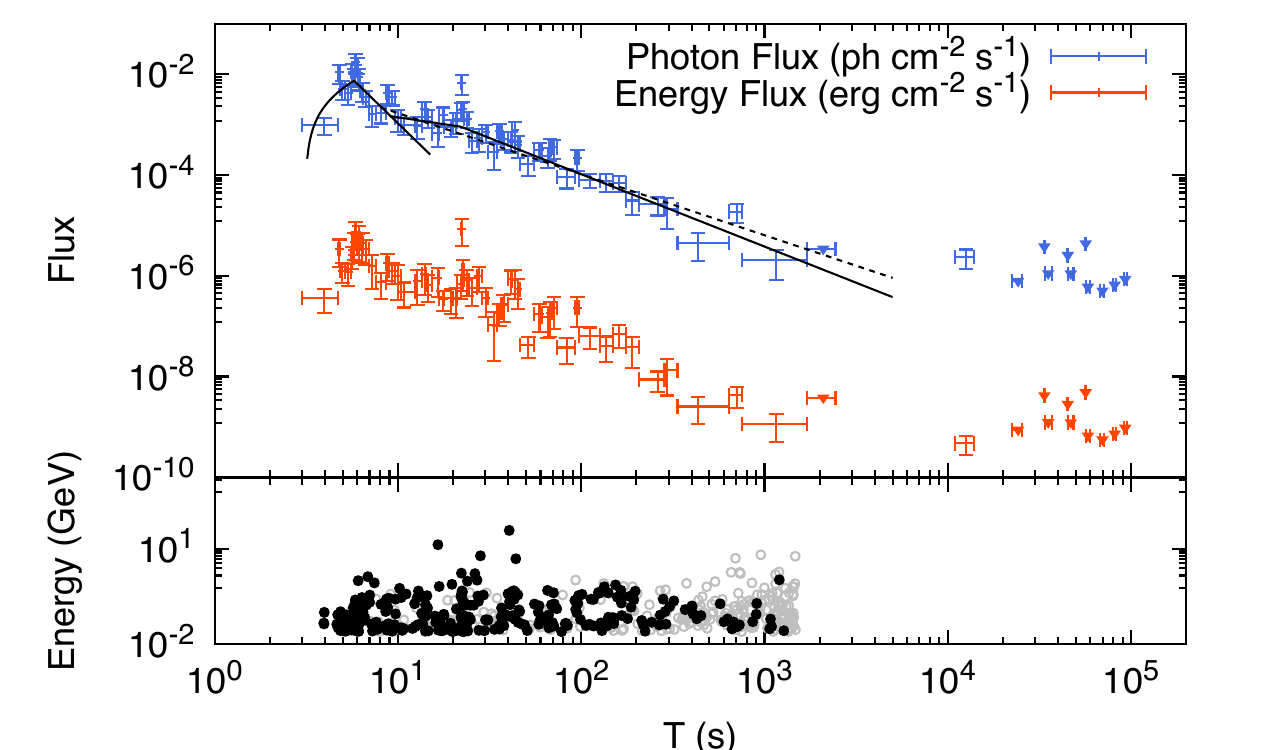}
\label{fig2a):GRB080916C}}
\subfloat[Sub b) GRB 090323][\centering{{\itshape Fermi}/LAT light curve for GRB 090323.}]{
\includegraphics[scale=0.39]{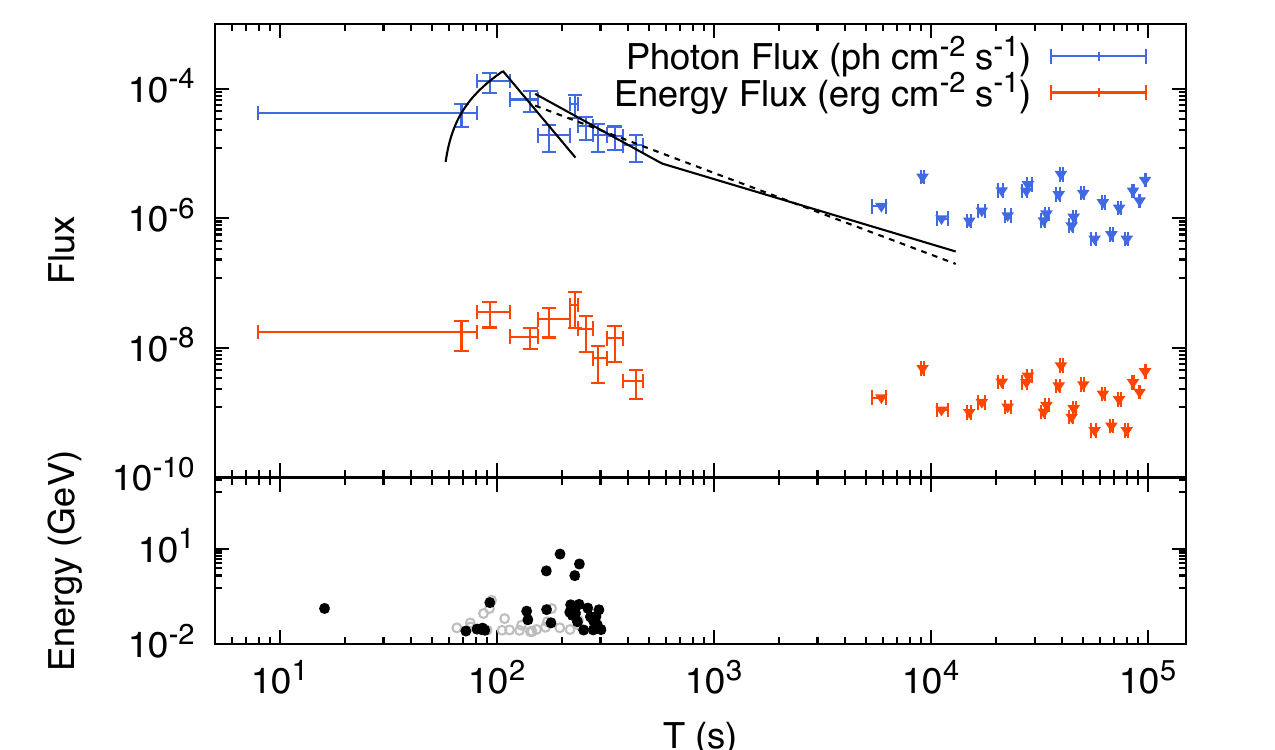}
\label{fig2b):GRB090323}}

\subfloat[Sub c) GRB 090902B][\centering{{\itshape Fermi}/LAT light curve for GRB 090902B.}]{
\includegraphics[scale=0.39]{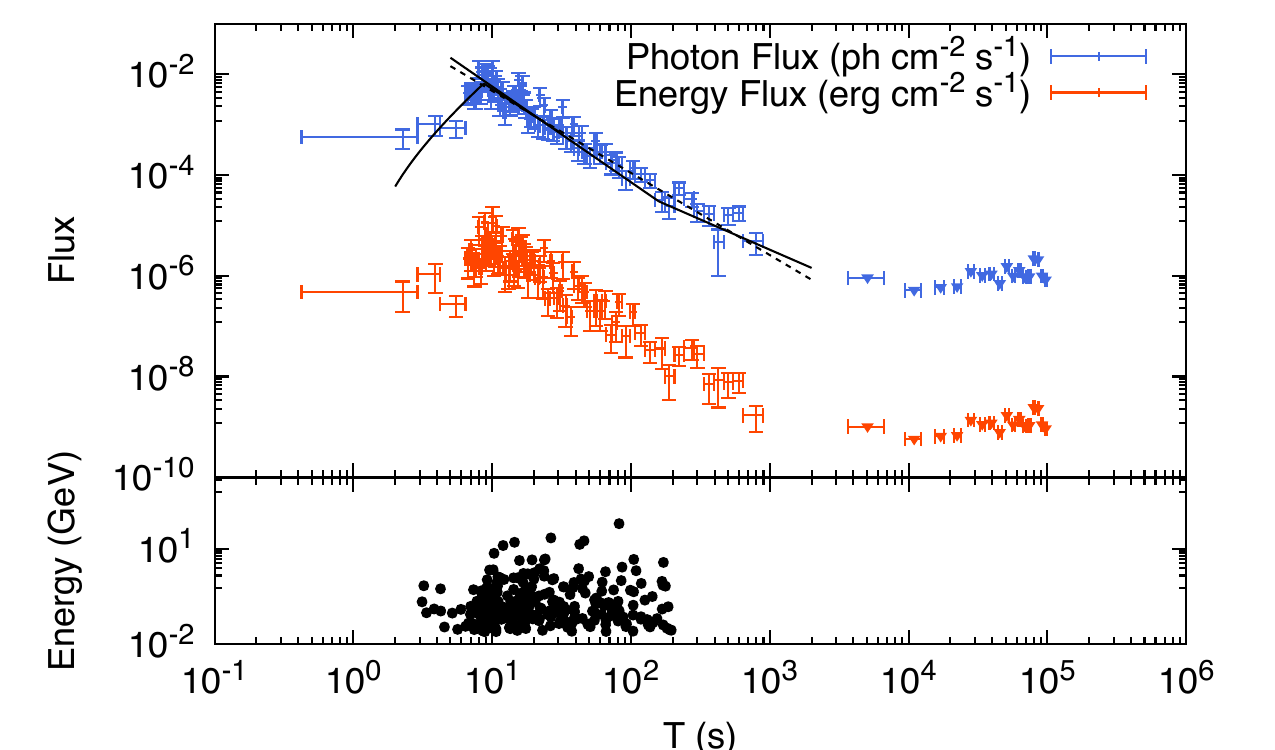}
\label{fig2d):GRB090926B}}
\subfloat[Sub d) GRB 090926A][\centering{{\itshape Fermi}/LAT light curve for GRB 090926A.}]{
\includegraphics[scale=0.39]{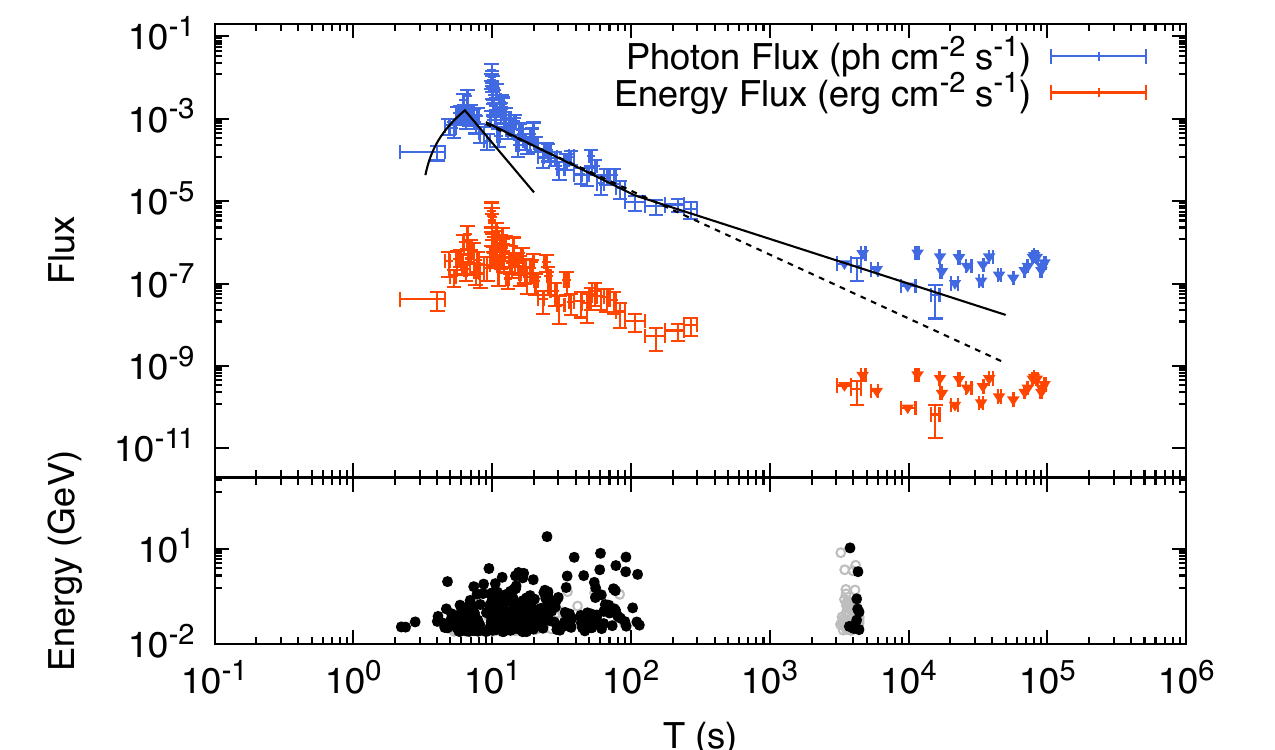}
\label{fig2e):GRB090926A}}

\subfloat[Sub e) GRB 110731A][\centering{{\itshape Fermi}/LAT light curve for GRB 110731A.}]{
\includegraphics[scale=0.39]{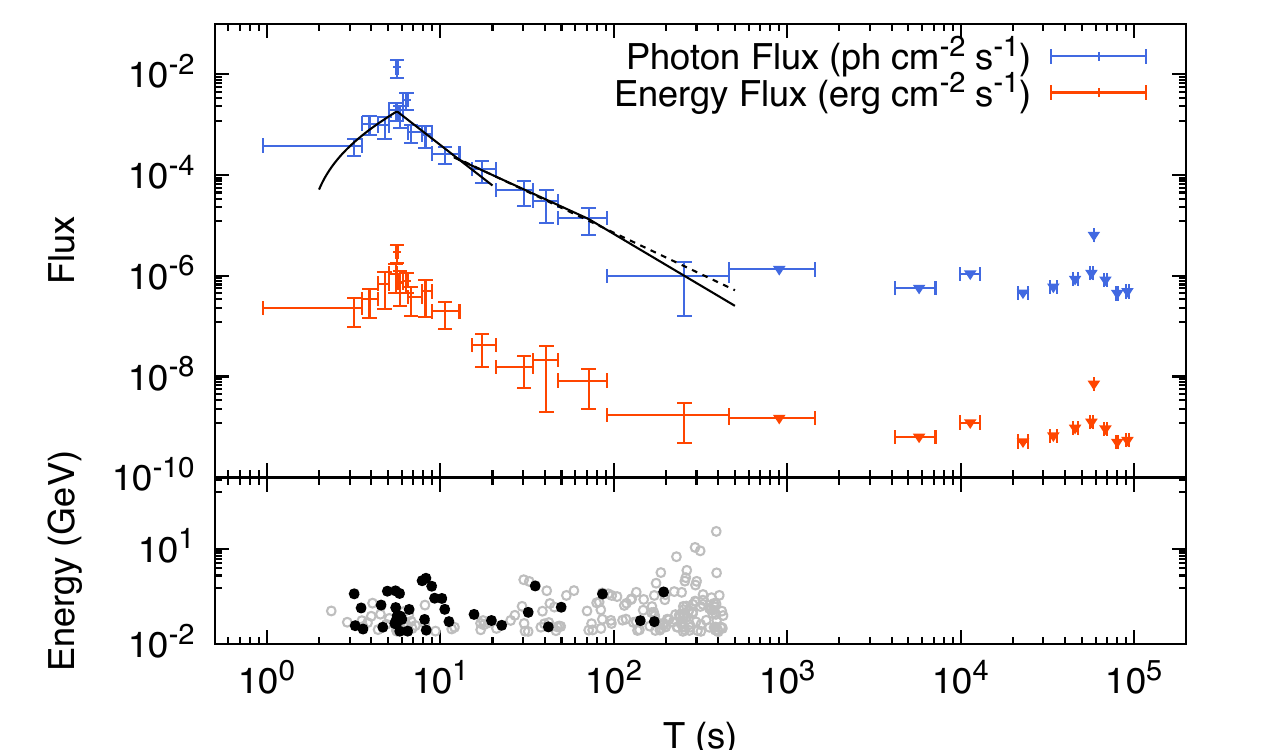}
\label{fig2f):GRB110731A}}
\subfloat[Sub f) GRB 130427A][\centering{{\itshape Fermi}/LAT light curve for GRB 130427A.}]{
\includegraphics[scale=0.39]{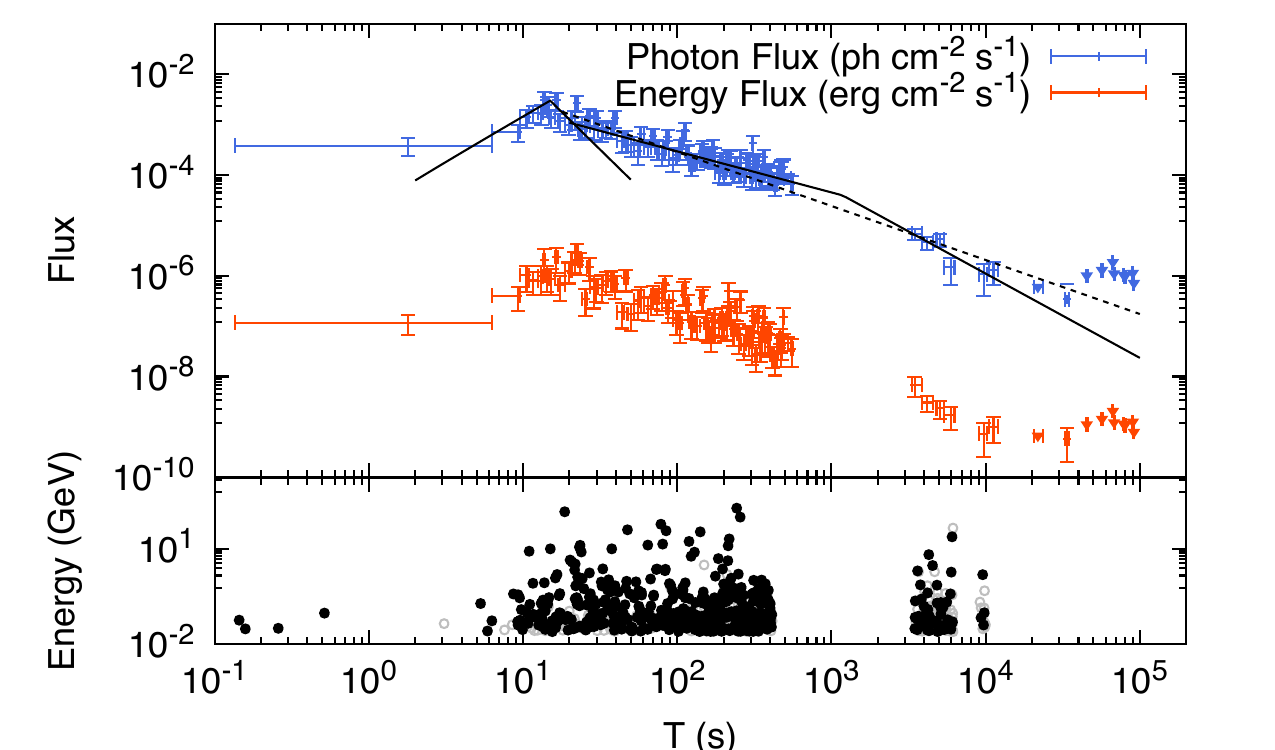}
\label{fig2g):GRB130427A}}

\subfloat[Sub g) GRB 160625B][\centering{{\itshape Fermi}/LAT light curve for GRB 160625B.}]{
\includegraphics[scale=0.39]{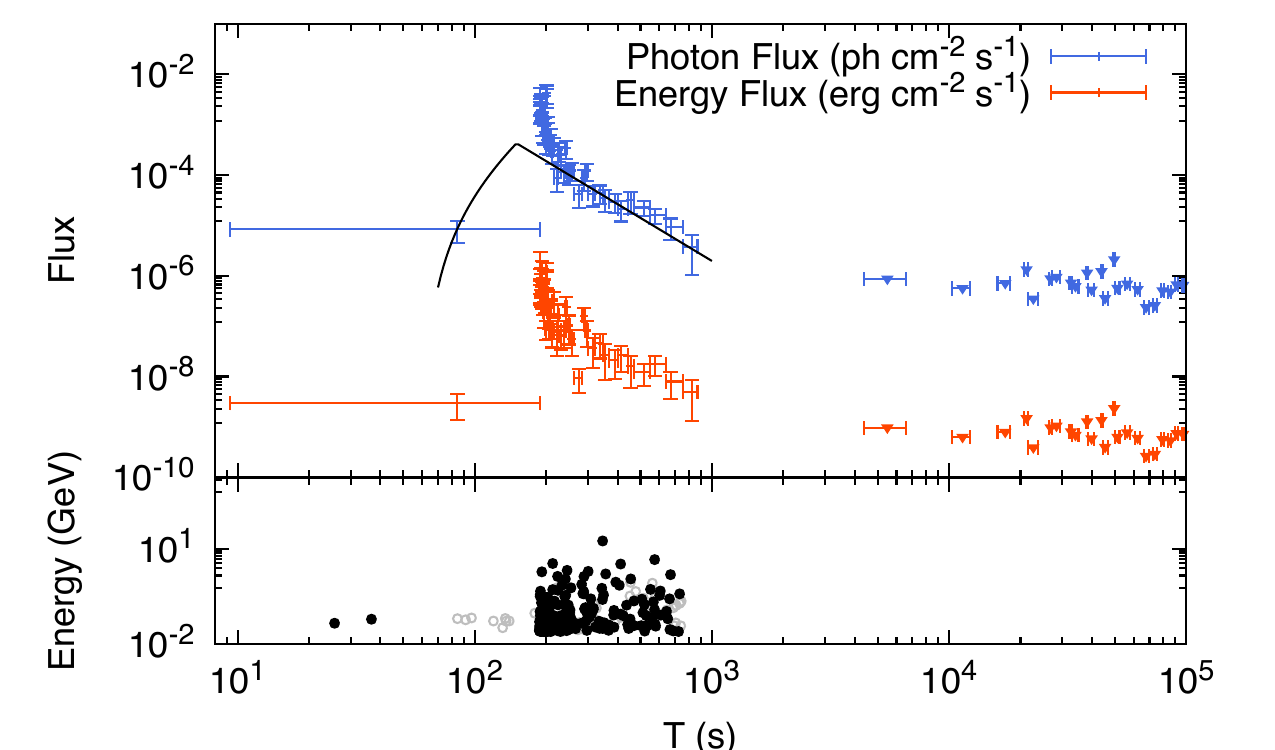}
\label{fig2h):GRB160625B}}
\subfloat[Sub h) GRB 180720B][\centering{{\itshape Fermi}/LAT light curve for GRB 180720B.}]{
\includegraphics[scale=0.39]{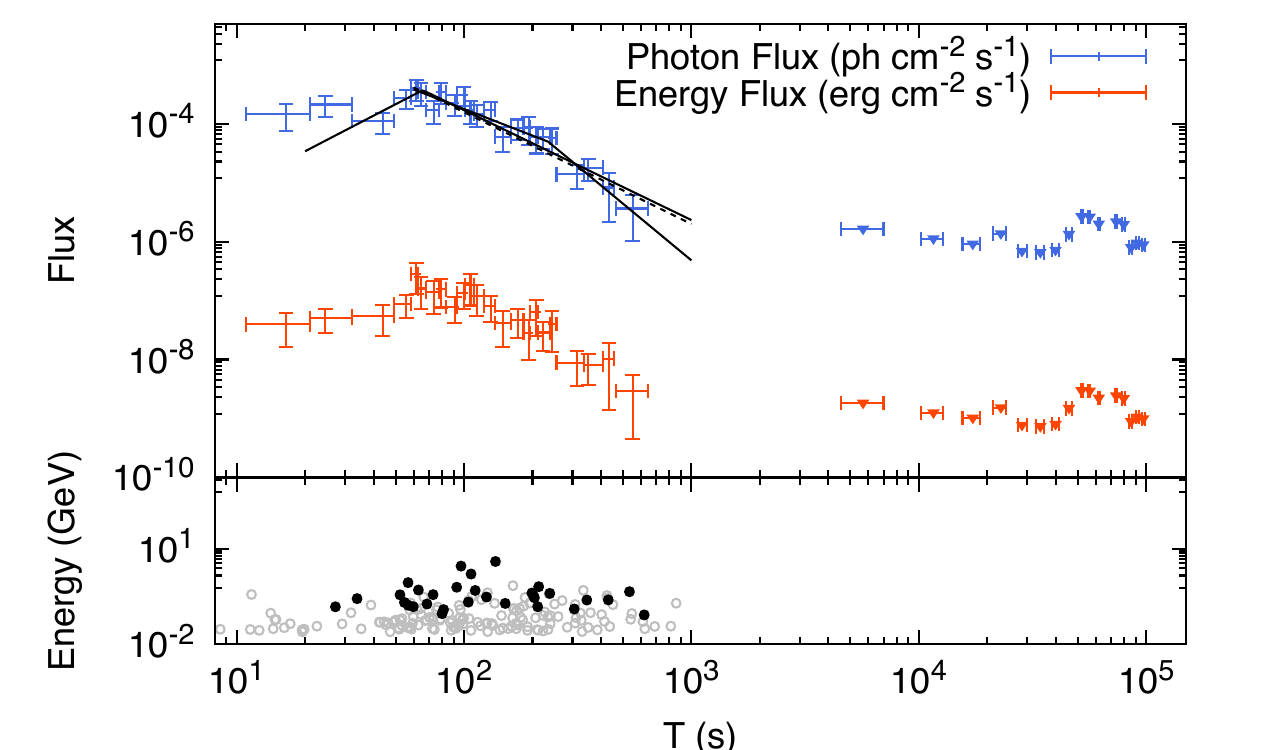}
\label{fig2i):GRB180720B}}

\caption{The photon flux (blue) and flux energy (red) detected by {\itshape Fermi}/LAT are shown in top panels. The photon energy during the entire time window are shown in bottom panels. The filled black dots correspond to a photon with a probability greater than 90\% of belonging to the GRB, while the open gray dots correspond to a photon with a probability lower than 90\% of belonging to the burst. } \label{fig2}
}
\end{figure}

The upper panels in each plot of Figure~\ref{fig2} show the {\itshape Fermi}/LAT energy flux (blue) and photon flux (red) light curves, and the lower panels exhibit all photons with energies $\geq 100$~MeV associated with GRB 080916C, GRB 090323, GRB 090902B, GRB 090926A, GRB 110731A, GRB 130427A, GRB 160625B and GRB 180720B. The filled circles in black correspond to the individual photons and their energies with probability $>90\%$ of being associated with the respective burst and the open circles in grey indicate the LAT gamma transient class photons.

In order to describe the short-lasting peak in the LAT observations, we use the function 
\begin{eqnarray}
\label{radio_s1_1}
F_{\rm L, se}(t)= A_{\rm se}  \begin{cases}
(t-T_a)^{-\alpha_{\rm r, bb}}\,\,\hspace{1cm}      t < t_{\rm r, br} , \cr
t^{-\alpha_{\rm r, ab}}\,\, \hspace{2cm}      t_{\rm r, br} <t\,,    \cr
\end{cases}
\end{eqnarray}
where $A_{\rm se}$ corresponds to the proportionality constant, $t_{\rm r,br}$ is the temporal break with the temporal  indexes before ($\alpha_{\rm r, bb}$) and after ($\alpha_{\rm r, ab}$) the break. The term $T_a$ is the starting time of the short-term component \citep{2007ApJ...655..973K, 2006Natur.442..172V}. For the long-lasting component, we use a PL function 
$F_{\rm L, ee}(t)=A_{ee}\,t^{-\alpha_{\rm f}}$ or BPL function 
\begin{eqnarray}
\label{radio_s1_2}
F_{\rm L, ee}(t)=A_{ee}  \begin{cases}
t^{-\alpha_{\rm f_j, bb}}\,\,\hspace{1.5cm}      t < t_{\rm f_j, br} , \cr
t^{-\alpha_{\rm f_j, ab}}\,\, \hspace{1.5cm}      t_{\rm f_j, br} <t\,,    \cr
\end{cases}
\end{eqnarray}
where $A_{\rm ee}$ corresponds to the proportionality constant, $t_{\rm f_j,br}$ is the temporal break with the temporal indexes before ($\alpha_{\rm f_j, bb}$) and after ($\alpha_{\rm f_j, ab}$) the break with ${\rm f=L}$, ${\rm X}$, ${\rm O}$ for LAT, X-ray and optical observations, respectively. The continuous and dashed lines on the photon energy flux correspond to the best-fit curves of the short- and long-lasting components, respectively. We fit each component of the energy flux in the LAT light curve with a series of broken and simple PLs. To find the best-fit values of the temporal indexes of PLs, we use the chi-square minimisation $\chi^2$ using the ROOT software package \citep{1997NIMPA.389...81B}. Table~\ref{Table4} lists the best-fit temporal indices of the short and long-lasting emissions, including the chi squares ($\chi^2$/n.d.f.).

Data sets from the X-ray Telescope (XRT) instrument on board the {\it Neil Gehrels Swift Observatory} were retrieved from the publicly available database on the official Swift website.\footnote{{\rm https://www.swift.ac.uk/burst\_analyser/00922968/}} The flux density at 10 keV is transformed to 1 keV using the conversion factor derived in \cite{2010A&A...519A.102E}.

\subsection{Afterglow evolution: LAT and multi-wavelength observations}

We report the best fit values for the short and long LAT components using ROOT Chi-square minimisation $\chi^2$ for the start time, the temporal breaks, and the indexes before and after the break in Table~\ref{Table4}. Additionally, with the chi-square minimisation $\chi^2$ of ROOT, we obtain the best-fit values of the X-ray and optical light curves, which we list in Table~\ref{Table5}.

\subsubsection{GRB 080916C}


 We note that the short-lasting component occurs during prompt emission ($t_{\rm x}< T_{90}$) because of the best-fit values of the starting time and the temporal break. Therefore, the RS evolves in the thick-shell regime. The best-fit value of the starting time indicates that the onset of the afterglow was very early. When comparing the best-fit values of the temporal indexes before and after the break of the SSC light curve, we can see that they are consistent with the cooling conditions $\nu^{\rm ssc}_{\rm c, r}<\nu_{\rm LAT}<\nu^{\rm ssc}_{\rm m, r}$ for $t< t_{\rm x}$ and $\nu^{\rm ssc}_{\rm m, r}<\nu_{\rm LAT}<\nu^{\rm ssc}_{\rm cut, r}$ for $t_{\rm x}< t$, for $p\approx 2.2$. It should be noted that the decay temporal index is consistent with the high-latitude emission ($\nu^{\rm ssc}_{\rm cut, r}<\nu_{\rm LAT}$); $\beta+2$. Other cooling conditions of the SSC light curve cannot explain the best-fit values of the temporal indexes.

\paragraph{High-energy events.}
At 3.94~s after the GBM trigger, the first high-energy photon was detected with a measured energy of 161.4~MeV. The energy range of the photons in this burst was extensive, with 283 photons over 100~MeV, 17 exceeding 1~GeV, and 2 exceeding 10~GeV. At 40.5~s after the GBM trigger, the highest-energy photon in the LAT data had a value of 27.43~GeV.

\paragraph{Multiwavelength afterglow analysis }

 Analysis of LAT, X-ray, and optical spectra with PL functions leads to spectral indexes of $\beta_{\rm LAT}=\Gamma_{\rm LAT} - 1= 1.20\pm0.06$ \citep{Ajello_2019}, $\beta_{\rm X}= \Gamma_{\rm X} - 1= 0.80\pm 0.40$\footnote{https://www.swift.ac.uk/xrt\_spectra/00020082/} and $\beta_{\rm Opt}=0.38\pm0.20$ \citep{2009A&A...498...89G}, respectively. Taking into consideration the LAT analysis of the long-lasting emission and the best-fit values of the temporal and spectral indexes of LAT, X-ray and optical observations, the observed fluxes evolve as {\small $F_{\rm \nu,L}\propto t^{-1.21\pm0.06}\,\nu^{-1.20\pm0.06}$, $F_{\rm \nu,X}\propto t^{-1.35\pm0.06}\,\nu^{-0.80\pm 0.30}$ and $F_{\rm \nu,Opt}\propto t^{-1.41\pm0.04}\,\nu^{-0.38\pm0.20}$}, respectively. The fact that the temporal (spectral) indexes for the optical and X-ray observations are larger (lower) than for the LAT observations suggests that the closure relations of the synchrotron FS model evolve in a slow-cooling regime through a wind-like medium ($k=2$) for $ p\approx 2.3\pm0.3$. Note that the spectral and temporal indices of the optical and X-ray observations are consistent with each other, so the synchrotron closure relations evolve under the condition $\nu^{\rm syn}_{\rm m, f}<\nu_{\rm Opt}< \nu_{\rm X} < \nu^{\rm syn}_{\rm c, f}$. Similarly, LAT observations indicate that the closure relations evolve under the cooling condition $\nu^{\rm syn}_{\rm c, f}< \nu_{\rm LAT}$.

\subsubsection{GRB 090323}

 We see that the short-lasting emission is present during the prompt episode ($t_{\rm x}< T_{90}$), so the RS occurs in the thick-shell regime. According to the best-fitting value of the starting time, the onset of the afterglow occurred at the end of the prompt episode. We can infer the evolution of the LAT frequency in the cooling conditions of the SSC light curve by comparing the best-fit values of the temporal indices before and after the break. In this case, $\nu_{\rm LAT}$ evolves as $\nu^{\rm ssc}_{\rm c, r}<\nu_{\rm LAT}<\nu^{\rm ssc}_{\rm m, r}$ for $t< t_{\rm x}$ and as $\nu^{\rm ssc}_{\rm m, r}<\nu_{\rm LAT}<\nu^{\rm ssc}_{\rm cut, r}$ for $t_{\rm x}< t$, for $p\approx 2.2$. The decay temporal index, $\beta+2$, agrees with the high-latitude emission ($\nu^{\rm ssc}_{\rm cut, r}<\nu_{\rm LAT}$), which is interesting to note.

\paragraph{High-energy events.}
The first high-energy photon, with a measured energy of 363.4~MeV, was detected 15.05~s after the GBM trigger.
In this burst, there were 32 photons with energy over 100~MeV and 5 with energies above 1~GeV. The highest energy photon detected in the LAT data was 7.40~GeV, which occurred 145.91~s after the GBM trigger.

\paragraph{Multiwavelength afterglow analysis }

We use the best-fit spectral indices presented in 2FLGC \citep[$\beta_{\rm LAT}= 1.30\pm0.20$;][]{Ajello_2019} and {\itshape Swift}/XRT repository ($\beta_{\rm X}= 0.80^{+0.34}_{-0.21}$)\footnote{https://www.swift.ac.uk/xrt\_spectra/00020102/} to evaluate the LAT and X-ray data, respectively. Given the GROND and TLS observations of the optical/NIR afterglow, \cite{2010A&A...516A..71M} found a spectral index of $\beta_{\rm Opt}=0.65\pm0.13$, considering the dust extinction similar to Small Magellanic Cloud (SMC) with $A^{\rm host}_{V}=0.14^{+0.04}_{-0.03}$. Using the LAT analysis of the log-lasting emission and the best-fit values of the temporal and spectral indexes of LAT, X-ray, and optical observations, the observed fluxes evolve as {\small $F_{\rm \nu,L}\propto t^{-1.26\pm0.40}\,\nu^{-1.30\pm0.10}$, $F_{\rm \nu,X}\propto t^{-1.58 \pm 0.08}\,\nu^{-0.80^{+0.34}_{-0.21}}$ and $F_{\rm \nu,Opt}\propto t^{-1.70\pm0.04}\,\nu^{-0.65\pm0.13}$}, respectively. The fact that the temporal (spectral) index for the optical and X-ray observations is larger (lower) than the LAT data suggests that the closure relations of the synchrotron FS model evolves in a slow cooling regime through a wind-like medium for $ p\approx 2.3\pm0.3$. It should be noted that the spectral and temporal indices of X-ray and optical observations are compatible with each other, and therefore the synchrotron closure relations evolve under the condition $\nu^{\rm syn}_{\rm m, f}<\nu_{\rm Opt}< \nu_{\rm X}<\nu^{\rm syn}_{\rm c, f}$. Moreover, LAT observations indicate that closure relations evolve under the condition $\nu^{\rm syn}_{\rm c, f} < \nu_{\rm LAT}$.

\subsubsection{GRB 090902B}

Comparing the best-fit values of the starting time and the temporal break, one can see that this component is observed during the prompt episode ($t_{\rm x} < T_{90}$), then the RS is in the thick-shell regime. The best fit values of the temporal indices before and after the break of the SSC light curve show that they are consistent with the cooling conditions $\nu^{\rm ssc}_{\rm c, r}<\nu_{\rm LAT}<\nu^{\rm ssc}_{\rm m, r}$ for $t< t_{\rm x}$ and $\nu^{\rm ssc}_{\rm m, r}<\nu_{\rm LAT}<\nu^{\rm ssc}_{\rm c, r}$ or $\nu^{\rm ssc}_{\rm cut, r}<\nu_{\rm LAT}$ for $t_{\rm x}< t$, for $p\approx 2.1$. The best-fit values of the temporal indices cannot be explained by any other cooling conditions of the SSC light curve.

\paragraph{High-energy events.}
At 1.86~s after the GBM trigger, the first high-energy photon was detected with a measured energy of 284.4~MeV. The energy range of the photons in this burst was extensive, with 469 photons over 100~MeV, 67 exceeding 1~GeV, and 7 exceeding 10~GeV. At 81.7~s after the GBM trigger, the highest-energy photon in the LAT data had a measured energy of 39.88~GeV.

\paragraph{Multiwavelength afterglow analysis }

We consider the best-fit spectral index reported in 2FLGC  \citep[$\beta_{\rm LAT}= 0.94\pm0.09$;][]{Ajello_2019}. \cite{2010ApJ...714..799P} analysed the afterglow X-ray and UV-optical IR observations of GRB 090902B. After the modelling of the broadband SED, the authors reported spectral indexes of $\beta_{\rm X}= 0.9\pm 0.1$ and $\beta_{\rm Opt}=0.68\pm 0.11$ for X-ray and UV-optical-IR observations, respectively, considering the SMC-like dust extinction with $A^{\rm host}_{V}=0.20\pm 0.06$. Given the analysis of the long-lasting LAT component and the best-fit values of the temporal and spectral indexes of LAT, X-ray, and optical observations, the observed fluxes evolve as {\small $F_{\rm \nu,L}\propto t^{-1.35\pm0.06}\,\nu^{-0.94\pm0.09}$, $F_{\rm \nu,X}\propto t^{-1.62\pm 0.15}\,\nu^{-0.80^{+0.34}_{-0.21}}$ and $F_{\rm \nu,Opt}\propto t^{-1.70\pm0.04}\,\nu^{-0.65\pm0.13}$}, respectively. The fact that the temporal (spectral) indexes for the optical and X-ray observations are larger (lower) than the LAT data suggests that the closure relations of the synchrotron FS model evolves in a slow cooling regime through a wind-like medium for $ p\approx 2.3\pm0.3$. It should be noted that the spectral and temporal indices of X-ray and optical observations are compatible with each other and, therefore, the synchrotron closure relations evolve under condition $\nu^{\rm syn}_{\rm m, f}<\nu_{\rm Opt}< \nu_{\rm X}<\nu^{\rm syn}_{\rm c, f}$, and the LAT observations evolve under condition $\nu^{\rm syn}_{\rm c, f} < \nu_{\rm LAT}$.

\subsubsection{GRB 090926A}
  Given the best-fit values of the onset time and the temporal break, we see that the short-lasting emission is present during the main episode ($t_{\rm x}< T_{90}$), so the RS happens in the thick-shell regime. According to the best-fitting value of the starting time, the onset of the afterglow occurred at the end of the prompt episode. We can infer the evolution of the LAT frequency in the cooling conditions of the SSC light curve by comparing the best-fit values of the temporal indices before and after the break. 
In this case, $\nu_{\rm LAT}$ evolves as $\nu^{\rm ssc}_{\rm c, r}<\nu_{\rm LAT}<\nu^{\rm ssc}_{\rm m, r}$ for $t< t_{\rm x}$ and as $\nu^{\rm ssc}_{\rm m, r}<\nu_{\rm LAT}<\nu^{\rm ssc}_{\rm c, r}$ for $t_{\rm x}< t$, for $p\approx 2.2$. We want to emphasise that the temporal index of decay agrees with the high-latitude emission that occurs under the cooling condition $\nu^{\rm ssc}_{\rm cut, r}<\nu_{\rm LAT}$.


\paragraph{High-energy events.}
A 130.6~MeV photon, the first of several high-energy photons, was detected 2.21~s after the GBM trigger.
There was a broad range of photon energies in this burst, with 339 having energies larger than 100~MeV, 31 having energies greater than 1~GeV, and 2 having energies greater than 10~GeV. About 24.84~s from the GBM trigger, the LAT data recorded a photon with a peak energy of 19.46~GeV.

\paragraph{Multiwavelength afterglow analysis }

We use the best-fit spectral indices presented in 2FLGC \citep[$\beta_{\rm LAT}= 1.14\pm0.05$;][]{Ajello_2019} and {\itshape Swift}/XRT repository ($\beta_{\rm X}= 0.98^{+0.15}_{-0.14}$)\footnote{https://www.swift.ac.uk/xrt\_spectra/00020113/} to evaluate the LAT and X-ray data, respectively. Using the afterglow observations of GROND and {\itshape Swift} / XRT in three different periods; $\approx T_0$ + 84 ks, $\approx T_0$ + 290~ks and $\approx T_0$ + 1.3 Ms, \cite{2010ApJ...720..862R} reported spectral indexes of $\beta_{\rm Opt}=1.02^{+0.03}_{-0.02}$, $1.05^{+0.04}_{-0.02}$ and $1.04^{+0.08}_{-0.06}$, respectively. Given the best-fit values of the temporal and spectral indices of LAT, X-ray, and optical observations, the observed fluxes evolve as {\small $F_{\rm \nu,L}\propto t^{-1.05\pm0.22}\,\nu^{-1.14\pm0.05}$, $F_{\rm \nu,X}\propto t^{-1.41\pm 0.03}\,\nu^{-0.98^{+0.15}_{-0.14}}$, and $F_{\rm \nu, Opt}\propto t^{-1.34\pm0.06}\,\nu^{-1.04^{+0.08}_{-0.06}}$}, respectively. The fact that the temporal (spectral) indexes for the optical and X-ray observations are larger (lower) than the LAT data suggests that the closure relations of the synchrotron FS model evolve in a slow cooling regime through a wind-like medium for $ p\approx 2.3\pm0.3$. It should be noted that the spectral and temporal indices of X-ray and optical observations are compatible with each other, and therefore the synchrotron closure relations evolve under condition $\nu^{\rm syn}_{\rm m, f}<\nu_{\rm Opt}< \nu_{\rm X}<\nu^{\rm syn}_{\rm c, f}$, and the LAT observations evolve under condition $\nu^{\rm syn}_{\rm c, f} < \nu_{\rm LAT}$.

\subsubsection{GRB 110731A}

We find that as short-lasting emission is present throughout the main episode, RS must occur in the thick-shell regime ($t_{\rm x}< T_{90}$), as shown by the best fits of the onset time and the temporal break. By comparing the best-fit values of the temporal indices before and after the break, we can deduce in which cooling conditions of the SSC light curve the LAT frequency evolves. For $t< t_{\rm x}$, $\nu_{\rm LAT}$ evolves as $\nu^{\rm ssc}_{\rm c, r}<\nu_{\rm LAT}<\nu^{\rm ssc}_{\rm m, r}$ when the onset time and the temporal break are not close to each other, and otherwise $\nu_{\rm LAT}$ could evolve $\nu_{\rm LAT}<\nu^{\rm ssc}_{\rm c, r}$ or
$\nu^{\rm ssc}_{\rm c, r} < \nu_{\rm LAT}$. For $t_{\rm x}< t$, $\nu_{\rm LAT}$ could evolve under the cooling conditions $\nu^{\rm ssc}_{\rm m, r}<\nu_{\rm LAT}<\nu^{\rm ssc}_{\rm c, r}$ for $p\approx 2.1$. The decay temporal index, $\beta+2$, agrees with the high-latitude emission, which is interesting to note.

\paragraph{High-energy events.}
The first high-energy photon, with a measured energy of 817.1~MeV, was observed 3.19~s after the GBM trigger. A large variety of photon energies were produced during this burst; 40 photons were more than 100~MeV and 4 more than 1~GeV. The highest energy photon recorded by the LAT occurred 1.93~s after the GBM trigger and had a value of 8.27~GeV.

\paragraph{Multiwavelength afterglow analysis }

We use the best-fit spectral indices presented in 2FLGC \citep[$\beta_{\rm LAT}= 1.3\pm0.2$;][]{Ajello_2019} to analyzed the LAT data. \cite{2013ApJ...763...71A} analysed the X-ray and UV/optical afterglow observations of GRB 110731A. After modelling the broadband SED, the authors reported at 550~s the spectral indexes of $\beta_{\rm X}= 0.95^{+0.07}_{-0.09}$ and $\beta_{\rm Opt}=0.45^{+0.07}_{-0.09}$ for X-ray and UV/optical observations, respectively. Given the LAT analysis and the best-fit values of the temporal and spectral indexes of LAT, late X-ray and optical observations, the closure relations are {\small $F_{\rm \nu,L}\propto t^{-1.14\pm0.23}\,\nu^{-1.3\pm0.2}$, $F_{\rm \nu,X}\propto t^{-1.30^ {+0.07}_{-0.05}}\,\nu^{-0.78^{+0.15}_{-0.13}}$, and $F_{\rm \nu,Opt}\propto t^{-1.38\pm0.09}\,\nu^{-0.45^{+0.07}_{-0.09}}$}, respectively. 
 The fact that the temporal (spectral) indexes for the optical and X-ray observations are larger (lower) than the LAT data suggests that the closure relations of the synchrotron FS model evolve in a slow cooling regime through a wind-like medium for $ p\approx 2.3\pm0.3$. It should be noted that the spectral and temporal indices of X-ray and optical observations are compatible with each other and, therefore, the synchrotron closure relations evolve under condition $\nu^{\rm syn}_{\rm m, f}<\nu_{\rm Opt}< \nu_{\rm X}<\nu^{\rm syn}_{\rm c, f}$, and the LAT observations evolve under condition $\nu^{\rm syn}_{\rm c, f} < \nu_{\rm LAT}$.

\subsubsection{GRB 130427A}

We find that as short-lasting emission is present throughout the main episode, RS must occur in the thick-shell regime ($t_{\rm x}< T_{90}$), as shown by the best fits of the onset time and the temporal break. By comparing the best-fit values of the temporal indices before and after the break, we can deduce in which cooling conditions of the SSC light curve the LAT frequency evolves. For $t< t_{\rm x}$, the {\itshape Fermi}/LAT band should evolve under the condition $\nu^{\rm ssc}_{\rm c, r}<\nu_{\rm LAT}<\nu^{\rm ssc}_{\rm m, r}$ when the onset time is near the temporal break. On the other hand, the {\itshape Fermi}/LAT band could evolve $\nu_{\rm LAT}<\nu^{\rm ssc}_{\rm c, r}$ or $\nu^{\rm ssc}_{\rm c, r} < \nu_{\rm LAT}$. For $t_{\rm x}< t$, the {\itshape Fermi} / LAT band could evolve under the cooling condition $\nu^{\rm ssc}_{\rm m, r}<\nu_{\rm LAT}<\nu^{\rm ssc}_{\rm c, r}$ for $p\approx 2.1$. We emphasise that the temporal index of decay agrees with the high-latitude emission and also with a contribution from the FS region.

\paragraph{High-energy events.}
With a measured energy of 189.8~MeV, the first high-energy photon was observed 0.14~s after the GBM trigger. Four hundred sixty-two photons in this burst had energies greater than 100~MeV, 78 were greater than 1~GeV, and 15 were greater than 10~GeV, indicating an extensive range of energies. The most energetic photon seen in the LAT data was clocked in at 94.11~GeV, 243.1~s after the GBM trigger.

\paragraph{Multiwavelength afterglow analysis }

According to the LAT and X-ray observations, we consider the best-fit spectral indexes reported in 2FLGC  \citep[$\beta_{\rm LAT}= 0.99\pm0.04$;][]{Ajello_2019} and Swift repository ($\beta_{\rm X}= 0.495\pm 0.016$ for $T_0+1196\,{\rm s}$ and $0.69\pm 0.08$ for $T_0+44907\,{\rm s}$)\footnote{https://www.swift.ac.uk/xrt\_spectra/00020113/}, respectively. Using UV/optical/IR afterglow observations collected by several ground telescopes, \cite{2014ApJ...781...37P} reported a spectral index evolution between $\beta_{\rm Opt}= 0.42^{+0.06}_{-0.06}$ and $0.48\pm 0.04$ for $t<$1 day. Given the best-fit values of the temporal and spectral indices of LAT, X-ray, and optical observations, the observed fluxes evolve as {\small $F_{\rm \nu,L}\propto t^{-1.02\pm0.2}\,\nu^{-0.99\pm0.04}$, $F_{\rm \nu,X}\propto t^{-1.30\pm 0.05}\,\nu^{-0.495^{+0.015}_{-0.016}}$, and $F_{\rm \nu,Opt}\propto t^{-1.33\pm0.04}\,\nu^{-0.48\pm 0.04}$}, respectively.  The fact that the temporal (spectral) index for optical and X-ray observations is larger (lower) than the LAT data suggests that the closure relations of the synchrotron FS model evolve in a slow cooling regime through a wind-like medium for $ p\approx 2.3\pm0.3$. It should be noted that the spectral and temporal indices of X-ray and optical observations are compatible with each other, and therefore the synchrotron closure relations evolve under the condition $\nu^{\rm syn}_{\rm m, f}<\nu_{\rm Opt}< \nu_{\rm X}<\nu^{\rm syn}_{\rm c, f}< \nu_{\rm LAT}$. 

\subsubsection{GRB 160602B}

 We see that the short-lived emission is present during the main episode ($t_{\rm x}< T_{90}$), so the RS evolves in the thick-shell regime. According to the best-fitting value of the starting time, the onset of the afterglow occurred at the end of the prompt episode. We can infer the evolution of the LAT frequency in the cooling conditions of the SSC light curve by comparing the best-fit values of the temporal indices before and after the break. For $t< t_{\rm x}$, the {\itshape Fermi}/LAT band should evolve under the condition $\nu^{\rm ssc}_{\rm c, r}<\nu_{\rm LAT}<\nu^{\rm ssc}_{\rm m, r}$ when the onset time is close to the temporal break. On the other hand, the {\itshape Fermi}/LAT band could evolve under conditions $\nu_{\rm LAT}<\nu^{\rm ssc}_{\rm c, r}$ or $\nu^{\rm ssc}_{\rm c, r} < \nu_{\rm LAT}$. For $t_{\rm x}< t$, $\nu_{\rm LAT}$ could evolve under the cooling condition $\nu^{\rm ssc}_{\rm m, r}<\nu_{\rm LAT}<\nu^{\rm ssc}_{\rm c, r}$ for $p\approx 2.6$. We note that the temporal index of decay agrees with the high-latitude emission $\nu^{\rm ssc}_{\rm c, r}<\nu_{\rm LAT}$.

\paragraph{High-energy events.}
The first high-energy photon was detected 25.6~s after the BAT trigger with an energy of 160.2~MeV. The energy range of the photons in this burst was very broad, with 255 photons over 100~MeV, 21 exceeding 1~GeV, and one over 10~GeV. The highest energy photon was observed 346.2~s after the BAT trigger by the LAT instrument and had an energy of 15.3~GeV.

\paragraph{Multiwavelength afterglow analysis }

We use the best-fit temporal and spectral index exhibited in {\itshape Swift}/XRT repository ($\alpha_{\rm X}=1.27\pm 0.30$ and $\beta_{\rm X}=0.66^{+0.21}_{-0.14}$)\footnote{https://www.swift.ac.uk/xrt\_spectra/00020113/} to evaluate the X-ray data. After modeling the broadband SED, \cite{2017ApJ...848...15F} and \cite{2017Natur.547..425T} reported spectral indexes of $\beta_{\rm Opt}=0.71\pm0.12$ and $\beta_{\rm Opt}=0.50\pm0.05$, after $\approx$ 3 and 8 hours for optical- NIR observations, respectively, after the trigger time. The late afterglow (the X-ray and optical fluxes) evolves as {\small $F_{\rm \nu,X}\propto t^{-1.27\pm 0.30}\,\nu^{-0.66^{+0.21}_{-0.14}}$ and $F_{\rm \nu,Opt}\propto t^{-0.921\pm0.163}\,\nu^{-0.71\pm 0.12}$}, respectively. Due to the fact that the temporal and spectral indexes for the late optical and X-ray observations are compatible with each other, the synchrotron closure relations can evolve through a wind-like medium for $ p\approx 2.3\pm0.3$ and under the condition $\nu^{\rm syn}_{\rm m, f}<\nu_{\rm Opt}< \nu_{\rm X}<\nu^{\rm syn}_{\rm c, f}$. 

\subsubsection{GRB 180720B}

This component occurs during the prompt emission because of the best-fit values of the starting time and the temporal break. Therefore, the RS evolves in the thick-shell regime ($t_{\rm x}< T_{90}$). We can infer the evolution of the LAT frequency in the cooling conditions of the SSC light curve by comparing the best-fit values of the temporal indices before and after the break. For $t< t_{\rm x}$, the {\itshape Fermi}/LAT band should evolve under the condition $\nu^{\rm ssc}_{\rm c, r}<\nu_{\rm LAT}<\nu^{\rm ssc}_{\rm m, r}$ when the onset time is close to the temporal break. On the other hand, the {\itshape Fermi}/LAT band could evolve $\nu_{\rm LAT}<\nu^{\rm ssc}_{\rm c, r}$ or $\nu^{\rm ssc}_{\rm c, r} < \nu_{\rm LAT}$. For $t_{\rm x}< t$, the {\itshape Fermi} / LAT band could evolve under the cooling condition $\nu^{\rm ssc}_{\rm m, r}<\nu_{\rm LAT}<\nu^{\rm ssc}_{\rm c, r}$ for $p\approx 2.6$. The decay temporal index, $\beta+2$, agrees with the high-latitude emission, which is interesting to note. This behavior could evolve with high-energy emission from the FS.

\paragraph{High-energy events.}
The first high-energy photon was detected 12.5~s after the BAT trigger with a measured energy of 175.2~MeV. The energy range of the photons in this burst was extensive, with 129 photons over 100~MeV and 8 exceeding 1~GeV. The highest energy photon was observed 142.4~s after the BAT trigger by the LAT instrument and had an energy of 4.9~GeV.

\paragraph{Multiwavelength afterglow analysis }

According to the LAT and X-ray observations, we consider the best-fit spectral indexes reported in 2FLGC  \citep[$\beta_{\rm LAT}= 1.23\pm0.10$;][]{Ajello_2019} and Swift repository ($\beta_{\rm X}= 0.697^{+0.010}_{-0.010}$)\footnote{https://www.swift.ac.uk/xrt\_spectra/00848890/}, respectively. Using optical observations in the R band collected by several ground telescopes, \cite{2019ApJ...885...29F} reported a spectral index evolution between $\beta_{\rm Opt}= 0.68\pm0.06$ and $0.70\pm 0.05$ corresponding to the intervals $\sim 2.0\times 10^2 - 2.5\times 10^3$ and $\sim 2.5\times 10^3 - 2.6\times 10^5$, respectively. Given the best-fit values of the temporal and spectral indexes of the X-ray and optical observations, while the X-ray flux varies from {\small $F_{\rm \nu,X}\propto t^{-0.79\pm 0.08}\,\nu^{-0.86\pm0.03}$} to {\small $F_{\rm \nu,X}\propto t^{-1.26\pm 0.06}\,\nu^{-0.697^{+0.010}_{-0.010}}$}, the optical flux evolves {\small $F_{\rm \nu,Opt}\propto t^{-1.22\pm0.02}\,\nu^{-0.68\pm 0.06}$} throughout the interval. Before $t\lesssim 2.5\times 10^3\,{\rm s}$, the temporal (spectral) index for optical observations is larger (lower) than X-ray data, suggesting that the closure relations of the synchrotron FS model evolve in slow-cooling regime through a wind-like medium for $ p\approx 2.0\pm0.25$. In this case, the synchrotron closure relations evolve under condition $\nu^{\rm syn}_{\rm m, f}<\nu_{\rm Opt} < \nu^{\rm syn}_{\rm c, f} < \nu_{\rm X}$. After $2.5\times 10^3\,{\rm s} \lesssim t$, the spectral and temporal indices of X-ray and optical observations are compatible with each other, and therefore the synchrotron closure relations evolve under condition $\nu^{\rm syn}_{\rm m, f}<\nu_{\rm Opt} < \nu_{\rm X} < \nu^{\rm syn}_{\rm c, f} $. We conclude that the temporal break at $\simeq 2.5\times 10^3\,{\rm s}$ to the passage of the forward shock cooling break through the {\itshape Swift}/XRT band.

\subsection{Results and Discussion}

\begin{figure}
    \centering{
    
        \subfloat[Sub a) GRB 080916C][\centering{ {\itshape Fermi}/LAT light curve for GRB 080916C.}]{\includegraphics[scale=0.65]{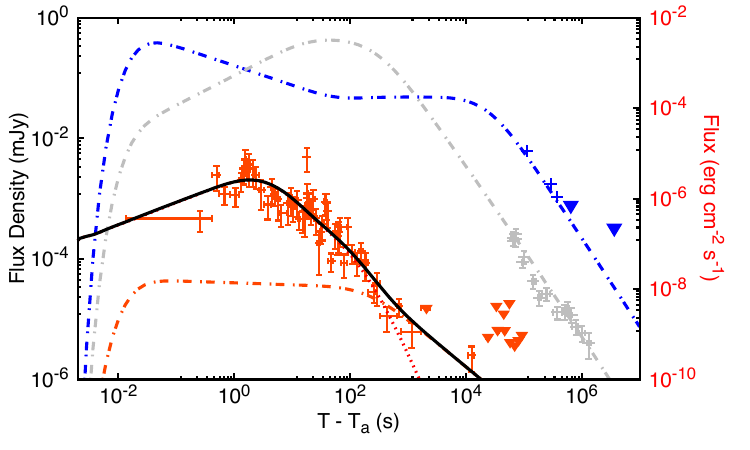}
        \label{fig5a):GRB080916C}}
        \subfloat[Sub b) GRB 090323][\centering{ {\itshape Fermi}/LAT light curve for GRB 090323.}]{\includegraphics[scale=0.65]{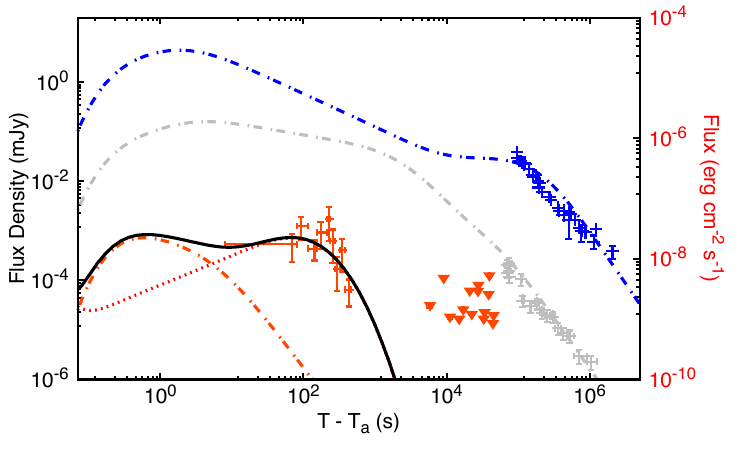}
        \label{fig5b):GRB090323}}

        \subfloat[Sub c) GRB 090902B][\centering{{\itshape Fermi}/LAT light curve for GRB 090902B.}]{\includegraphics[scale=0.65]{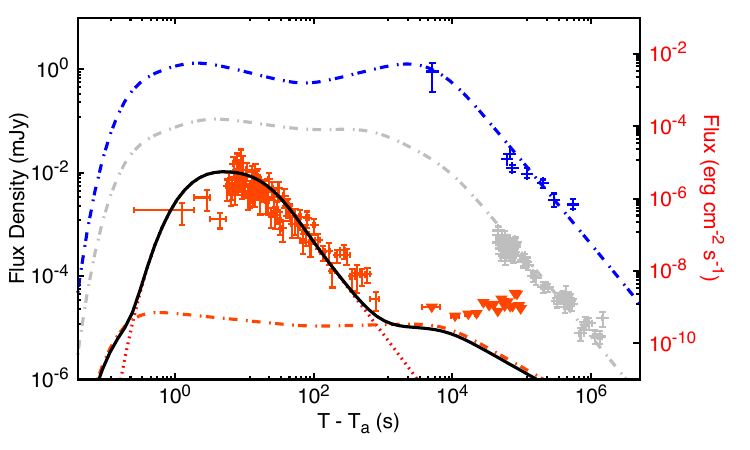}
        \label{fig5d):GRB090902B}}
        \subfloat[Sub d) GRB 090926A][\centering{{\itshape Fermi}/LAT light curve for GRB 090926A.}]{\includegraphics[scale=0.65]{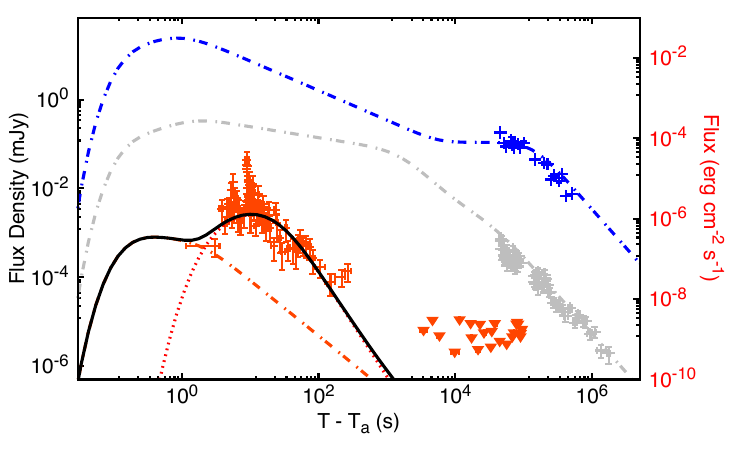}
        \label{fig5e):GRB090926A}}

        \subfloat[Sub e) GRB 110731A][\centering{{\itshape Fermi}/LAT light curve for GRB 110731A.}]{\includegraphics[scale=0.65]{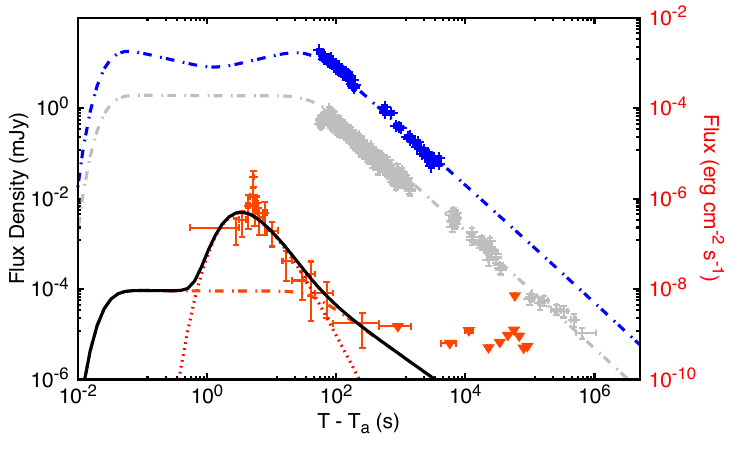}
        \label{fig5f):GRB110731A}}
        \subfloat[Sub f) GRB 130427A][\centering{{\itshape Fermi}/LAT light curve for GRB 130427A.}]{\includegraphics[scale=0.65]{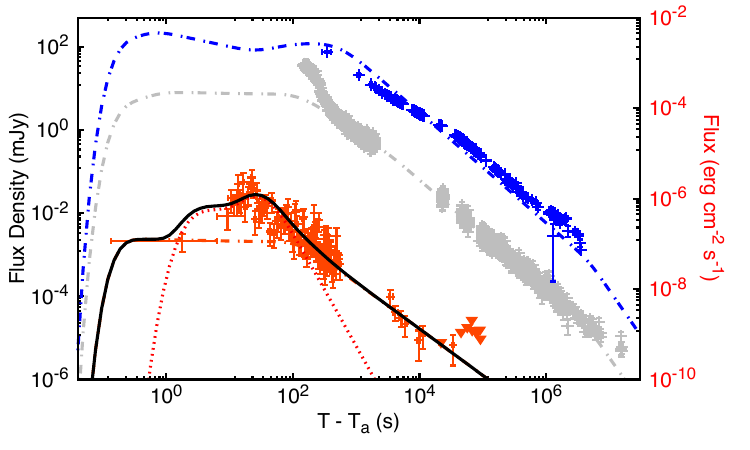}
        \label{fig5g):GRB130427A}}

        \subfloat[Sub g) GRB 160625B][\centering{{\itshape Fermi}/LAT light curve for GRB 160625B.}]{\includegraphics[scale=0.65]{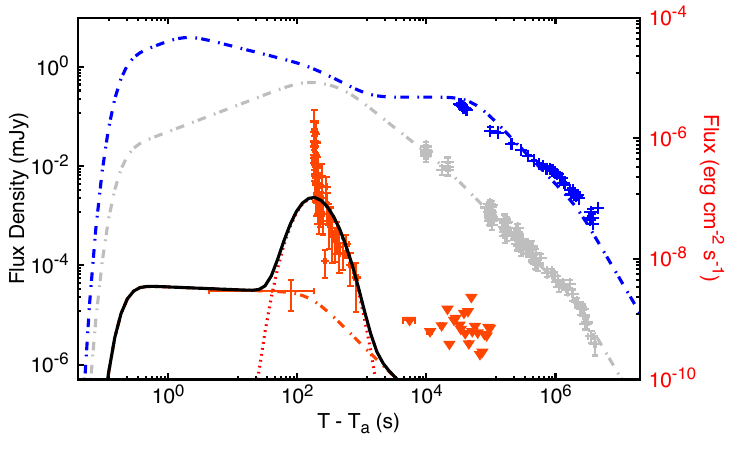}
        \label{fig5h):GRB160625B}}
        \subfloat[Sub h) GRB 180720B][\centering{{\itshape Fermi}/LAT light curve for GRB 180720B.}]{\includegraphics[scale=0.65]{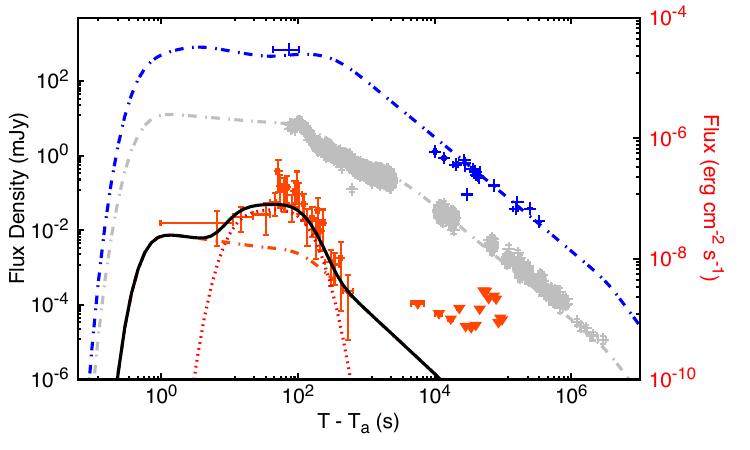}
        \label{fig5i):GRB180720B}}
    }
    \caption{The LAT, X-ray and optical observations of our sample of GRBs with the best-fit curve of the short- (dotted) and long- (dotted-dashed) lasting components. The total emission is displayed with a solid line and $T_a$ corresponds to the starting time obtained from the best-fit parameters listed in Table \ref{Table4}.}\label{fig5}
\end{figure}

\begin{figure}
    \centering
    \includegraphics[scale=0.35]{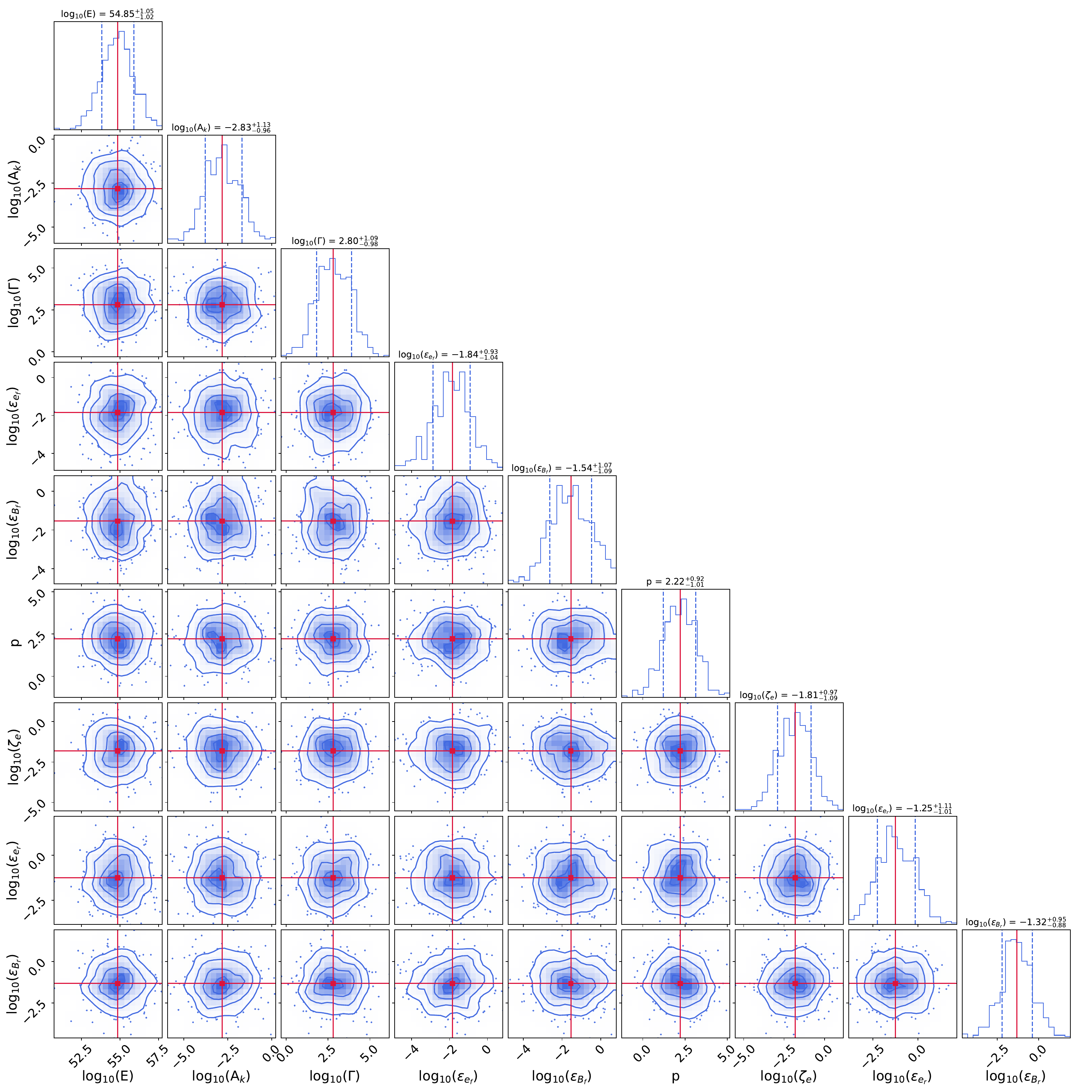}
    \caption{Results of our MCMC parameter estimation are shown in the corner plot for the synchrotron FS and SSC RS model in GRB 080916C. The marginalized posterior densities for each parameter are shown by the histograms on the diagonal, the contours are shown for 1$\sigma$, 2$\sigma$ and 3$\sigma$, and the median values in red lines.}
    \label{fig:CornerGRB080916C}
\end{figure}

\begin{figure}
    \centering
    \includegraphics[scale=0.35]{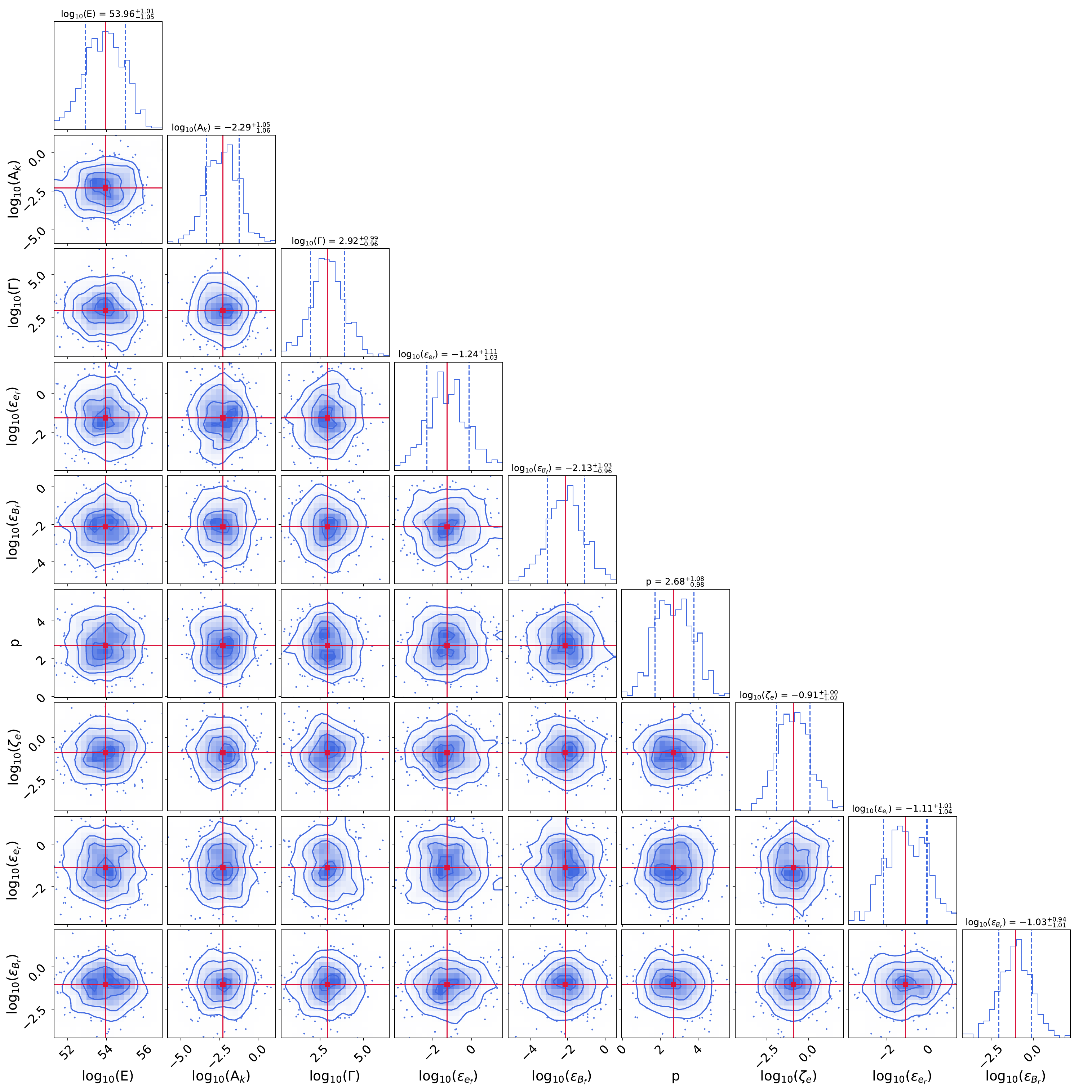}
    \caption{Same as Figure~\ref{fig:CornerGRB080916C}, but for GRB 090323.}
    \label{fig:CornerGRB090323}
\end{figure}

\begin{figure}
    \centering
    \includegraphics[scale=0.35]{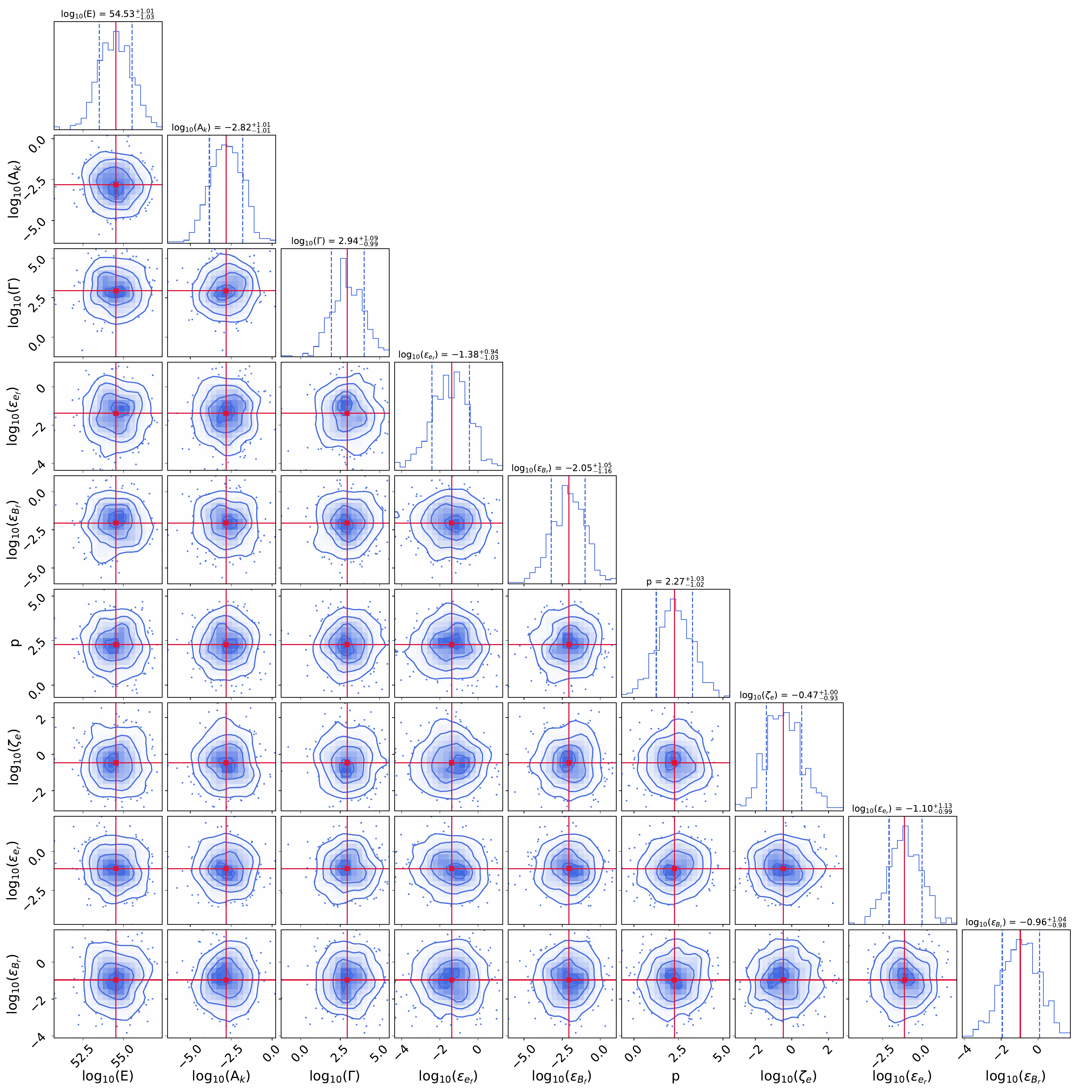}
    \caption{Same as Figure~\ref{fig:CornerGRB080916C}, but for GRB 090902B.}
    \label{fig:CornerGRB090902B}
\end{figure}

\begin{figure}
    \centering
    \includegraphics[scale=0.35]{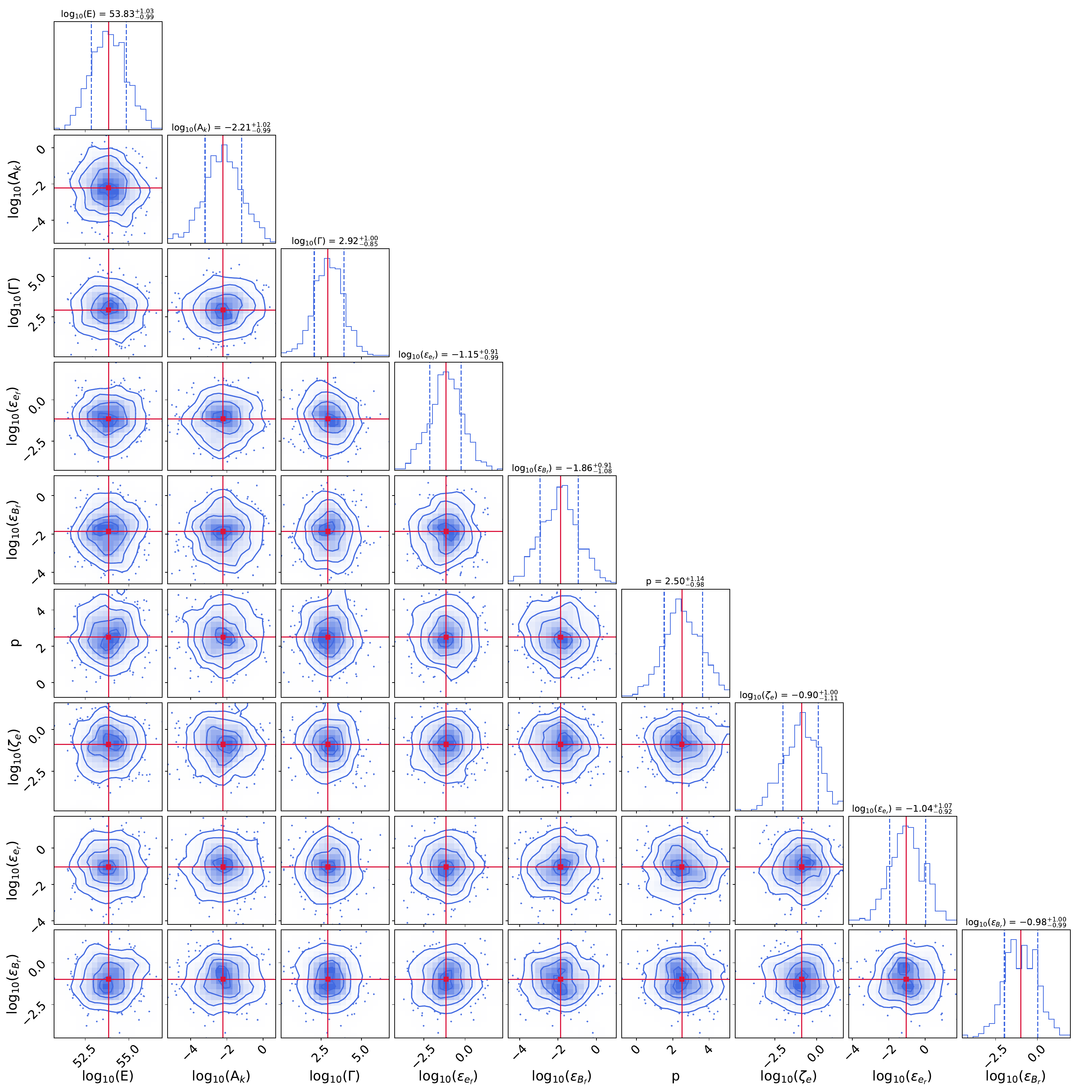}
    \caption{Same as Figure~\ref{fig:CornerGRB080916C}, but for GRB 090926A}
    \label{fig:CornerGRB090926A}
\end{figure}

\begin{figure}
    \centering
    \includegraphics[scale=0.35]{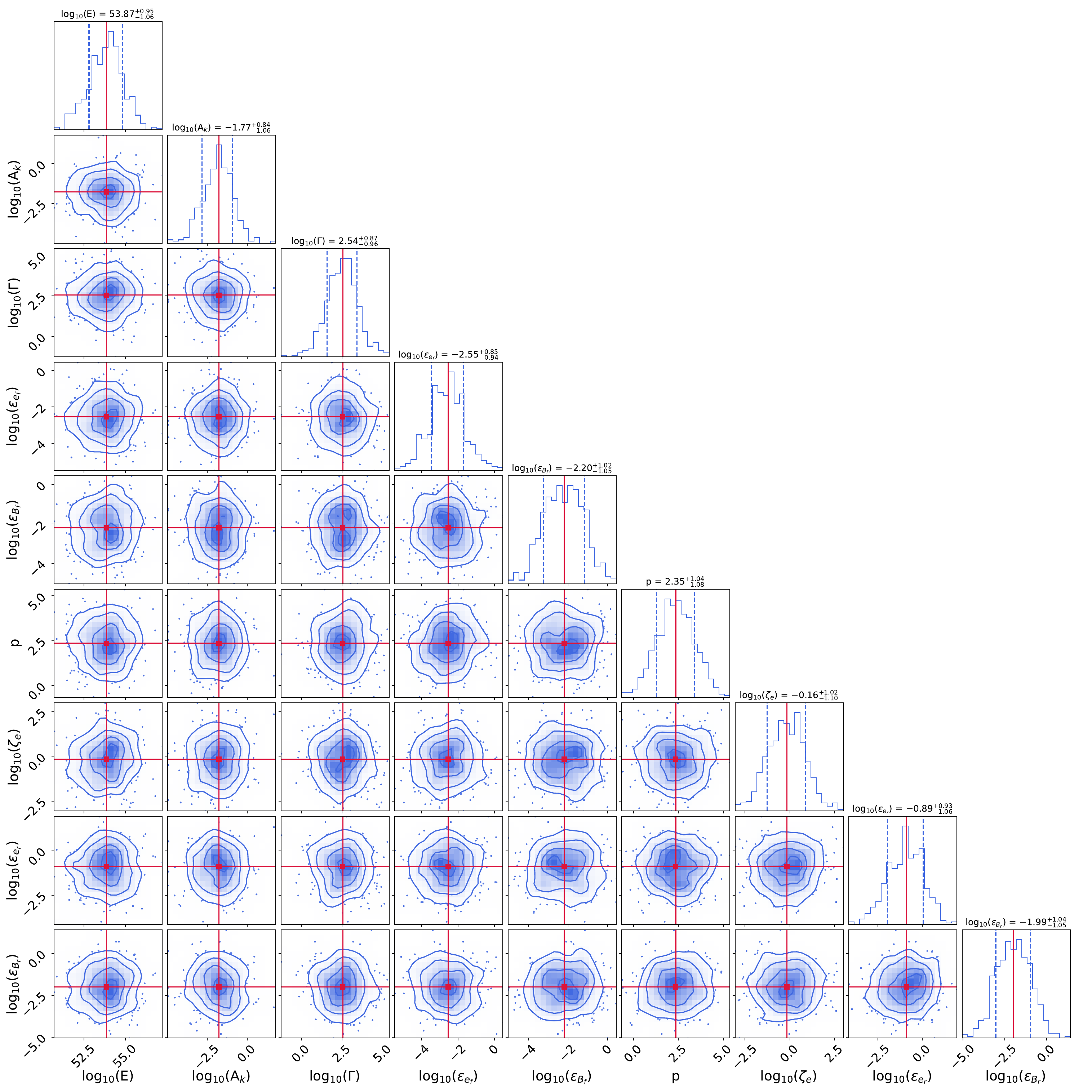}
    \caption{Same as Figure~\ref{fig:CornerGRB080916C}, but for GRB 110731A.}
    \label{fig:CornerGRB110731A}
\end{figure}

\begin{figure}
    \centering
    \includegraphics[scale=0.35]{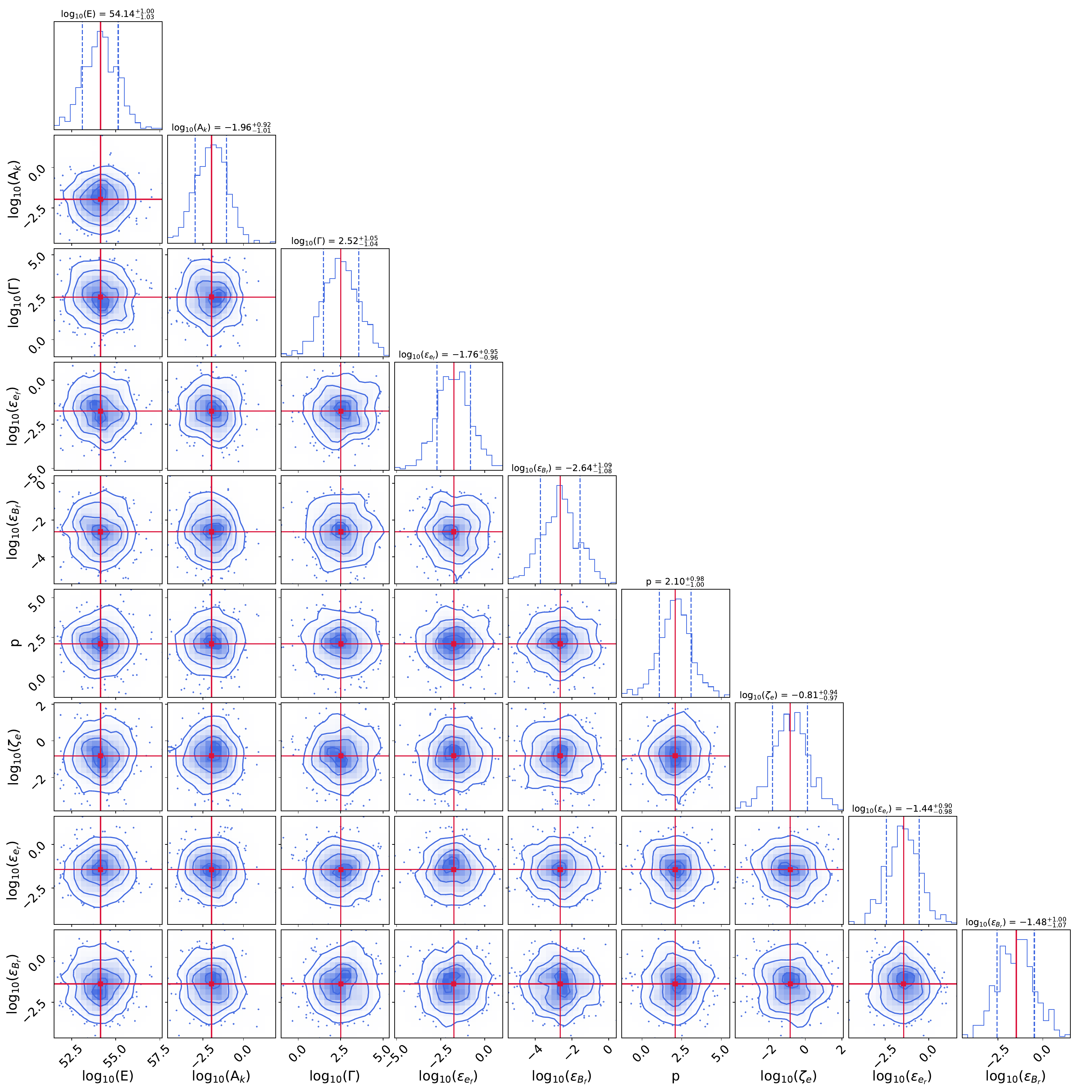}
    \caption{Same as Figure~\ref{fig:CornerGRB080916C}, but for GRB 130427A.}
    \label{fig:CornerGRB130427A}
\end{figure}

\begin{figure}
    \centering
    \includegraphics[scale=0.35]{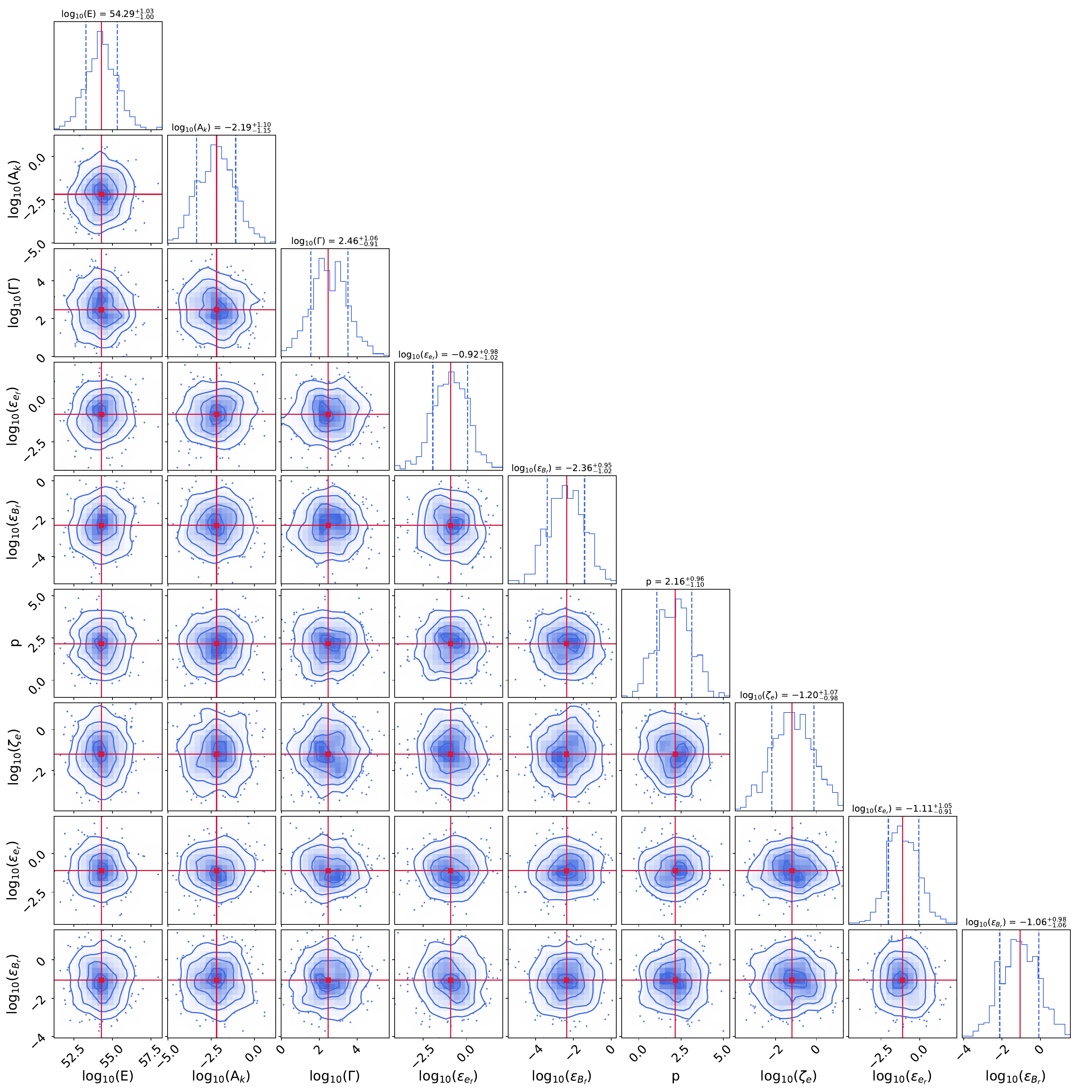}
    \caption{Same as Figure~\ref{fig:CornerGRB080916C}, but for GRB 160625B.}
    \label{fig:CornerGRB160625B}
\end{figure}

\begin{figure}
    \centering
    \includegraphics[scale=0.35]{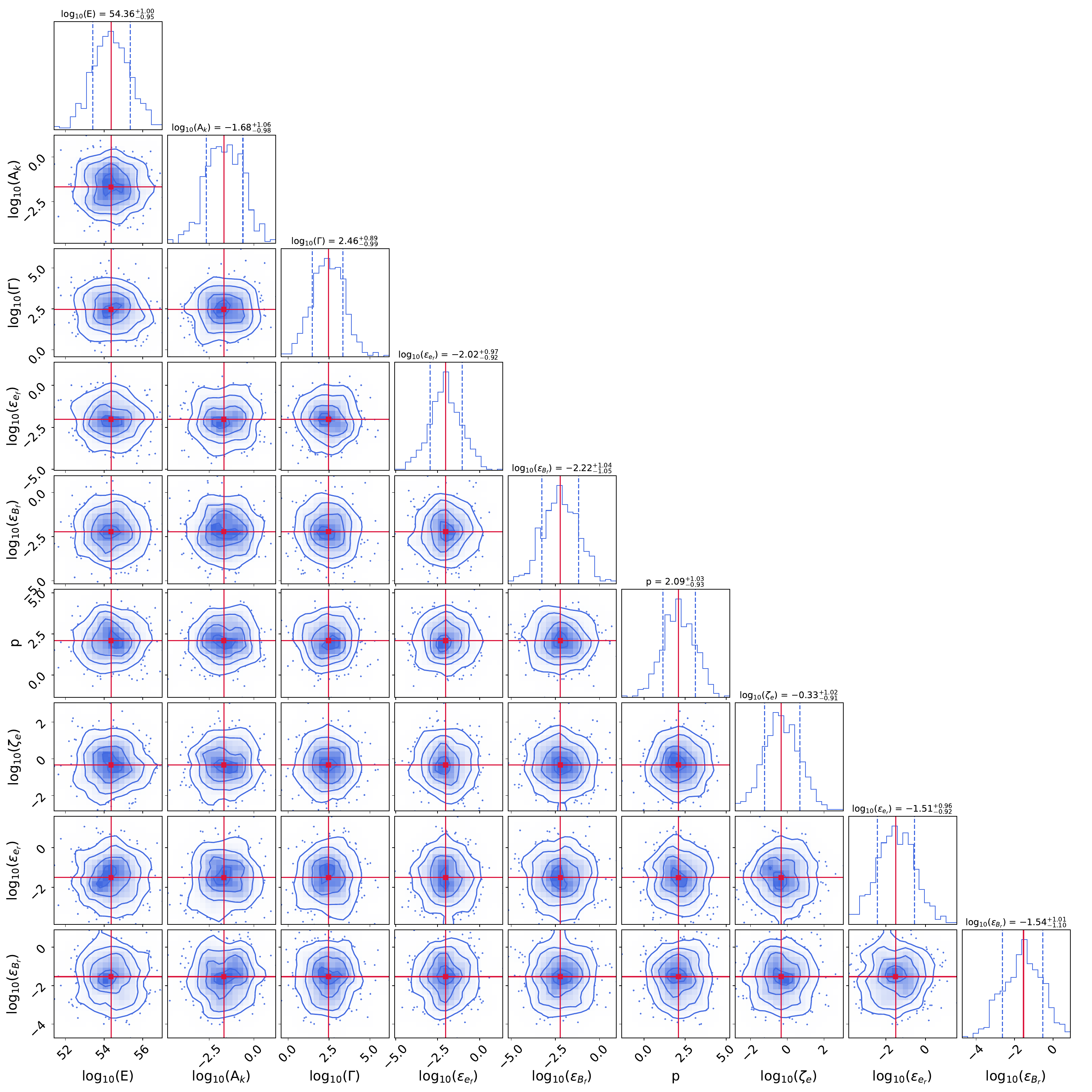}
    \caption{Same as Figure~\ref{fig:CornerGRB080916C}, but for GRB 180720B.}
    \label{fig:CornerGRB180720B}
\end{figure}

\begin{figure}
    \centering{
    
        \subfloat[Sub a) GRB 080916C][\centering{ {\itshape Fermi}/LAT light curve for GRB 080916C.}]{\includegraphics[scale=0.39]{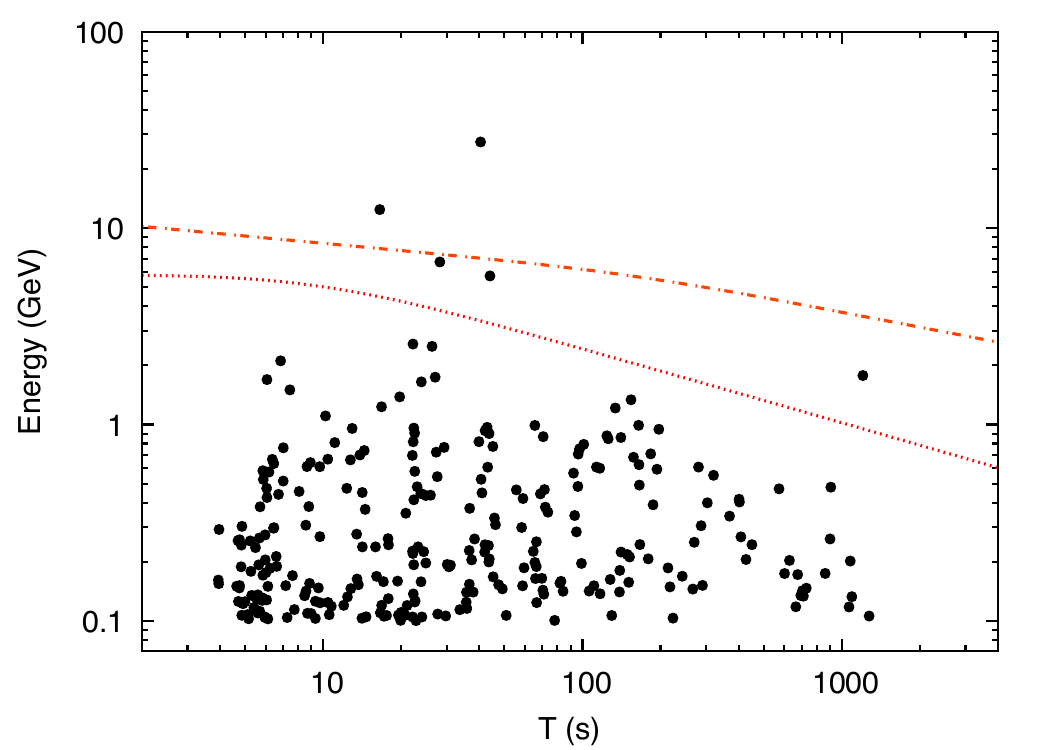}
        \label{fig4a):GRB080916C}}
        \subfloat[Sub b) GRB 090323][\centering{ {\itshape Fermi}/LAT light curve for GRB 090323.}]{\includegraphics[scale=0.39]{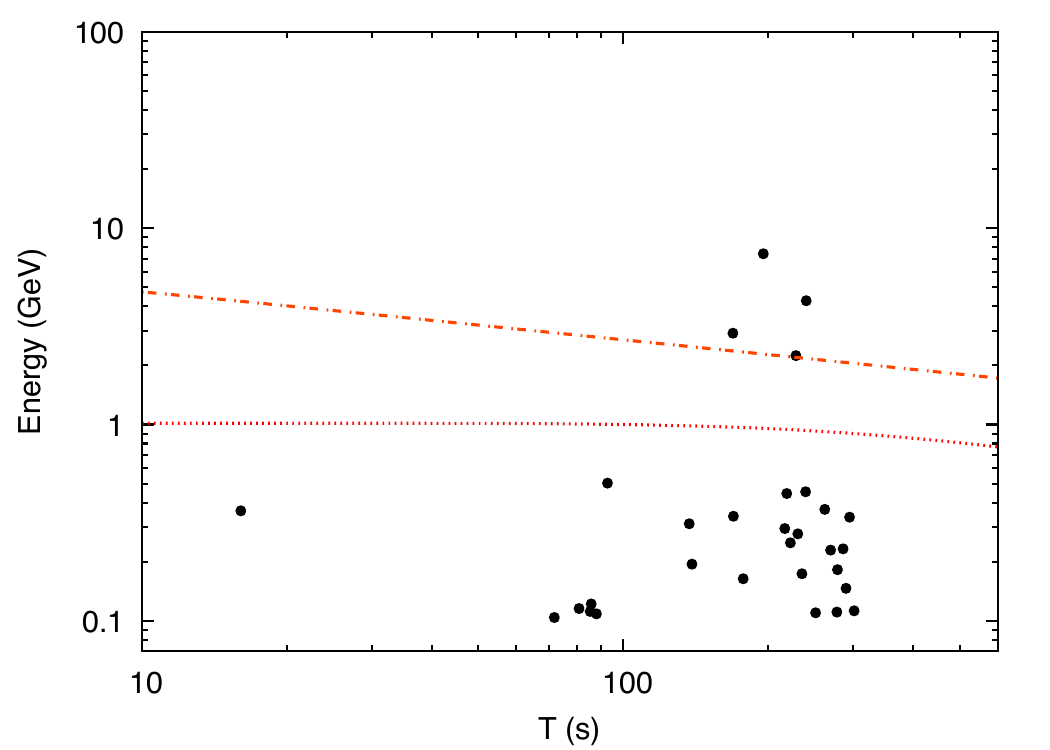}
        \label{fig4b):GRB090323}}

        \subfloat[Sub c) GRB 090902B][\centering{{\itshape Fermi}/LAT light curve for GRB 090902B.}]{\includegraphics[scale=0.39]{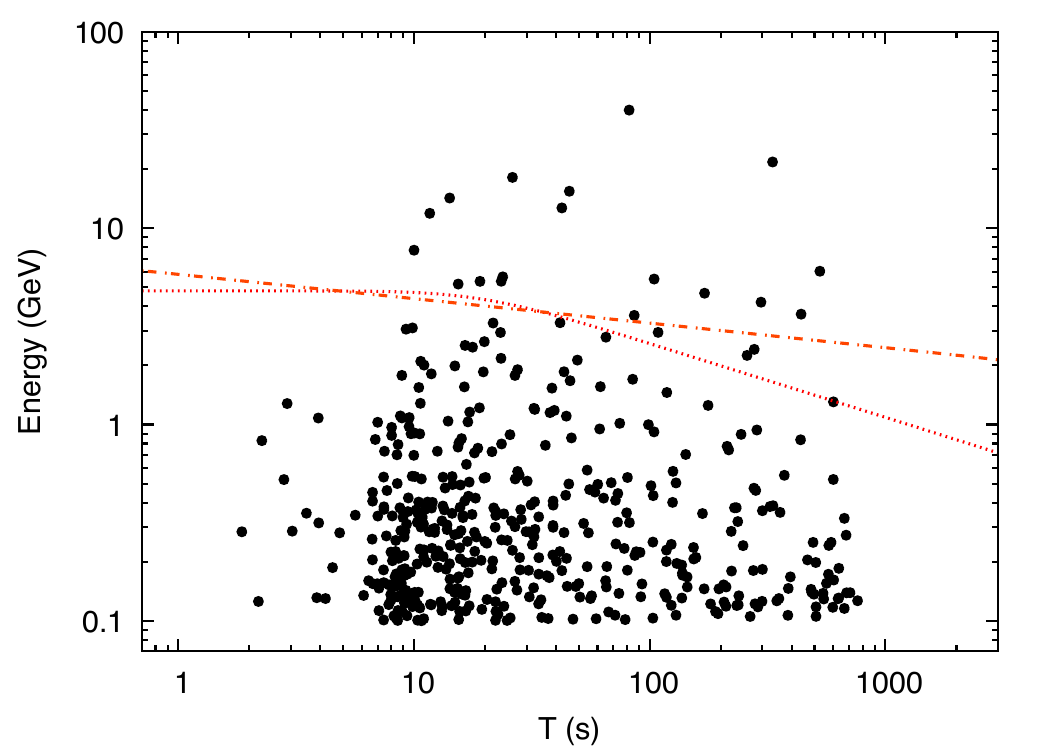}
        \label{fig4d):GRB090902B}}
        \subfloat[Sub d) GRB 090926A][\centering{{\itshape Fermi}/LAT light curve for GRB 090926A.}]{\includegraphics[scale=0.39]{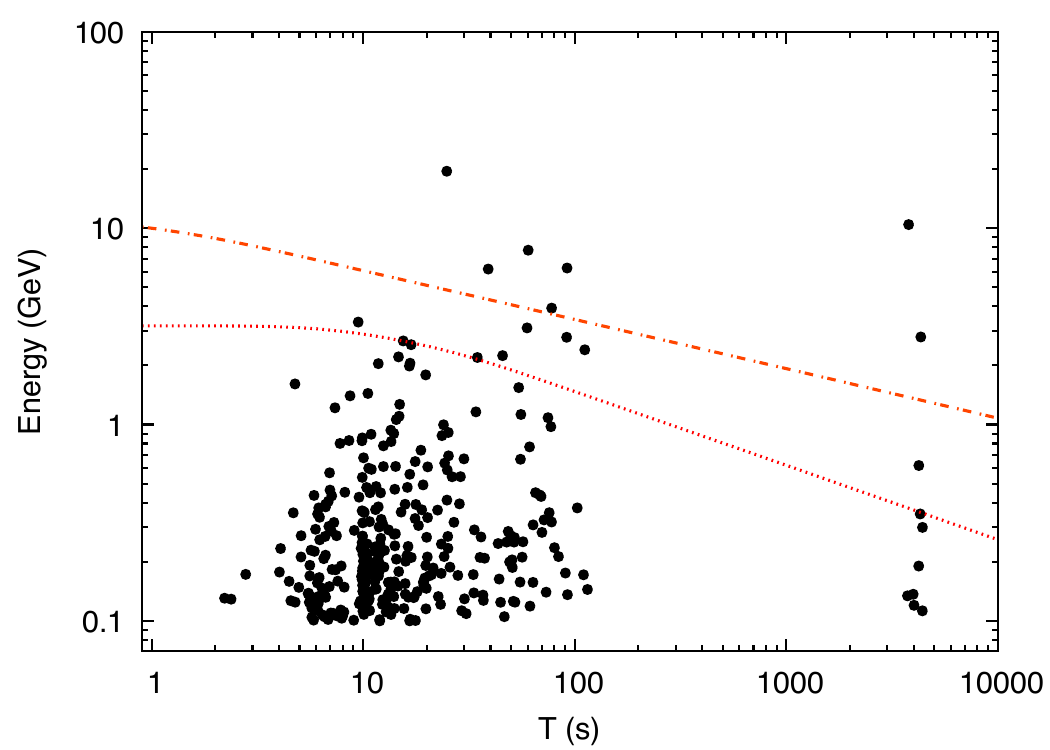}
        \label{fig4e):GRB090926A}}

        \subfloat[Sub e) GRB 110731A][\centering{{\itshape Fermi}/LAT light curve for GRB 110731A.}]{\includegraphics[scale=0.39]{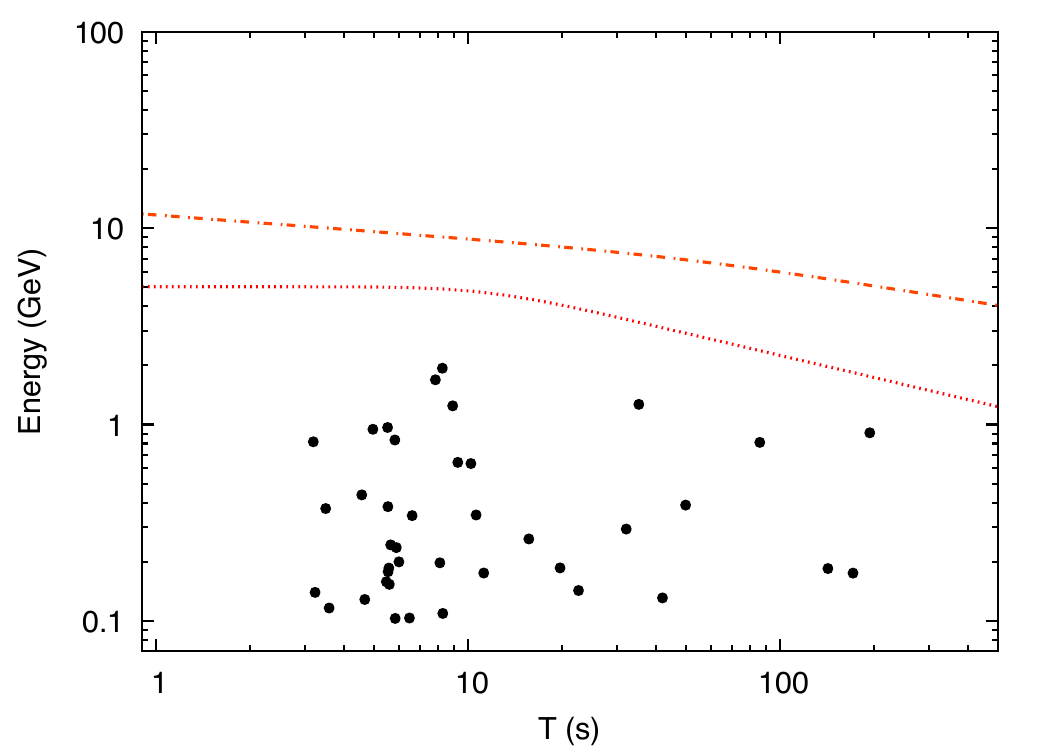}
        \label{fig4f):GRB110731A}}
        \subfloat[Sub f) GRB 130427A][\centering{{\itshape Fermi}/LAT light curve for GRB 130427A.}]{\includegraphics[scale=0.39]{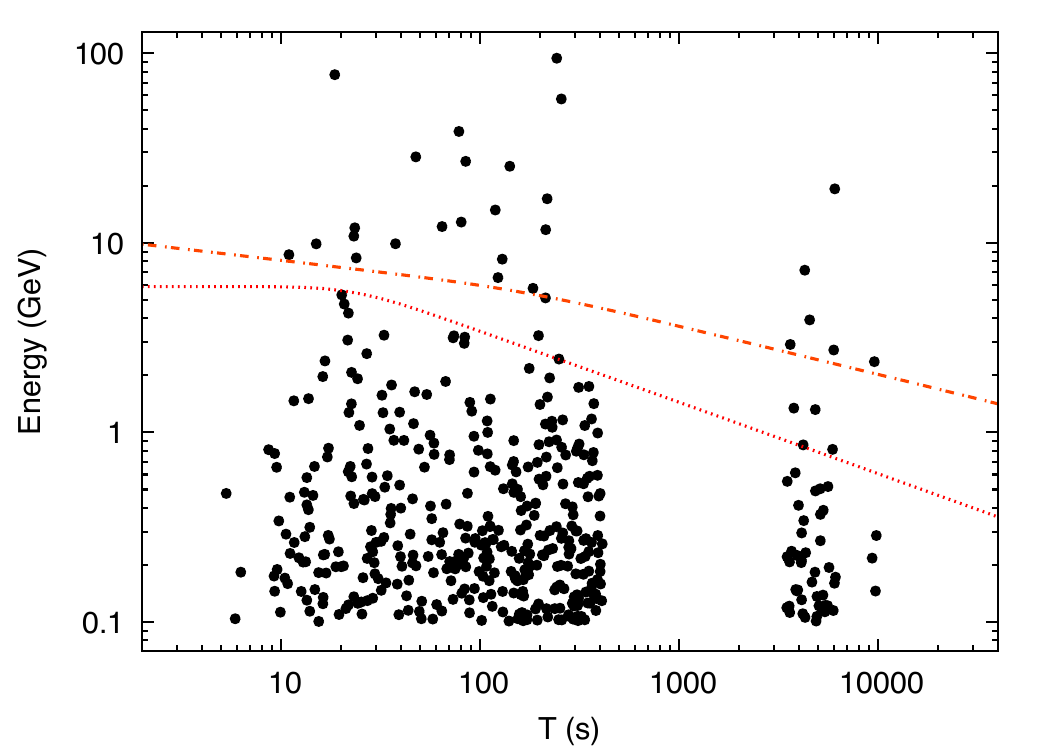}
        \label{fig4g):GRB130427A}}

        \subfloat[Sub g) GRB 160625B][\centering{{\itshape Fermi}/LAT light curve for GRB 160625B.}]{\includegraphics[scale=0.39]{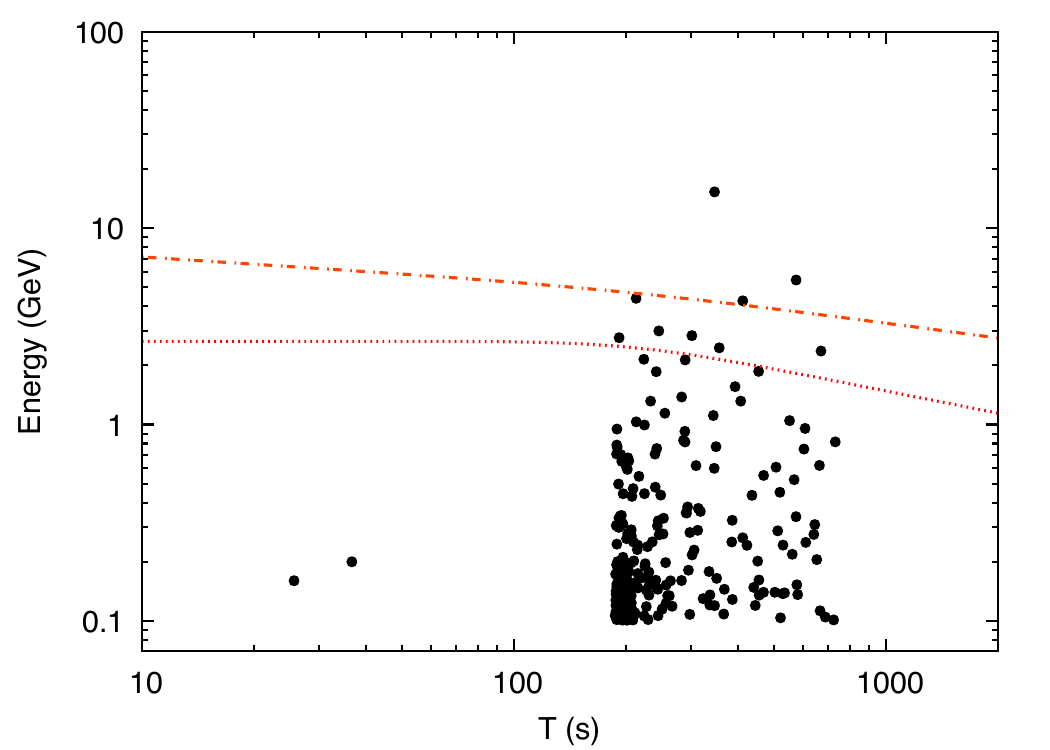}
        \label{fig4h):GRB160625B}}
        \subfloat[Sub h) GRB 180720B][\centering{{\itshape Fermi}/LAT light curve for GRB 180720B.}]{\includegraphics[scale=0.39]{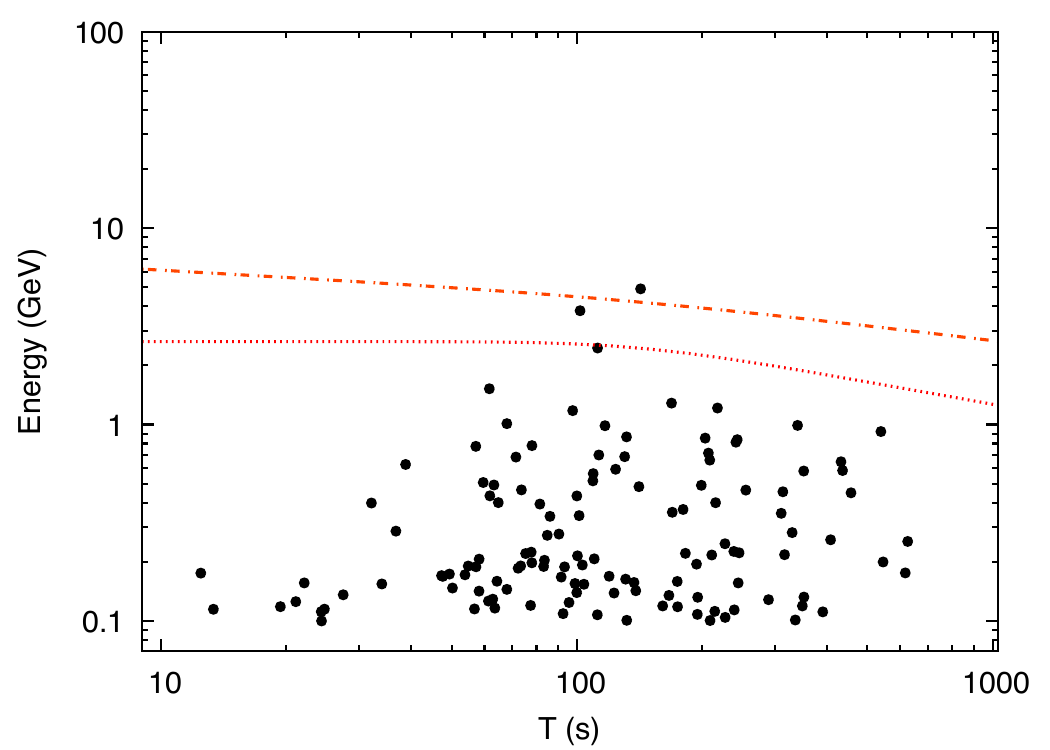}
        \label{fig4h):GRB180720B}}
    }
    \caption{All the photons with energies $> 100$~MeV and probabilities $>90$\% of being associated with each burst in our sample. The red lines correspond to the maximum photon energies released by the SSC (dotted) and synchrotron (dashed) from the reverse and forward afterglow model.}\label{fig4}
\end{figure}

All panels in Figure~\ref{fig5} show the LAT, X-ray and optical observations of GRB 080916C, GRB 090323, GRB 090902B, GRB 090926A, GRB 110731A, GRB 130427A, GRB 160625B and GRB 180720B with the best-fit curve of the short- (dotted) and long- (dotted-dashed) lasting components. Total emission is displayed with a solid line. We used Markov-Chain Monte Carlo (MCMC) simulations with the eight parameters used for the complete sample of GRBs to find the best-fit values that describe the multiwavelength observations with the SSC and synchrotron from the reverse and forward shock models. To represent all the data in this case, a total of 15,900 samples and 4,400 tuning steps are used. Figures~\ref{fig:CornerGRB080916C} - \ref{fig:CornerGRB180720B} show the best-fit values and the median of the posterior distributions of the parameters. In Table~\ref{Table6}, the best-fit values are shown in red, and the median of the posterior distributions is presented. Table~\ref{Table3} shows the SSC spectral breaks from reverse shocks, which are estimated with the best-fit values reported in Table~\ref{Table6}. The SSC spectral breaks from the RS region are estimated for GRB 080916C, 090323, 090902B, 090926A, 110731A, 130427A, 160625B, and 180720B at $4\,(30)\,{\rm s}$, $100\,(315)\,{\rm s}$, $3\,(16)\,{\rm s}$,  $6\,(20)\,{\rm s}$, $6\,(20)\,{\rm s}$, $6\,(32)\,{\rm s}$, $100\,(316)\,{\rm s}$ and $63\,(200)\,{\rm s}$ for $t< t_{\rm x} (t_{\rm x}< t)$, respectively.  The synchrotron spectral breaks from the FS region are estimated to be $10^5\,{\rm s}$.  Figure~\ref{fig4} exhibits all photons with energies $>100$~MeV and probabilities $>90$\% of being associated with each burst. Additionally, we show in red lines the maximum photon energies released by the SSC (dotted) and synchrotron (dashed) from the reverse and forward afterglow models, respectively, estimated with the best-fit values reported in Table~\ref{Table6}.

\subsubsection{Microphysical parameters}

\paragraph{$\varepsilon_{\rm B}$-parameter}
The best-fit values of the magnetic microphysical parameter from the RS lie in the range of $0.01\lesssim \varepsilon_{\rm B_r}\lesssim 0.1$. Taking into account the values of the microphysical parameters, the magnetisation parameter is estimated to be in the range $0.01\lesssim \sigma\lesssim 0.1$, indicating that the ejecta are moderately magnetic and as a result the RS is expected to be successful. If this was not the case, the acceleration of the particles in RS would be ineffective and RS would not have occurred \citep{2005ApJ...628..315Z, 2004A&A...424..477F}. The microphysical parameter of the magnetic field in the FS and RS regions is different. The ratio of these parameters lie in the range of $1\lesssim {\mathcal R_B} \lesssim 20$.\footnote{${\mathcal R_B}$ is defined as $\varepsilon_{\rm B_r}/\varepsilon_{\rm B_f}$.}    It has been suggested that the high-energy emission seen in the brightest LAT-detected bursts could be explained by Poynting flux-dominated models with arbitrary magnetization \citep[e.g., see][]{2014NatPh..10..351U,2011ApJ...726...90Z}. The values of the magnetisation parameter suggest that a substantial amount of Poynting flux was released by the ejecta during the prompt emission phase, with the internal collision-induced magnetic reconnection and turbulence (ICMART) event being the most likely mechanism that accounts for this behaviour \citep{2011ApJ...726...90Z}.

\paragraph{$\varepsilon_{\rm e}$-parameter}

The best-fit values of the $\varepsilon_{\rm e}$ microphysical parameter in the forward- and RS model are presented in the 7$^{\rm th}$ and 8$^{\rm th}$ rows of Table~\ref{Table6}, respectively. For the forward shock, we find $2.8\times10^{-3}\lesssim\varepsilon_{\rm e_f}\lesssim0.12$, while for the RS we have $0.03\lesssim\varepsilon_{\rm e_r}\lesssim0.13$. As a general rule, we notice that $\epsilon_{\rm e_f}<\epsilon_{\rm e_r}$. Moreover, there is a large variation in the ratio of both quantities between each burst. For example, in the case of GRB 110731A we find the largest disparity $\mathcal{R}_e\equiv \epsilon_{\rm e_r}/\epsilon_{\rm e_f}\approx46$, while for GRB 160625B $\mathcal{R}_e\approx0.7$, so the variation can occur over several orders of magnitude. It should be noted that an RS is suppressed for $\mathcal{R}_e\ll1$ \citep{2020ApJ...905..112F}.

Although it is usually assumed that the electron microphysical parameter in both the RS and the FS has a similar value \citep{2005ApJ...628..315Z,2015ApJ...810..160G}, the possibility of them being different has been explored in previous work, such as the one by \cite{2004MNRAS.353..511P}, in which the authors noted that in the case of GRB 990123 for a $k=0$ model $\epsilon_{\rm e_r}<0.1\epsilon_{\rm e_f}$, which is not in agreement with our results. However, given the current uncertainties on the acceleration process that energises electrons, it proves difficult to determine a relation between this parameter in both regions.

\subsubsection{The post jet-break decay phase}
After about~$T+1$~day \citep{Racusin2009}, during the post jet break decay phase, the multi-wavelength light curves would evolve as $F^{\rm syn}_\nu \propto t^{-p}$, for $\nu^{\rm syn}_{\rm m, f} < \nu < \nu^{\rm syn}_{\rm c, f}$ or ${\rm max}\{\nu^{\rm syn}_{\rm m, f}, \nu^{\rm syn}_{\rm c, f}\}<\nu$ \citep[see e.g.][]{Fraija2022}, which are different from the temporal decay indexes found for all bursts \citep{Pereyra2022,Becerra2019}, except GRB 160625B \citep{Fraija2017}. This means that, apart from GRB 160625B, they were likely released from a broad outflow with a large half-opening angle, as shown by multi-wavelength measurements which show no indication of late steep decays. Based on the best-fit values of the circumburst density and the kinetic equivalent energies, we find that the jet opening angles become $\gtrsim 10^\circ$, and for GRB 160625B, the value of the jet opening angle is $\theta_j\approx 2.3^\circ$, which is in the typical interval for $\theta_j=2^\circ-10^\circ$ \citep{Bloom2001}.

\subsubsection{Efficiency of equivalent kinetic energy}

The efficiency provides crucial information on the gamma-ray emitting process. The best fit values of the equivalent kinetic energies $10^{53.96^{+1.01}_{-1.05}}\leq E\leq 10^{54.85^{+1.05}_{-1.02}}\,{\rm erg}$ and the isotropic energies in gamma-rays reported by the GBM instrument during the prompt episode in the range of $(0.39 \pm 0.09)\times 10^{52}\leq E_{\rm \gamma, iso}\leq (1.7\pm 0.1)\times 10^{54}\,{\rm erg}$ \citep{Ajello_2019} lead to kinetic efficiencies in the range of $0.01 \lesssim \eta \lesssim 0.36$, which are typical compared to those values reported in the literature \citep{2001ApJ...557..399G, 2007ApJ...655..989Z, 2015PhR...561....1K}, and a kinetic efficiency of $\eta\approx 0.02$ for GRB 180720B \citep{Abdalla2019,  2019ApJ...885...29F}, which is very low. The atypical value of efficiency for GRB~180720B was calculated considering the isotropic energy reported by GBM during the interval 4.4-53.2~s. If we have considered the isotropic energy reported by the LAT instrument throughout the interval 11.8-625.0 s, the kinetic efficiency would have been $\eta\approx 0.002$. Similarly, the second-lowest efficiency, which corresponds to $\eta=0.01$. If we had considered the isotropic energy reported by the LAT in the interval 0.1 - 34366.2~s instead of the GBM instrument, the kinetic efficiency would be $\eta\approx 0.06$. 


\subsubsection{The profile of the circumburst environment}
 The best-fit values of the wind parameter lie in the range of $ 10^{-3}\lesssim A_{\rm k} \lesssim 0.02$, typical for GRBs identified as powerful bursts \citep{2013ApJ...763...71A, 2014ApJ...781...37P, 2014Sci...343...38V, 2012ApJ...751...33F, 2008Natur.455..183R,Fraija2017,Becerra2017}. Recently, \cite{2023Galax..11...25D} considered the GRBs reported in the 2FLGC and analysed the evolution of the relationship between the spectral and temporal indices using closure relations in a stratified environment. They found that the afterglow model without energy injection is preferred over that with energy injection with a clear preference for the cooling condition $\nu >$ max\{$\nu_c,\nu_m$\}, as derived (see Table~\ref{Table3}). Furthermore, the authors reported that the density profiles $r^{-k}$ with $k=2$ have a higher rate of occurrence. 

 The best-fit values of the initial bulk Lorentz factor lie in the range $290\lesssim \Gamma\lesssim 10^3$, which is similar to the values found in other bursts detected by the LAT instrument and those expected in numerical simulations \citep{Tchekhovskoy2008}. Since our GRB sample displayed the most energetic photons, the values of the bulk Lorentz factor are predicted to fall within the same range as the bursts observed by the LAT that are the brightest \citep{2011ApJ...729..114A, 2012ApJ...755...12V, 2013ApJ...763...71A, 2009ApJ...706L.138A, 2010ApJ...716.1178A, 2014Sci...343...42A, 2019ApJ...879L..26F, 2019ApJ...885...29F}, as found. Given the best-fit values,  the critical Lorentz factors are consistent with the thick-shell scenario.

\subsubsection{Fraction of the shock-accelerated electrons}
In the standard GRB afterglow model, it is assumed that the fraction of electrons accelerated by shock is $\zeta_e\sim1$. However, the validity of this assumption has been questioned by several authors. For example, \cite{1996ApJ...461L..37B} considered the increase in the electron Lorentz factor taking a value $\zeta_e\sim10^{-3}$, which would lead to a change in the synchrotron spectrum. On the other hand, \cite{2005ApJ...627..861E} concluded that, while observations suggest that $\zeta_e\sim1$, the range $m_e/m_p\approx5\times10^{-4}\leq\zeta_e\leq1$ cannot be ruled out because models with different values of $\zeta_e$ in the aforementioned range lead to very similar predictions. 

In the last row of Table~\ref{Table6}, we present the parameter values of $\zeta_e$ that we obtained for our sample of GRBs through our MCMC estimation. We notice that our lowest value is for GRB 080916C with $\zeta_e\approx0.02$, while the largest corresponds to GRB 180720B with $\zeta_e\approx0.62$. This means that all estimates are within the allowed range of \cite{2005ApJ...627..861E}.

\subsubsection{The highest energy photons}

Figure~\ref{fig4} shows that with the exception of GRB 110731A, the SSC and synchrotron models radiated in the RS and FS regions cannot explain the highest-energy photons exhibited. Therefore, a different process needs to be invoked to interpret these photons. The number of photons ($N_{\gamma}$) with energy $h\nu$ that reach the {\itshape Fermi}-LAT instrument during a specific time interval ($\Delta t$) can be estimated considering the effective area of LAT ($A$) and the observed flux at the determined time. The number of photons is
{\small
\begin{eqnarray}\label{vhe-photon}
N_{\gamma}\sim 1\,{\rm ph} \left(\frac{h\nu}{10\,{\rm~GeV}}\right) \left(\frac{10\,{\rm s}}{\Delta t} \right) \left(\frac{F^{\rm ssc}_{\nu,j}}{10^{-9}\,{\rm \frac{erg}{cm^{2}\,s}}}\right) \left(\frac{10^4\,{\rm cm^{2}}}{A}\right)\,.
\end{eqnarray}
}
In order to estimate the number of photons we discuss hadronic and SSC scenarios. In the hadronic scenarios, high-energy gamma-ray emission has been interpreted via photo-hadronic interactions; ultrarelativistic protons accelerated in the jet with internal synchrotron photons \citep{2009ApJ...705L.191A, 2000ApJ...537..255D}, inelastic proton-neutron collisions \citep{2000ApJ...541L...5M}, and relativistic neutrons with seed photons coming from the outflow \citep{2004A&A...418L...5D, 2004ApJ...604L..85A}. Even though GRBs are among the most plausible candidates to accelerate cosmic rays up to ultra-high energies \citep[$\gtrsim 10^{18}$ eV; ][]{1995PhRvL..75..386W, 1995ApJ...453..883V} and thus, potential candidates  for  neutrino  detection, the IceCube collaboration reported  no coincidences between neutrinos and GRBs after analyzing years of data \citep{2022arXiv220511410A,2012Natur.484..351A, 2016ApJ...824..115A, 2015ApJ...805L...5A}.  Because of this, we rule out hadronic models as an explanation for the observed properties of GRBs and conclude that the number of hadrons is too small for hadronic interactions to efficiently generate observable gamma-ray signals in GRBs. On the other hand, the SSC scenario has been successfully applied to interpret the highest-energy photons \citep[e.g., see][]{2019ApJ...885...29F, 2019ApJ...884..117W, 2019arXiv191109862Z, 2019ApJ...883..162F,2021ApJ...918...12F, 2017ApJ...848...94F}. In this scenario, the same electron population can up-scatter synchrotron photons up to higher energies as $h\nu^{\rm ssc}_{\rm i, f}\sim\gamma^2_{\rm i, f} h\nu^{\rm syn}_{\rm i, f}$, reaching a maximum flux of $F^{\rm ssc}_{\rm max,f}\sim\, \frac{4\sigma_T n r}{3 g(p)}  \,F^{\rm syn}_{\rm max,f}$ with $r\simeq 4c \Gamma^2 t/(1+z)$. The spectral breaks  and the maximum flux for SSC emission in the forward shock can be expressed as
{\small
\bary\label{ssc_br-h}
h\nu^{\rm ssc}_{\rm m,f}&\simeq& 4.3\times 10^2\,{\rm~MeV}\, \left(\frac{1+z}{2} \right)^{\frac{10-3k}{2(4-k)}}\,g^4(2.2)\,\zeta_e^{-4}\,\varepsilon_{\rm e_f,-1}^{4}\,\varepsilon_{\rm B_f,-2}^{\frac12}\,A^{-\frac{1}{4-k}}_{\rm k,-1}\,E_{53}^{\frac{6-k}{2(4-k)}}t^{-\frac{18-5k}{2(4-k)}}_2,\cr
h\nu^{\rm ssc}_{\rm c,f}&\simeq& 4.6\times 10^2 {\rm eV} \left(\frac{1+z}{2} \right)^{-\frac{3(2+k)}{2(4-k)}}\left(\frac{1+Y_{\rm f}}{2} \right)^{-4}\varepsilon_{\rm B_f,-2}^{-\frac72}A^{-\frac{9}{4-k}}_{\rm k,-1}\,E_{53}^{-\frac{10-7k}{2(4-k)}}\,t^{-\frac{2-5k}{2(4-k)}}_2\,,\cr
F^{\rm ssc}_{\rm max, f}&\simeq& 8.1\times 10^{-3}\,{\rm mJy}\left(\frac{1+z}{2} \right)^{\frac{14-k}{2(4-k)}}\,g^{-1}(2.2)\,\zeta_e\,\varepsilon_{\rm B_f,-2}^{\frac12}\,A^{\frac{5}{4-k}}_{\rm k,-1}\,d^{-2}_{\rm z, 28.3}\,E_{53}^{\frac{5(2-k)}{2(4-k)}}\,t^{\frac{2-3k}{2(4-k)}}_2\,.
\eary
}
%
%
In the fast- and slow-cooling regime the SSC light curve are given by  \citep{2001ApJ...548..787S}
{\footnotesize
\begin{eqnarray}
\label{ssc_ism1}
F^{\rm ssc}_{\nu,f}\propto  \begin{cases}
t^{\frac{2-k}{4(4-k)}} \nu^{-\frac12}, \hspace{2.8cm} \nu^{\rm ssc}_{\rm c,f}<\nu <\nu^{\rm ssc}_{\rm m,f},\hspace{.25cm}\cr
t^{\frac{2(10-3k)+p(5k-18)}{4(4-k)}}\,\nu^{-\frac{p}{2}},\,\,\,\, \hspace{1.1cm}  \nu^{\rm ssc}_{\rm m,f} <\nu\,, \cr
\end{cases}
\end{eqnarray}
}
and
{\footnotesize
\begin{eqnarray}
\label{ssc_ism2}
F^{\rm ssc}_{\nu,f}\propto  \begin{cases}
t^{\frac{11(2-k)+p(5k-18)}{4(4-k)}} \nu^{-\frac{p-1}{2}}, \hspace{1.2cm} \nu^{\rm ssc}_{\rm m,f}<\nu <\nu^{\rm ssc}_{\rm c,f},\hspace{.25cm}\cr
t^{\frac{2(10-3k)+p(5k-18)}{4(4-k)}}\,\nu^{-\frac{p}{2}},\,\,\,\, \hspace{1.1cm}  \nu^{\rm ssc}_{\rm c,f} <\nu\,, \cr
\end{cases}
\end{eqnarray}
}
respectively. For $\nu^{\rm ssc}_{\rm c, f}<\nu^{\rm ssc}_{\rm m, f}$ and $\nu^{\rm ssc}_{\rm m, f}<\nu^{\rm ssc}_{\rm c, f}$, the spectral breaks due to the KN effect are

{
\small
\bary
h \nu^{\rm ssc}_{\rm KN,m,f}&\simeq& 1.2\times10^2\,{\rm~GeV} \left(\frac{1+z}{2} \right)^{-\frac{1}{4-k}}g(2.2)\zeta_e^{-1} \varepsilon_{\rm e_f,-1}\,A^{-\frac{1}{4-k}}_{\rm k,-1} \,E^{\frac{1}{4-k}}_{53}\,t^{-\frac{3-k}{4-k}}_2\,,
\eary
}

{
\small
\bary
h \nu^{\rm ssc}_{\rm KN,c,f}&\simeq& 3.8\,{\rm~GeV} \left(\frac{1+z}{2} \right)^{-\frac{3}{4-k}} \left(\frac{1+Y_{\rm f}}{2} \right)^{-1} \varepsilon_{\rm B_f,-2}^{-1}\,A^{-\frac{3}{4-k}}_{\rm k,-1} \,E^{-\frac{1-k}{4-k}}_{53}\,t^{-\frac{1-k}{4-k}}_2\,.
\eary
}
The maximum energy radiated by the SSC process is estimated by $h\nu^{\rm ssc}_{\rm max,f}\simeq \gamma^2_{\rm max, f} (h\nu^{\rm syn}_{\rm max,f})$.

Given the best-fit values listed in Table~\ref{Table6}, the number of photons at the highest energies are: 0.9, 0.2, 1.2, 1.6, 31.2, 0.6 and 1.7 for GRB 080916C, 090323, 090902B, 090926A, 130427A, 160625B and 180720B, respectively. Therefore, although there is an excess from GRB 130427A due to the spectral index, the most energetic photons could be explained by the SSC mechanism from the forward shock.

\hspace{1cm}

\subsubsection{Spectral breaks in the Klein-Nishina regime}

The values of the spectral breaks derived from the best fit parameters and listed in Table~\ref{Table3} show that, with the exception of GRB 090323 and GRB 090926A, the KN effects are neglected.  For GRB 090323, the spectral breaks are in the order $ \nu^{\rm syn}_{\rm c} <  h\nu^{\rm syn}_{\rm KN}(\gamma_*)< \nu^{\rm syn}_{\rm m} < \nu^{\rm syn}_{\rm KN}(\gamma_{\rm m})$ with $h\nu^{\rm syn}_{\rm m,r}=1.1\times 10^3\,{\rm eV}$, $h\nu^{\rm syn}_{\rm c,r}=8.2\,{\rm eV}$, $h\nu^{\rm syn}_{KN,m,r}=1.2\times 10^5\,{\rm eV}$, $h\nu^{\rm syn}_{KN,c,r}=6.3\times 10\,{\rm eV}$ and $h\nu^{\rm syn}_{KN,r}(\gamma_*)=3.5\times 10^2\,{\rm eV}$. In this case, the Compton parameter was recalculated \citep[for details, see][]{2010ApJ...712.1232W}. A similar procedure was developed for GRB 090926A.
 
\vspace{0cm} 
\section{Summary}\label{sec4}

We have extended the SSC RS model in a medium of constant density initially proposed to explain the LAT GeV flare observed in GRB 160509A \citep{2020ApJ...905..112F}. In the current model, we have generalised and derived the SSC light curves from the RS region in the thick and thin shell scenarios for a stratified environment with a density profile $ \propto r^{\rm -k}$ with $0\leq {\rm k}< 3 $, including the maximum energy radiated by this process in both regimes.   In order to apply our theoretical model to a sample of eight bursts (GRB 080916C, 090323, 090902B, 090926A, 110731A, 130427A, 160625B and 180720B) reported in the 2FLGC \citep{Ajello_2019}, which exhibited interesting short-lasting bright peaks, we have obtained the Fermi-LAT light curves together with the photons with energies $\geq 100$~ MeV associated with each burst.   We have used the multiwavelength observations to constrain the parameters in the model and fitted the LAT light curves of the sample of GRBs with a joint model which considers synchrotron and SSC emission from both the FS and the RS through MCMC simulations.  We have shown that the emission in the thick-shell scenario could describe the short-lasting bright peaks exhibited in our GRB sample. Because the shock crossing time is smaller than the duration of the prompt emission in the thick-shell regime, the bright peak in this regime is expected at the beginning of the FS emission. By contrast, if the shock crossing time is longer than the duration of the prompt emission in the thin-shell scenario, the bright peak is separated from the prompt emission and expected during the FS emission. 

We show that at the beginning of the LAT observations, the SSC flux from the RS region is larger than the synchrotron flux from the FS. Therefore, the closure relations of the synchrotron FS model are not expected when the LAT light curves are a superposition of the SSC and synchrotron from RS and FS.  It is worth highlighting that, depending on the parameter values, the short-lasting bright peaks could be hidden by the long-lasting emission.

The best-fit values of the magnetic microphysical parameter for our sample of bursts were consistent with moderate magnetic ejecta and $\mathcal{R}_e\ll 1$, which indicated that the RS was successful. With respect to the electron microphysical parameter, we noticed that there was a large variation between the values of the FS and the RS, which suggests that the acceleration mechanisms are substantially different.  We found that in all bursts except GRB 160625B, the radiation was most likely released from a broad outflow with a large opening angle. We calculated the kinetic efficiency for every burst in our sample and found that most of the values are consistent with those of previous studies. However, GRB 180720B presented a value that was far below the standard. We have calculated the best-fit values of the wind parameter for our sample of bursts. We have found that all of the results are consistent with powerful GRBs. This conclusion is further strengthened by the fact that our sample displayed the most energetic photons, which are also expected from the most powerful bursts. We tested the assumption that the fraction of accelerated electrons in the shock is close to unity. We have found that this is not the case in all the bursts of our sample. However, an accurate determination of this parameter is difficult, as different values lead to similar predictions.

Given the best-fit values, we have shown that the first high-energy photons are consistent with the starting time of the short-lasting emission. Similarly, we have shown that, unlike GRB 110731A, the highest energy photon cannot be described neither by the standard synchrotron emision from RS nor from the FS, and a different process from the ones considered have to be invoked to interpret these photons. We show that unlike GRB 090323 the highest-energy photons are consistent with SSC from the FS region.   We have shown that in all but two bursts from our sample, KN effects are neglected. However, in the case of GRB 090323 and GRB 090926A, they are important, and so we have recalculated the Compton parameter appropriately in both cases.

\section*{Acknowledgements}
NF  acknowledges  financial  support  from UNAM-DGAPA-PAPIIT  through  grant  IN106521. M.G.D. acknowledges funding from the AAS Chretienne Fellowship and the MINIATURA2 grant. R.L.B. acknowledges support from the CONAHCyT postdoctoral fellowship and the financial  support  from UNAM-DGAPA-PAPIIT through  grant IN105921.
\clearpage


\section*{Data availability}
The data underlying this article will be shared on reasonable request to the corresponding author.\\



\bibliographystyle{mnras}
\bibliography{main} 



\newpage

\appendix

\begin{table}
\centering \renewcommand{\arraystretch}{0.67}\addtolength{\tabcolsep}{3pt}
\caption{The best-fit values for the modeling of the short and long-lasting components of the LAT light curves of our GRB sample using PL functions.}
\label{Table4}
\begin{tabular}{l  c  c c  c }
 \hline \hline
\scriptsize{Event} & \scriptsize{LAT} &  \scriptsize{Parameter}  &\hspace{0.5cm}   \scriptsize{Best-fit value}    & \hspace{0.5cm} \scriptsize{ $\chi^2$/ndf} \\ 
\hline \hline
GRB 080916C & 	        &  \scriptsize{$\alpha_{\rm r_L, bb}$}  &\hspace{0.5cm} \scriptsize{$-1.05\pm 0.84$}		&\hspace{0.5cm}  \scriptsize{$1.22$}\\

&  	   \scriptsize{Short-lasting component}      &  \scriptsize{$\alpha_{\rm r_L, ab}$}  &\hspace{0.5cm} \scriptsize{$3.53\pm 0.70$}		&\hspace{0.5cm}  \scriptsize{$$}\\ 
&           &  \scriptsize{$T_{\rm a}$ (s)}  &\hspace{0.5cm} \scriptsize{$2.40\pm 2.20$}		&\hspace{0.5cm}  \scriptsize{$$}\\ 
&                  	        &  \scriptsize{$t_{\rm r, pk}$ (s)}  &\hspace{0.5cm} \scriptsize{$5.76 \pm 0.84$}		&\hspace{0.5cm}  \scriptsize{$$}\\ 
\cline{2-5}                  	        
&\scriptsize{}        &  \scriptsize{$\alpha_{L,1}$}  &\hspace{0.5cm} \scriptsize{$0.54 \pm 0.27$}		&\hspace{0.5cm}  \scriptsize{$1.18$}\\	
&  \scriptsize{Long-lasting component}  &  \scriptsize{$\alpha_{L,2}$}  &\hspace{0.5cm} \scriptsize{$1.47 \pm 0.10$}		&\hspace{0.5cm}  \scriptsize{$$}\\	
&   		 	        & \scriptsize{$t_{\rm f_L, br}$ (s)}  &\hspace{0.5cm} \scriptsize{$22.08 \pm 0.02$}		&\hspace{0.5cm}  \scriptsize{$$}\\	
\cline{3-5}
&   		 	        &  \scriptsize{$\alpha_{L}$}  &\hspace{0.5cm} \scriptsize{$1.21 \pm 0.06$}		&\hspace{0.5cm}  \scriptsize{$1.12$}\\			        
\hline \hline
GRB 090323 & 	        &  \scriptsize{$\alpha_{\rm r_L, bb}$}  &\hspace{0.5cm} \scriptsize{$-1.11\pm 0.74$}		&\hspace{0.5cm}  \scriptsize{$0.96$}\\

&  	   \scriptsize{Short-lasting component}      &  \scriptsize{$\alpha_{\rm r_L, ab}$}  &\hspace{0.5cm} \scriptsize{$4.00\pm 1.90$}		&\hspace{0.5cm}  \scriptsize{$$}\\ 
&           &  \scriptsize{$T_{\rm a}$ (s)}  &\hspace{0.5cm} \scriptsize{$55.19\pm 50.25$}		&\hspace{0.5cm}  \scriptsize{$$}\\ 
&                  	        &  \scriptsize{$t_{\rm r, pk}$ (s)}  &\hspace{0.5cm} \scriptsize{$106.60 \pm 72.82$}		&\hspace{0.5cm}  \scriptsize{$$}\\ 
\cline{2-5}            
&\scriptsize{}        &  \scriptsize{$\alpha_{L,1}$}  &\hspace{0.5cm} \scriptsize{$1.84 \pm 1.12$}		&\hspace{0.5cm}  \scriptsize{$0.86$}\\	
&  \scriptsize{Long-lasting component}  &  \scriptsize{$\alpha_{L,2}$}  &\hspace{0.5cm} \scriptsize{$1.00 \pm 0.70$}		&\hspace{0.5cm}  \scriptsize{$$}\\	
&   		 	        & \scriptsize{$t_{\rm f_L, br}$ (s)}  &\hspace{0.5cm} \scriptsize{$576.10 \pm 798.60$}		&\hspace{0.5cm}  \scriptsize{$$}\\	
\cline{3-5}
&   		 	        &  \scriptsize{$\alpha_{L}$}  &\hspace{0.5cm} \scriptsize{$1.26 \pm 0.40$}		&\hspace{0.5cm}  \scriptsize{$0.67$}\\			        
\hline \hline
GRB 090902B & 	        &  \scriptsize{$\alpha_{\rm r_L, bb}$}  &\hspace{0.5cm} \scriptsize{$-2.00\pm 1.86$}		&\hspace{0.5cm}  \scriptsize{$0.92$}\\

&  	   \scriptsize{Short-lasting component}      &  \scriptsize{$\alpha_{\rm r_L, ab}$}  &\hspace{0.5cm} \scriptsize{$1.77\pm 0.39$}		&\hspace{0.5cm}  \scriptsize{$$}\\ 
&           &  \scriptsize{$T_{\rm a}$ (s)}  &\hspace{0.5cm} \scriptsize{$4.52 \pm 1.22$}		&\hspace{0.5cm}  \scriptsize{$$}\\ 
&                  	        &  \scriptsize{$t_{\rm r, pk}$ (s)}  &\hspace{0.5cm} \scriptsize{$8.74 \pm 0.52$}		&\hspace{0.5cm}  \scriptsize{$$}\\ 
\cline{2-5}
&\scriptsize{}        &  \scriptsize{$\alpha_{L,1}$}  &\hspace{0.5cm} \scriptsize{$1.90 \pm 0.2$}		&\hspace{0.5cm}  \scriptsize{$1.45$}\\	
&  \scriptsize{Long-lasting component}  &  \scriptsize{$\alpha_{L,2}$}  &\hspace{0.5cm} \scriptsize{$1.30 \pm 0.01$}		&\hspace{0.5cm}  \scriptsize{$$}\\	
&   		 	        & \scriptsize{$t_{\rm f_L, br}$ (s)}  &\hspace{0.5cm} \scriptsize{$100 \pm 0.08$}		&\hspace{0.5cm}  \scriptsize{$$}\\	
\cline{3-5}
&   		 	        &  \scriptsize{$\alpha_{L}$}  &\hspace{0.5cm} \scriptsize{$1.35 \pm 0.06$}		&\hspace{0.5cm}  \scriptsize{$1.32$}\\			        
\hline \hline
GRB 090926A & 	        &  \scriptsize{$\alpha_{\rm r_L, bb}$}  &\hspace{0.5cm} \scriptsize{$-1.37\pm 0.96$}		&\hspace{0.5cm}  \scriptsize{$0.92$}\\

&  	   \scriptsize{Short-lasting component}      &  \scriptsize{$\alpha_{\rm r_L, ab}$}  &\hspace{0.5cm} \scriptsize{$4.00\pm 1.82$}		&\hspace{0.5cm}  \scriptsize{$$}\\ 
&           &  \scriptsize{$T_{\rm a}$ (s)}  &\hspace{0.5cm} \scriptsize{$3.00\pm 1.81$}		&\hspace{0.5cm}  \scriptsize{$$}\\ 
&                  	        &  \scriptsize{$t_{\rm r, pk}$ (s)}  &\hspace{0.5cm} \scriptsize{$6.35 \pm 0.23$}		&\hspace{0.5cm}  \scriptsize{$$}\\ 
\cline{2-5}            
&\scriptsize{}        &  \scriptsize{$\alpha_{L,1}$}  &\hspace{0.5cm} \scriptsize{$1.65 \pm 0.10$}		&\hspace{0.5cm}  \scriptsize{$1.43$}\\	
&  \scriptsize{Long-lasting component}  &  \scriptsize{$\alpha_{L,2}$}  &\hspace{0.5cm} \scriptsize{$1.08 \pm 0.10$}		&\hspace{0.5cm}  \scriptsize{$$}\\	
&   		 	        & \scriptsize{$t_{\rm f_L, br}$ (s)}  &\hspace{0.5cm} \scriptsize{$106.5 \pm 0.08$}		&\hspace{0.5cm}  \scriptsize{$$}\\	
\cline{3-5}
&   		 	        &  \scriptsize{$\alpha_{L}$}  &\hspace{0.5cm} \scriptsize{$1.05 \pm 0.22$}		&\hspace{0.5cm}  \scriptsize{$1.38$}\\			        
\hline \hline
GRB 110731A & 	        &  \scriptsize{$\alpha_{\rm r_L, bb}$}  &\hspace{0.5cm} \scriptsize{$-1.50\pm 1.05$}		&\hspace{0.5cm}  \scriptsize{$1.34$}\\

&  	   \scriptsize{Short-lasting component}      &  \scriptsize{$\alpha_{\rm r_L, ab}$}  &\hspace{0.5cm} \scriptsize{$2.67 \pm 0.062$}		&\hspace{0.5cm}  \scriptsize{$$}\\ 
&           &  \scriptsize{$T_{\rm a}$ (s)}  &\hspace{0.5cm} \scriptsize{$2.63\pm 1.75$}		&\hspace{0.5cm}  \scriptsize{$$}\\ 
&                  	        &  \scriptsize{$t_{\rm r, pk}$ (s)}  &\hspace{0.5cm} \scriptsize{$5.62 \pm 0.03$}		&\hspace{0.5cm}  \scriptsize{$$}\\ 
\cline{2-5}            
&\scriptsize{}        &  \scriptsize{$\alpha_{L,1}$}  &\hspace{0.5cm} \scriptsize{$1.57 \pm 0.40$}		&\hspace{0.5cm}  \scriptsize{$0.79$}\\	
&  \scriptsize{Long-lasting component}  &  \scriptsize{$\alpha_{L,2}$}  &\hspace{0.5cm} \scriptsize{$2.04 \pm 1.34$}		&\hspace{0.5cm}  \scriptsize{$$}\\	
&   		 	        & \scriptsize{$t_{\rm L, br}$ (s)}  &\hspace{0.5cm} \scriptsize{$71.64 \pm 0.54$}		&\hspace{0.5cm}  \scriptsize{$$}\\	
\cline{3-5}
&   		 	        &  \scriptsize{$\alpha_{L}$}  &\hspace{0.5cm} \scriptsize{$1.14 \pm 0.23$}		&\hspace{0.5cm}  \scriptsize{$1.76$}\\			        
\hline \hline
GRB 130427A & 	        &  \scriptsize{$\alpha_{\rm r_L, bb}$}  &\hspace{0.5cm} \scriptsize{$-0.71\pm 0.28$}		&\hspace{0.5cm}  \scriptsize{$1.07$}\\

&  	   \scriptsize{Short-lasting component}      &  \scriptsize{$\alpha_{\rm r_L, ab}$}  &\hspace{0.5cm} \scriptsize{$2.00\pm 0.60$}		&\hspace{0.5cm}  \scriptsize{$$}\\ 
&           &  \scriptsize{$T_{\rm a}$ (s)}  &\hspace{0.5cm} \scriptsize{$1.01 \pm 0.76$}		&\hspace{0.5cm}  \scriptsize{$$}\\ 
&                  	        &  \scriptsize{$t_{\rm r, pk}$ (s)}  &\hspace{0.5cm} \scriptsize{$19.78 \pm 0.01$}		&\hspace{0.5cm}  \scriptsize{$$}\\ 
\cline{2-5}            
&\scriptsize{}        &  \scriptsize{$\alpha_{L,1}$}  &\hspace{0.5cm} \scriptsize{$0.81 \pm 0.05$}		&\hspace{0.5cm}  \scriptsize{$0.89$}\\	
&  \scriptsize{Long-lasting component}  &  \scriptsize{$\alpha_{L,2}$}  &\hspace{0.5cm} \scriptsize{$1.67 \pm 0.52$}		&\hspace{0.5cm}  \scriptsize{$$}\\	
&   		 	        & \scriptsize{$t_{\rm f_L, br}$ (s)}  &\hspace{0.5cm} \scriptsize{$1182.00 \pm 1071.00$}		&\hspace{0.5cm}  \scriptsize{$$}\\	
\cline{3-5}
&   		 	        &  \scriptsize{$\alpha_{L}$}  &\hspace{0.5cm} \scriptsize{$1.27 \pm 0.20$}		&\hspace{0.5cm}  \scriptsize{$1.08$}\\			        
\hline \hline
GRB 160625B & 	        &  \scriptsize{$\alpha_{\rm r_L, bb}$}  &\hspace{0.5cm} \scriptsize{$-0.94\pm 0.45$}		&\hspace{0.5cm}  \scriptsize{$0.87$}\\

&  	   \scriptsize{Short-lasting component}      &  \scriptsize{$\alpha_{\rm r_L, ab}$}  &\hspace{0.5cm} \scriptsize{$2.50\pm 1.73$}		&\hspace{0.5cm}  \scriptsize{$$}\\ 
&           &  \scriptsize{$T_{\rm a}$ (s)}  &\hspace{0.5cm} \scriptsize{$63.20 \pm 11.96$}		&\hspace{0.5cm}  \scriptsize{$$}\\ 
&                  	        &  \scriptsize{$t_{\rm r, pk}$ (s)}  &\hspace{0.5cm} \scriptsize{$130.20 \pm 18.96$}		&\hspace{0.5cm}  \scriptsize{$$}\\ 
\hline \hline

GRB 180720B & 	        &  \scriptsize{$\alpha_{\rm r_L, bb}$}  &\hspace{0.5cm} \scriptsize{$-0.71\pm 0.28$}		&\hspace{0.5cm}  \scriptsize{$1.15$}\\

&  	   \scriptsize{Short-lasting component}      &  \scriptsize{$\alpha_{\rm r_L, ab}$}  &\hspace{0.5cm} \scriptsize{$1.92\pm 0.20$}		&\hspace{0.5cm}  \scriptsize{$$}\\ 
&           &  \scriptsize{$T_{\rm a}$ (s)}  &\hspace{0.5cm} \scriptsize{$31.03 \pm 4.16$}		&\hspace{0.5cm}  \scriptsize{$$}\\ 
&                  	        &  \scriptsize{$t_{\rm r, pk}$ (s)}  &\hspace{0.5cm} \scriptsize{$65.78 \pm 0.21$}		&\hspace{0.5cm}  \scriptsize{$$}\\ 
\cline{3-5}            
&  	   \scriptsize{}      &  \scriptsize{$\alpha_{\rm r_L, ab,1}$}  &\hspace{0.5cm} \scriptsize{$1.420\pm 0.60$}		&\hspace{0.5cm}  \scriptsize{$1.18$}\\ 
&  	   \scriptsize{}      &  \scriptsize{$\alpha_{\rm r_L, ab,2}$}  &\hspace{0.5cm} \scriptsize{$3.15\pm 0.60$}		&\hspace{0.5cm}  \scriptsize{$$}\\ 
&                  	        &  \scriptsize{$t_{\rm r, ab}$ (s)}  &\hspace{0.5cm} \scriptsize{$281.84 \pm 2.24$}		&\hspace{0.5cm}  \scriptsize{$$}\\ 
%
\hline \hline

\end{tabular}
\end{table}

\newpage
\begin{table}
\centering \renewcommand{\arraystretch}{1.2}\addtolength{\tabcolsep}{1pt}
\caption{The best-fit values found from modeling the X-ray and optical light curves using PL functions from our GRB sample.}
\label{Table5}
\begin{tabular}{l  c  c c  c }
 \hline \hline
Event & Band &  Parameter  &\hspace{0.5cm}  Best-fit value    & \hspace{0.5cm} $\chi^2$/ndf \\ 
\hline \hline
GRB 080916C &    X-rays     & $\alpha_{\rm X}$  &  \hspace{0.5cm} {$1.35 \pm 0.06$}     &  \hspace{0.5cm}  {$1.16$}\\
\cline{2-5}
            &    Optical    & {$\alpha_{\rm opt}$}  &  \hspace{0.5cm} {$1.41\pm 0.04$}      &  \hspace{0.5cm}  {$1.21$}\\      
\hline
\hline
GRB 090323 &    X-rays     & {$\alpha_{\rm X}$}  &  \hspace{0.5cm} {$1.58 \pm 0.08$}     &  \hspace{0.5cm}  {$1.08$}\\
\cline{2-5}
            &    Optical    & {$\alpha_{\rm opt}$}  &  \hspace{0.5cm} {$1.70\pm 0.04$}      &  \hspace{0.5cm}  {$1.15$}\\
\hline
\hline            

GRB 090902B &    X-rays     & {$\alpha_{\rm X}$}  &  \hspace{0.5cm} {$1.62\pm 0.15$}     &  \hspace{0.5cm}  {$0.91$}\\
\cline{2-5}
            &    Optical    & {$\alpha_{\rm opt}$}  &  \hspace{0.5cm} {$1.70\pm0.04$}      &  \hspace{0.5cm}  {$1.23$}\\

\hline
\hline 
GRB 090926A   &    X-Rays  & {$\alpha_{\rm X}$}  &  \hspace{0.5cm} {$1.41\pm 0.03$}     &  \hspace{0.5cm}  {$1.47$}\\
\cline{2-5}
              &    Optical  & {$\alpha_{\rm opt}$}  &  \hspace{0.5cm} {$1.34\pm 0.06$}     &  \hspace{0.5cm}  {$1.31$}\\
\hline
\hline
GRB 110731A   &    X-Rays  & {$\alpha_{\rm X,1}$}  &  \hspace{0.5cm} {$1.15\pm 0.02$}     &  \hspace{0.5cm}  {$1.61$}\\
               &           & {$\alpha_{\rm X,2}$}  &  \hspace{0.5cm} {$1.30\pm 0.03$}     &  \hspace{0.5cm}  {$$}\\
              &            & {$t_{\rm f_X,br}$ (s)}  &  \hspace{0.5cm} {$(3.00 \pm 0.98)\times10^{3}$}     &  \hspace{0.5cm}  {$$}\\               
\cline{2-5}
            &    Optical    & {$\alpha_{\rm opt}$}  &  \hspace{0.5cm} {$1.38\pm 0.09$}      &  \hspace{0.5cm}  {$1.54$}\\
\hline
\hline
GRB 130427A  &    X-Rays  & {$\alpha_{\rm X,1}$}  &  \hspace{0.5cm} {$1.23\pm 0.04$}     &  \hspace{0.5cm}  {$2.23$}\\
               &           & {$\alpha_{\rm X,2}$}  &  \hspace{0.5cm} {$1.30\pm 0.05$}     &  \hspace{0.5cm}  {$$}\\
              &            & {$t_{\rm f_X,br}$ (s)}  &  \hspace{0.5cm} {$(2.36 \pm 0.72)\times10^{4}$}     &  \hspace{0.5cm}  {$$}\\
\cline{2-5}
              &    Optical    & {$\alpha_{\rm opt,1}$}  &  \hspace{0.5cm} {$1.16\pm 0.12$}      &  \hspace{0.5cm}  {$1.63$}\\
              &               & {$\alpha_{\rm opt,2}$}  &  \hspace{0.5cm} {$0.89\pm 0.01$}      &  \hspace{0.5cm}  {$$}\\              
              &               & {$\alpha_{\rm opt,3}$}  &  \hspace{0.5cm} {$1.33\pm 0.04$}      &  \hspace{0.5cm}  {$$}\\
              &            & {$t_{\rm f_O,br_1}$ (s)}  &  \hspace{0.5cm} {$(1.78 \pm 0.02)\times10^{3}$}     &  \hspace{0.5cm}  {$$}\\              
              &            & {$t_{\rm f_O,br_2}$ (s)}  &  \hspace{0.5cm} {$(4.50 \pm 0.01)\times10^{4}$}     &  \hspace{0.5cm}  {$$}\\              
\hline
\hline
GRB 160625B   &    X-Rays  & {$\alpha_{\rm X,1}$}  &  \hspace{0.5cm} {$1.15\pm 0.13$}     &  \hspace{0.5cm}  {$1.65$}\\
               &           & {$\alpha_{\rm X,2}$}  &  \hspace{0.5cm} {$1.27\pm 0.30$}     &  \hspace{0.5cm}  {$$}\\
              &            & {$t_{\rm f_X,br}$ (s)}  &  \hspace{0.5cm} {$(3.12 \pm 0.91)\times10^{3}$}     &  \hspace{0.5cm}  {$$}\\               
\cline{2-5}

            &    Optical    & {$\alpha_{\rm opt}$}  &  \hspace{0.5cm} {$0.921\pm 0.163$}      &  \hspace{0.5cm}  {$1.21$}\\
\hline
\hline
GRB 180720B   &    X-Rays  & {$\alpha_{\rm X,1}$}  &  \hspace{0.5cm} {$0.79\pm 0.08$}     &  \hspace{0.5cm}  {$0.92$}\\
               &           & {$\alpha_{\rm X,2}$}  &  \hspace{0.5cm} {$1.26\pm 0.06$}     &  \hspace{0.5cm}  {$$}\\
              &            & {$t_{\rm f_X,br}$ (s)}  &  \hspace{0.5cm} {$(2.36 \pm 0.72)\times10^{3}$}     &  \hspace{0.5cm}  {$$}\\
\cline{2-5}
              &    Optical    & {$\alpha_{\rm opt,1}$}  &  \hspace{0.5cm} {$1.16\pm 0.12$}      &  \hspace{0.5cm}  {$1.62$}\\
              &               & {$\alpha_{\rm opt,2}$}  &  \hspace{0.5cm} {$0.89\pm 0.01$}      &  \hspace{0.5cm}  {$$}\\              
              &               & {$\alpha_{\rm opt,3}$}  &  \hspace{0.5cm} {$1.22\pm 0.02$}      &  \hspace{0.5cm}  {$$}\\
              &            & {$t_{\rm f_O, br_1}$ (s)}  &  \hspace{0.5cm} {$(1.78 \pm 0.02)\times10^{3}$}     &  \hspace{0.5cm}  {$$}\\              
              &            & {$t_{\rm f_O, br_2}$ (s)}  &  \hspace{0.5cm} {$(4.50 \pm 0.01)\times10^{4}$}     &  \hspace{0.5cm}  {$$}\\              
\hline
\hline
\end{tabular}
\end{table}

\newpage
\begin{table}
\centering \renewcommand{\arraystretch}{1.6}\addtolength{\tabcolsep}{-1pt}
\caption{Median values of parameters of our GRB sample with symmetrical quantiles. The SSC and synchrotron models from reverse and forward region, respectively are used to constrain the parameters.}
\label{Table6}
\begin{tabular}{ l c c c c c c c c}
\hline
\hline
{\large  GRB }	& {080916C} & {090323} &  {090902B} & {090926A} & {110731A} & {130427A} & {160625B} & {180720B}   		 \\ 
%
\hline \hline
\\
\small{$\rm{log_{10}}(E/{\rm erg})$}	\hspace{1.5cm}&      \small{$54.85_{-1.02}^{+1.05}$} & \small{$53.96_{-1.05}^{+1.01}$} &    \small{$54.53_{-1.03}^{+1.01}$} &     \small{$53.83_{-0.99}^{+1.03}$}&     \small{$53.87_{-1.06}^{+0.95}$}&     \small{$54.14_{-1.03}^{+1.00}$} &     \small{$54.29_{-1.00}^{+1.03}$}  &    \small{$54.36_{-0.95}^{+1.00}$} \\

\small{$\rm{log_{10}}(A_{\rm k})\, $}	\hspace{1.5cm}&      \small{$-2.83_{-0.96}^{+1.13}$}& \small{$-2.29_{-1.06}^{+1.05}$}&     \small{$-2.82_{-1.01}^{+1.01}$}&     \small{$-2.21_{-0.99}^{+1.02}$}&     \small{$-1.77_{-1.06}^{+0.84}$}&     \small{$-1.96_{-1.01}^{+0.92}$}&     \small{$-2.19_{-1.15}^{+1.10}$}	&     \small{$-1.68_{-0.98}^{+1.06}$} \\

\small{$\rm{log_{10}}(\Gamma)$}	\hspace{1.5cm}&     \small{$2.80_{-0.98}^{+1.09}$} & \small{$2.92_{-0.96}^{+0.99}$}&   \small{$2.94_{-0.99}^{+1.09}$}&     \small{$2.92_{-0.85}^{+1.00}$}&     \small{$2.54_{-0.96}^{+0.87}$}&     \small{$2.52_{-1.04}^{+1.05}$} &     \small{$2.46_{-0.91}^{+1.06}$} &     \small{$2.46_{-0.99}^{+0.89}$}	\\

\small{$p$}	\hspace{1.5cm}&     \small{$2.22_{-1.01}^{+0.92}$}& 
\small{$2.68_{-0.98}^{+1.08}$}&   \small{$2.27_{-1.02}^{+1.03}$}&     \small{$2.50_{-0.98}^{+1.14}$}&     \small{$2.35_{-1.08}^{+1.04}$}&     \small{$2.10_{-1.00}^{+0.98}$} &     \small{$2.16_{-1.10}^{+0.96}$} &     \small{$2.09_{-0.93}^{+1.03}$}	\\

\small{$\rm{log_{10}}(\varepsilon_{\rm B_f})$}	\hspace{1.5cm}&     \small{$-1.54_{-1.09}^{+1.07}$}&  \small{$-2.13_{-0.96}^{+1.03}$}&     \small{$-2.05_{-1.16}^{+1.05}$}&     \small{$-1.86_{-1.08}^{+0.91}$}&     \small{$-2.20_{-1.05}^{+1.02}$}&     \small{$-2.64_{-1.08}^{+1.09}$} &     \small{$-2.36_{-1.02}^{+0.95}$}	&     \small{$-2.22_{-1.05}^{+1.04}$}  \\

\small{$\rm{log_{10}}(\varepsilon_{\rm B_r})$}	\hspace{1.5cm}&      \small{$-1.32_{-0.88}^{+0.95}$}& \small{$-1.03_{-1.01}^{+0.94}$}&   \small{$-0.96_{-0.98}^{+1.04}$}&     \small{$-0.98_{-0.99}^{+1.00}$}&     \small{$-1.99_{-1.05}^{+1.04}$}&     \small{$-1.48_{-1.07}^{+1.00}$}&     \small{$-1.06_{-1.06}^{+0.98}$}	&     \small{$-1.54_{-1.10}^{+1.01}$}  \\

\small{$\rm{log_{10}}(\varepsilon_{\rm e_f})$}	\hspace{1.5cm}&     \small{$-1.84_{-1.04}^{+0.93}$}&  \small{$-1.24_{-1.03}^{+1.11}$}&    \small{$-1.38_{-1.03}^{+0.94}$}&     \small{$-1.15_{-0.99}^{+0.91}$}&     \small{$-2.55_{-0.94}^{+0.85}$}&     \small{$-1.76_{-0.96}^{+0.95}$} &     \small{$-0.92_{-1.02}^{+0.98}$}	&     \small{$-2.02_{-0.92}^{+0.97}$}	 \\

\small{$\rm{log_{10}}(\varepsilon_{\rm e_r})$}	\hspace{1.5cm}&      \small{$-1.25_{-1.01}^{+1.11}$}& \small{$-1.11_{-1.04}^{+1.01}$}&    \small{$-1.10_{-0.99}^{+1.13}$}&     \small{$-1.04_{-0.92}^{+1.07}$}&     \small{$-0.89_{-1.06}^{+0.93}$}&     \small{$-1.44_{-0.98}^{+0.90}$}&     \small{$-1.11_{-0.91}^{+1.05}$}	&     \small{$-1.51_{-0.92}^{+0.96}$}	 \\

\small{$\rm{log_{10}}(\zeta_{e})$}	\hspace{1.5cm}&     \small{$-1.81_{-1.09}^{+0.97}$}&  \small{$-0.91_{-1.02}^{+1.00}$}&     \small{$-0.47_{-0.93}^{+1.00}$}&     \small{$-0.90_{-1.11}^{+1.00}$}&     \small{$-0.16_{-1.10}^{+1.02}$}&     \small{$-0.81_{-0.97}^{+0.94}$}&     \small{$-1.20_{-0.98}^{+1.07}$}	&     \small{$-0.33_{-0.91}^{+1.02}$}	 \\

\hline
\end{tabular}
\end{table}

\begin{table}
\centering \renewcommand{\arraystretch}{1.6}\addtolength{\tabcolsep}{-1pt}
\caption{Derived quantities from the best-fit parameter values of the reverse shock.}
\label{Table3}
\begin{tabular}{ l c c c c c c c c}
\hline
\hline
{\large  GRB }	& {080916C} & {090323} &  {090902B} & {090926A} & {110731A} & {130427A} & {160625B} & {180720B}   		 \\ 
%
\hline \hline

& & & & $(t< t_{\rm x})$ & & & & \\


$h\nu^{\rm ssc}_{\rm m,r}\,({\rm~GeV})$	\hspace{1.5cm}&      \small{$7.7$} & \small{$4.1 \times 10^2$} &    \small{$1.8 \times 10^{-1}$} &     \small{$5.9 \times 10^3$}&     \small{$1.1 \times 10^{-8}$}&     \small{$9.1 \times 10^{-8}$} &     \small{$5.8 \times 10^{-4}$}  &    \small{$2.9 \times 10^{-7}$} \\
$h\nu^{\rm ssc}_{\rm c,r}\,({\rm~GeV})$	\hspace{1.5cm}&      \small{$9.3\times 10^{-2}$}& \small{$1.2\times 10^{-3}$}&  \small{$2.9 \times 10^{-5}$}& \small{$2.2 \times 10^{-6}$}&     \small{$1.1$}&     \small{$5.5 \times 10^{-6}$}&     \small{$1.8 \times 10^{-2}$}	&     \small{$1.1 \times 10^{-4}$} \\
$F^{\rm ssc}_{\rm max,r}\,({\rm mJy})$	\hspace{1.5cm}&      \small{$1.6\times 10^{-3}$} & \small{$2.0\times 10^{-4}$} &    \small{$1.3 \times 10^{-1}$} &     \small{$2.3 \times 10^{-2}$}&     \small{$3.6 \times 10^{-1}$}&     \small{$2.7 \times 10$} &     \small{$4.7 \times 10^{-3}$}  &    \small{$3.7$} \\
$h\nu^{\rm ssc}_{\rm KN, m,r}\,({\rm~GeV})$	\hspace{1.5cm}&      \small{$4.8\times 10^2$} & \small{$7.2 \times 10^2$} &    \small{$9.9 \times 10$} &     \small{$1.0 \times 10^3$}&     \small{$8.2$}&     \small{$3.2$} &     \small{$2.8 \times 10$}  &    \small{$3.4$} \\
$h\nu^{\rm ssc}_{\rm KN, c,r}\,({\rm~GeV})$	\hspace{1.5cm}&      \small{$2.4\times 10^2$}& \small{$1.5\times 10$}&     \small{$1.2 \times 10$}&     \small{$4.5$}&     \small{$6.9 \times 10^2$}&     \small{$8.9$}&     \small{$6.7 \times 10$}	&     \small{$1.5 \times 10$} \\

\cline{1-1}

 & & & & $(t_{\rm x}< t)$ & & & & \\

$h\nu^{\rm ssc}_{\rm m,r}\,({\rm~GeV})$	\hspace{1.5cm}&      \small{$7.4\times 10^{-2}$} & \small{$6.0\times 10^2$} &    \small{$1.1 \times 10^{-1}$} &     \small{$5.8 \times 10^2$}&     \small{$1.1 \times 10^{-9}$}&     \small{$8.3 \times 10^{-9}$} &     \small{$7.3 \times 10^{-5}$}  &    \small{$4.0 \times 10^{-8}$} \\
$h\nu^{\rm ssc}_{\rm cut,r}\,({\rm~GeV})$	\hspace{1.5cm}&      \small{$1.6$}& \small{$8.1 \times 10^{-2}$}&     \small{$6.1 \times 10^{-3}$}&     \small{$8.8 \times 10^{-6}$}&     \small{$1.2 \times 10^2$}&     \small{$2.0 \times 10^{-4}$}&     \small{$1.2 \times 10^{-1}$}	&     \small{$1.2 \times 10^{-3}$} \\
$F^{\rm ssc}_{\rm max,r}\,({\rm mJy})$	\hspace{1.5cm}&      \small{$1.2\times 10^{-4}$} & \small{$5.3\times 10^{-5}$} &    \small{$2.1 \times 10^{-2}$} &     \small{$5.4 \times 10^{-3}$}&     \small{$8.3 \times 10^{-2}$}&     \small{$4.4$} &     \small{$4.7 \times 10^{-3}$}  &    \small{$9.7 \times 10^{-1}$} \\
$h\nu^{\rm ssc}_{\rm KN, m,r}\,({\rm~GeV})$	\hspace{1.5cm}&      \small{$1.4\times 10^2$} & \small{$6.3 \times 10^2$} &    \small{$9.5 \times 10$} &     \small{$6.1 \times 10^2$}&     \small{$1.8$}&     \small{$2.3$} &     \small{$2.8 \times 10$}  &    \small{$2.3$} \\
$h\nu^{\rm ssc}_{\rm KN, c,r}\,({\rm~GeV})$	\hspace{1.5cm}&      \small{$8.8\times 10^2$}& \small{$3.8 \times 10$}&     \small{$7.7 \times 10$}&     \small{$1.1 \times 10$}&     \small{$2.6 \times 10^3$}&     \small{$3.8 \times 10$}&     \small{$6.7 \times 10$}	&     \small{$4.2 \times 10$} \\

\hline
\end{tabular}
\end{table}
%


%
%
%



\bsp	
\label{lastpage}
\end{document}